\documentclass[review]{elsarticle}

\usepackage[a4paper, total={6in, 9in}]{geometry}

\usepackage{lineno,hyperref}

\usepackage{lipsum}
\makeatletter
\def\ps@pprintTitle{%
 \let\@oddhead\@empty
 \let\@evenhead\@empty
 \def\@oddfoot{}
 \let\@evenfoot\@oddfoot}

\usepackage{pdflscape}

\usepackage{chngcntr}

\usepackage[usenames] {color}
\definecolor{webblue} {rgb} {0,0,.5}

\usepackage{amsmath,amsthm,amssymb}
\usepackage{gensymb}

\usepackage{graphics,graphicx,float}
\usepackage{epsfig}
\graphicspath{{/Users/aam58/PROYECTOS_FINALIZADOS/Jupiter_radiance_at_5microns/Jupiter_brightness_variability/Brightness_variability_paper/}}

\usepackage[font=footnotesize,labelfont=bf] {caption}
\usepackage{multirow}
\usepackage{array}
\usepackage{xtab}
\usepackage{tabularx}

\usepackage{setspace}

\usepackage{makeidx}
\makeindex

\usepackage{natbib}

\biboptions{authoryear}

\begin{document}

\begin{frontmatter}

\title{Jupiter's Atmospheric Variability from Long-Term Ground-Based Observations at 5 microns}

\author[*,a]{Arrate Antu\~{n}ano\corref{cor1}}
 \ead{aam58@leicester.ac.uk}
 \author[a]{Leigh N. Fletcher}
 \author[b]{Glenn S. Orton}
 \author[a]{Henrik Melin}
  \author[a]{Steve Milan}
 \author[c]{John Rogers}
 \author[d]{Thomas Greathouse}
 \author[e]{Joseph Harrington}
 \author[a]{Padraig T. Donnelly}
  \author[d]{Rohini Giles}
  
\address[a]{Department of Physics \& Astronomy, University of Leicester, University Road, Leicester LE1 7RH, UK}
\address[b]{Jet Propulsion Laboratory, California Institute of Technology, 4800 Oak Grove Drive, Pasadena, CA 91109, USA}
\address[c]{British Astronomical Association, London, UK}
\address[d]{Department of Space Science, Southwest Research Institute, San Antonio, TX 78228, USA}
\address[e]{Planetary Sciences Group, Department of Physics, University of Central Florida, Orlando, FL 32816-2385, USA}


\begin{abstract}
Jupiter's banded structure undergoes strong temporal variations, changing the visible and infrared appearance of the belts and zones in a complex and turbulent way due to physical processes that are not yet understood. In this study we use ground-based 5-$\mu$m infrared data captured between 1984 and 2018 by 8 different instruments mounted on the Infrared Telescope Facility in Hawai'i and on the Very Large Telescope in Chile to analyze and characterize the long-term variability of Jupiter's cloud-forming region at the 1-4 bar pressure level. The data show a large temporal variability mainly at the equatorial and tropical latitudes, with a smaller temporal variability at mid-latitudes. We also compare the 5-$\mu$m-bright and -dark regions with the locations of the visible zones and belts and we find that these regions are not always co-located, specially in the southern hemisphere. We also present Lomb-Scargle and Wavelet Transform analyzes in order to look for possible periodicities of the brightness changes that could help us understand their origin and predict future events. We see that some of these variations occur periodically in time intervals of 4-8 years. The reasons of these time intervals are not understood and we explore potential connections to both convective processes in the deeper weather layer and dynamical processes in the upper troposphere and stratosphere. Finally we perform a Principal Component analysis to reveal a clear anticorrelation on the 5-$\mu$m brightness changes between the North Equatorial Belt and the South Equatorial Belt, suggesting a possible connection between the changes in these belts.\\
\end{abstract}

\begin{keyword}
Jupiter, Jupiter's atmosphere, Infrared ground-based data\\

\end{keyword}

\end{frontmatter}


\section{Introduction}

Jupiter's dominant appearance at visible wavelengths is a banded structure formed by low-albedo brownish belts and high-albedo whitish zones alternating in latitude. This banded structure, which appears to be bounded by the eastward and westward zonal tropospheric jets on Jupiter, undergoes dramatic planetary-scale variations over short-time scales, where the belts can expand, contract, or disappear entirely in a complex and turbulent way \citep[e.g.][]{Rogers_1995, Sanchez_Lavega_1996, Fletcher_2017a}. During these events, Jupiter's upper tropospheric temperatures, aerosols and cloud structures change due to physical processes that are not yet well understood.  \\

Some of these variations occur randomly, like the impressive changes observed at the Southern Equatorial Belt (SEB), located between 7$^\circ$ S and 17$^\circ$ S planetocentric latitude (all latitudes in this study are planetocentric). This belt changes from being the darkest jovian belt at visible wavelengths (bright at 5 $\mu$m, sensitive to thermal emission from Jupiter's 1-4 bar region) to a whitish (5-$\mu$m-dark) zone-like band in a timescale of months \citep{Peek_1958, Sanchez_Lavega_1989, Rogers_1995, Sanchez_Lavega_1996, Fletcher_2011, Perez_Hoyos_2012}. This SEB fade and revival cycle, the last one occurring in 2009-2011 \citep{Fletcher_2011, Perez_Hoyos_2012, Fletcher_2017}, follows a repeatable pattern: first, a cessation of the turbulent rifting and convective events located northwest of the Great Red Spot (GRS) gives way to the formation of a cloud-covered whitish band (dark at 5 $\mu$m) at the SEB (at $\sim$12$^\circ$ S) fading this belt. After 1 to 3 years, a sudden and violent eruption of a white (high-opacity) plume at the SEB starts the revival process \citep{Sanchez_Lavega_1996, Rogers_2003, Simon_Miller_2012}. Finally, a series of convective plumes, all apparently emanating from the same source region, encircle the planet restoring the brown color of the SEB. The temporal intervals between the revival and the next fading event are unpredictable, but sometimes could be as short as 3 years \citep{Rogers_1995, Rogers_2017b, Rogers_2017c}. \\

A similar process has also been observed at the fastest jovian jet at $\sim$21$^\circ$ N in the southern edge of the North Temperate Belt (NTBs) since the early 20\textsuperscript{th} century \citep{Peek_1958, Rogers_1995, Sanchez_Lavega_1991, Sanchez_Lavega_2008, Sanchez_Lavega_2017}, where an eruption of a high-albedo plume or plumes \citep{Rogers_1995, Sanchez_Lavega_2017} generates  a complex pattern of low- and high-albedo features encircling the planet and finally, forming a new NTB south belt (NTBS) \citep{Sanchez_Lavega_2017}. However, unlike the variations observed at the SEB, these NTB outbreaks tend to occur quasi-periodically at intervals of $\sim$5 years \citep{Rogers_1995}, the last event occurring in 2016 \citep{Sanchez_Lavega_2017}. Additionally, the NTB is also observed to move northward, creating a new North Temperate Zone Belt (NTZB) \citep{Rogers_1995}, every $\sim$10 years \citep{Rogers_1995}. \\

Similarly, the North Equatorial Belt (NEB), usually located between $\sim$7$^\circ$ N and $\sim$17$^\circ$ N, and the Equatorial Zone (EZ) at  $\sim$7$^\circ$ S - $\sim$7$^\circ$ N, also undergo periodic/cyclic variations. The NEB is observed to expand and contract in latitude every 3-5 years since 1987 \citep{Rogers_1995, Simon_Miller_2001, Garcia_Melendo_2001, Rogers_2004, Rogers_2013, Rogers_2017, Fletcher_2017b}, usually due to decreases on the albedo of the southern edge of the North Tropical Zone (NTrZ (S)) at $\sim$18 - 21$^\circ$ N \citep{Rogers_1995, Fletcher_2017b}. The EZ, however, changes completely its appearance, both at visible and 5 $\mu$m wavelength, every 6-8 or 13-14 years, changing from a visibly white to a darker and/or strongly colored state, which overlaps with a change at 5 $\mu$m from a cloud-covered to a cloud-free region on time-scales of months \citep{Antunano_2018}. \\

So far, the origin and nature of these changes, as well as their relationship to the three-dimensional atmospheric dynamics are not well understood. Additionally, each of the studies described above have dealt with snapshots of observations at one epoch in time, usually as a result of an interesting or unexpected event. Here we seek to analyze these changes in a systematic fashion, testing the robustness of the periodicities cited in the literature. In this study, we use a large dataset of ground-based observations at infrared wavelengths spanning around 4 decades (1984-2018) to analyze the relationship between the belt/zone structure in the visible and 5 $\mu$m wavelength (section 3) and to study the temporal brightness variability of Jupiter's belts and zones at 5 $\mu$m between  $\pm$50$^\circ$ latitude (section 4 and 5). This is the first time that all the available ground-based 5 $\mu$m data is used to analyze Jupiter's long-term variability at these latitudes, as \cite{Antunano_2018} focused on the Equatorial Zone. The timescales for changes are studied via both a Lomb-Scargle analysis and Wavelet Transform analysis (section 6), which could allow us predict future atmospheric changes in Jupiter's belts/zone structure, essential for current and future missions. Finally, we also compute a Principal Component Analysis to analyze the correlation of the 5 $\mu$m brightness variations at different belts and zones (section 6). Such correlations have been suggested in the literature, and are known as 'global upheavals', suggesting that there is a predictability to the sequence of events happening in very different locations on the planet.  We seek to test that suggestion with our data in a systematic fashion. We summarize the conclusions in section 7. 

\section{Observations and Methodology}

\subsection{Observations} \index{Observations} \indent

In this study, we use images captured between June 1984 and July 2018 at wavelengths between 4.76 and 5.18 $\mu$m by 8 different ground-based instruments, mounted at the Infrared Telescope Facility (IRTF) on Maunakea, Hawai'i, and at the Very Large Telescope (VLT) in Chile. The 5-$\mu$m radiation, originated in Jupiter's tropospheric cloud decks, mainly reveals the thermal emission from 1-4 bar level (below the ammonia ice clouds). The use of a single narrow-band filter does not allow us to distinguish reflected sunlight from thermal emission, however, the reflected sunlight makes only a small contribution to the 5-$\mu$m radiance \citep{Giles_2015}. At this wavelength cloudy regions appear dark at 5 $\mu$m, while cloud-free regions appear bright, due to the high aerosol opacities blocking the light. A summary of the dataset used in shown in Table \ref{tab:dataset} and a brief description of the eight instruments used is given below:

\begin{itemize}

\item  \textbf{BOLO-1}: The 2-30 $\mu$m Bolometer ("BOLO") mounted at the Infrared Telescope Facility (IRTF) on Maunakea, Hawaii, provided observations of Jupiter between 1980 and 1993, first by central meridian scans and later by raster scans, sampling at 1 arcsecond intervals in right ascension and declination \citep{Orton_1994}.  In this study, M-band filter (4.80 $\mu$m) data covering Jupiter's full disk is used to expand Jupiter's brightness variability analysis to the 1980s (June 1984 - February 1991). These data have not previously been studied and we followed the same techniques as in \cite {Orton_1991} (for the stratosphere) and \cite{Orton_1994} (for the troposphere).

\item  \textbf{ProtoCam}: The infrared array camera ProtoCam was a 62x58 pixel InSb camera \citep{Toomey_1990} mounted at the 3-meter Infrared Telescope Facility in Hawaii, sensitive to wavelengths between 1 $\mu$m and 5 $\mu$m. Its plate scale varied between 0.14 and 0.35 arc seconds and due to the low number of pixels on the CCD, images of Jupiter's whole disk had to be acquired by mosaicking \citep{Harrington_1996, Harrington_1996b}. Mosaicked ProtoCam data captured at 4.85 $\mu$m from January-April 1992 \citep{Harrington_1996, Harrington_1996b} are used in this study to analyze, in particular, Jupiter's Equatorial Zone disturbance \citep{Antunano_2018}.

\item \textbf{NSFCam and NSFCam2}: The NSFCam \citep{Shure_1994} was a 1-5.5 $\mu$m camera with a 256x256 InSb detector mounted at the IRTF between September 1993 and April 2004. This camera captured Jupiter images with a very high temporal resolution, enabling us to analyze the brightness variability at the cloud-forming region in 4.78 and 4.85 $ \mu$m in detail between April 1994 and February 2004, with pixel scales of 0.15" and a field of view of 37.9" (smaller than Jupiter's disk at opposition). In 2004, the NSFCam imager was upgraded with a 2048x2048 Hawaii 2RG detector array, giving way to NSFCam2. The pixel scale of the NSFCam2 is 0.04"/pixel with a field of view of 81x81", providing higher spatial resolution images of Jupiter's full disk. In this study, NSFCam2 images captured at 4.78 $\mu$m between April 2006 and October 2008 are used. These two cameras, like SpeX, MIRSI and VISIR instruments described below, captured data in a chopping plus nodding mode, where at each nod position a chop of the secondary mirror by a few arcseconds is performed in order to subtract the background and sky emission and detect Jupiter's emission \citep{Fletcher_2009}.  

\item \textbf{MIRSI}: The Mid-Infrared Imager and Spectrometer (MIRSI) \citep{Deutsch_2003} is a 2-20 $\mu$m camera and grism spectrograph with a 320x240 Si:As array detector. Its plate scale of 0.27"/pixel provides a field of view of 85x64", where Jupiter's full disk can be captured albeit with a lower spatial resolution than data obtained with NSFCam, NSFCam2, SpeX, TEXES and VISIR. In this study, MIRSI images captured between January 2005 and April 2006 at 4.9 $\mu$m are essential to be able to fill the temporal gap between the higher resolution images captured by NSFCam and NSFCam2.  

\item \textbf{SpeX}:  The SpeX 0.7-5.3 $\mu$m spectrograph \citep{Rayner_2003} mounted at the Infrared Telescope Facility in Hawaii since May 2000, is used in this study for images captured at 4.76 and 5.1 $\mu$m between July 2009 and July 2018. This guide-camera, which was upgraded in 2014, used a Raytheon Aladdin 3 1024x1024 InSb array in the spectrograph with a pixel scale of 0.15" before the upgrade and a Teledyne 2048x2048 Hawaii-2RG array with spectrograph pixel scale of 0.12" after the upgrade. 

\item \textbf{TEXES}: The Texas Echelon Cross Echelle Spectrograph (TEXES, \cite{Lacy_2002}) is a cross-dispersed grating spectrograph mounted on the IRTF able to obtain spectra between 5 and 24 $\mu$m. In this study, we use data captured at the M band (5 $\mu$m) between December 2014 and July 2018 using the medium (R $\sim$15000) and low (R $\sim$2000) spectral resolutions. Unlike the rest of the instruments' data used in this study, these maps are produced by scanning the TEXES slit repeatedly across Jupiter's disk. Detailed technical information of the instrument and the observations can be found in \cite{Fletcher_2016}.
    
\item \textbf{VISIR}: The Very Large Telescope (VLT) Imager and Spectrometer for the mid-infrared (VISIR, \cite{Lagage_2004}) provides diffraction-limited imaging between 5 and 20 $\mu$m with a pixel-size of 0.0453"/pixel  and a 38x38" field of view, smaller than Jupiter's disc at opposition. In this study, we use the data captured at 5$\mu$m between December 2016 and July 2017 \citep{Fletcher_2018}.

\end{itemize}

\begin{table} [h!]
	\centering
	\begin{spacing}{0.9}
	\begin{tabular}{p{3.3cm} p{2cm} p{2.6cm} p{3.6cm}}
         &&\\
	\hline
        &&\\
	\textbf{Date} & \textbf{Instrument} &  \textbf{Configuration} &  \textbf{References}\\
	&&\\
	\hline
	&&\\
	Jun 1984 - Feb 1991 &   BOLO-1 &     Raster-scanned  & \cite{Orton_1994} \\
	Jan 1992 - Apr 1992  &   ProtoCam & Mosaicked & \cite{Harrington_1996, Harrington_1996b} \\
         Apr 1994 - Feb 2004  &  NSFCam  &  Full-frame & \cite{Ortiz_1998} \\
         Jan 2005 - Apr 2006  & MIRSI  &   Full-frame & \cite{Fletcher_2009} \\
         Apr 2006 - Oct 2008  &  NSFCam2 &   Full-frame & \cite{Fletcher_2010} \\
         Jul 2009 - Aug 2017 & SpeX &  Full-frame & - \\
         Dec 2014 - Jul 2018  & TEXES &  Scanned & \cite{Fletcher_2016} \\
         Dec 2016 - Jul 2017	& VISIR & Full-frame & \cite{Fletcher_2018} \\
        
         &&\\
	\hline
	
	\end{tabular}
	\begin{quote}
	\caption[Dataset]{Summary of the observations used in this study}
	\label{tab:dataset}
	\end{quote}
	
\end{spacing}
    \end{table}

The data mentioned above were acquired under a vast array of weather conditions. In order to omit the poor quality observations in our large dataset, we analyze each of the data carefully, removing those of particularly poor quality. When data were acquired under average seeing conditions of $\sim 1''$ , the minimum spatial resolution of our data on Jupiter's equator is $\sim 2850$ km (2.3$^\circ$ longitude) at opposition and $\sim$3800 km (3.1$^\circ$ longitude) at quadrature (corresponding to the 3m-IRTF data as the 8m-VLT data display a higher spatial-resolution), sufficient to perform detailed analysis of the temporal variability of the cloud morphology and dynamics of Jupiter's zone and belts.  \\    
    
Images captured with NSFCam, NSFCam2, MIRSI, SpeX and VISIR are reduced following \cite{Fletcher_2009}, by subtracting the sky emission using the chop-nodded images, flat fielding, navigating them by limb fitting and projecting the data in cylindrical maps. TEXES data are reduced using the Pipeline Data Reduction software \citep{Lacy_2002}, which performs the sky subtraction, flat fielding and radiometric calibration automatically. Information for the raster-scanning method used in BOLO-1 observations is described in \cite{Orton_1994}. Finally, we use the mosaicked ProtoCam data from \cite{Harrington_1996, Harrington_1996b}. We made no attempt to calibrate the data due to the very different atmospheric conditions of the observations made during almost 40 years, which often were carried out in the absence of a reliable standard star. However, a relative calibration, described in section 2.3, is performed in order to be able to compare the data from different epochs.  

\subsection{Longitudinally-averaged brightness} \index{Longitudinally-averaged brightness} \indent

When analyzing the temporal variability of the latitudinal brightness of Jupiter's atmosphere at 5 $\mu$m, it is essential to take into account that the brightness is not homogeneously distributed in longitude, varying with emission angle and displaying limb-darkening. Therefore, understanding the dependency of the observed brightness with the emission angle is essential to choose the right average value for each latitude and date. In order to do so, we follow \cite{Antunano_2018} and compute the average brightness by binning the brightness corresponding to emission angles smaller than 75$^\circ$ of all images captured in a single observation night, in latitudinal bins of 1$^\circ$ between $\pm$70$^\circ$ latitude. Brightnesses corresponding to larger emission angles are excluded from the zonal mean in order to avoid most of the effect of limb-darkening.  \\

This technique is then compared to the average brightness obtained by performing a second order polynomial fitting of the brightness vs emission angle at each latitude and date, where the average (fitted) brightness corresponds to the smallest emission angle (i.e. the brightness value less affected by the dependency of the emission angle). Once the average brightness is scaled to a quiescent region (see below), both techniques show a similar temporal variability of the brightness, with only small differences of less than 5$\%$ of their peak values.  \\ 

\subsection{Scaling} \index{Scaling} \indent

As mentioned above, due to the very different atmospheric conditions at which the data were captured and the large temporal coverage of the dataset used in this study, we did not calibrate the data before computing the average brightness. However, in order to be able to compare Jupiter's brightness at 5 $\mu$m from different epochs, a scaling of the average brightness to a quiescent latitudinal region must be performed \citep{Fletcher_2011, Fletcher_2017}. As Jupiter's zones are usually less variable than belts, they are better scaling regions, although the lower signal/noise ratio could be troublesome. Previous studies scaled Jupiter's brightness to the typically-dark Equatorial Zone (EZ) \citep{Fletcher_2011, Fletcher_2017}. However, the Equatorial Zone disturbance observed at 5 $\mu$m and reported by \cite{Antunano_2018}, completely changes the appearance of the EZ during these events, making this region unsuitable for this purpose. \\

In order to identify the optimal scaling region (i.e. the most quiescent zone on Jupiter during the epochs under study), we first omitted the data corresponding to the dates where an EZ disturbance is present (i.e. January-April 1992, August 1999-August 2000 and April 2006-August 2007). We then scale the average brightness to the temporal average brightness of the equator ($\pm$5$^\circ$ latitude of the equator, which is known to be a 5-$\mu$m-dark region away from the EZ disturbances), by applying the following equation:
 
 \begin{equation}
L_{\phi,t} = counts_{\phi,t}\frac{<d>}{d}
 \end{equation} 

\noindent where L$_{\phi,t}$ is the scaled brightness at each latitude $\phi$ and date $t$, counts$_{\phi,t}$ is the zonal mean brightness at each latitude and date, $d$ is averaged brightness over the range of latitudes $\pm 5^\circ$ and the $< >$ indicate the average over all the dates.\\

\begin{figure}[H]
	\centering
		\includegraphics[width=0.99\textwidth]{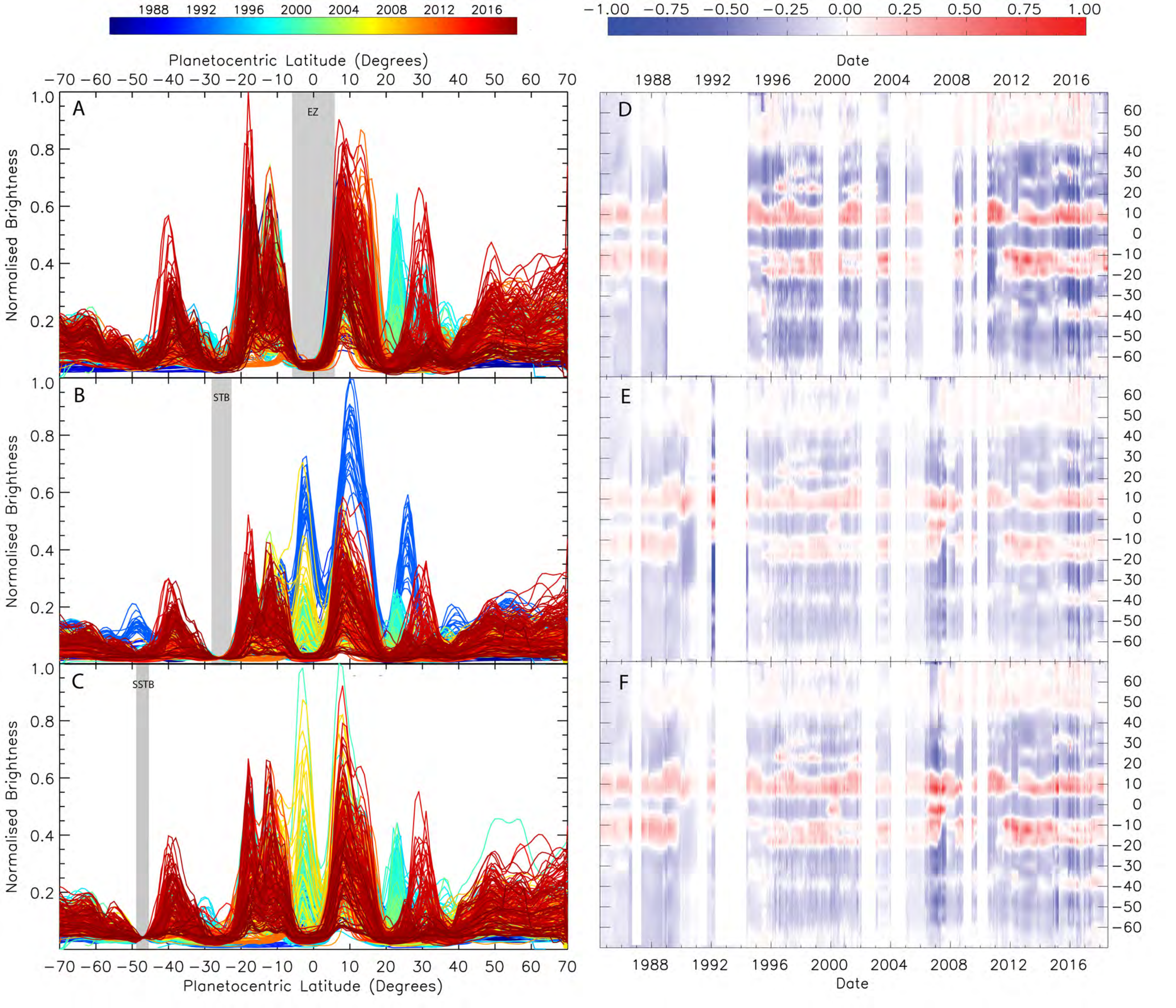}
	\begin{quote}
	\caption[Scaling Region]{Normalized average brightness scans at 5 $\mu$m between 1984 and 2018 as a function of latitude (A-C) and mean-centre brightness (see text in section 2.3) as a function of time (D-F), scaled at the equator between 7$^\circ$ S and 3$^\circ$ N (A, D), at the South Temperate Belt between 24$^\circ$ S and 28$^\circ$ S (B, E) and at the South South South Temperate Zone between $\sim 46^\circ$ S and $48^\circ$ S (C, F), showing the brightness variability of Jupiter's zones and belts relative to the scaling regions, which are assumed to stay invariant during the studied epoch. Different color lines in (A-C) correspond to different dates and the grey shadowed regions indicate the location of the scaling regions. The white region in (D-F) correspond to gaps in the sampling larger or equal to 9 months.}
	\label{fig:scaling}
	\index{Scaling Region}
	\end{quote}
\end{figure}

The results are shown in Figure \ref{fig:scaling}a, where the normalized average brightness is shown against latitude (different colors corresponding to different dates and the shadowed regions indicating Jupiter's zones) for the dates where no EZ disturbance at 5 $\mu$m was present. In Figure \ref{fig:scaling}d we show the normalized mean-centered brightness as a function of time, showing Jupiter's 5-$\mu$m appearance when scaling the zonally-averaged brightness to the equator. The normalized mean-centered brightness allows us to identify the 5-$\mu$m changes more clearly and it is computed by subtracting the mean of zonally-averaged brightness to the data and then normalizing the maximum value to 1 (-1 in the case of negative values). Here red represents 5-$\mu$m-bright regions (belts), while blue represents 5-$\mu$m-dark regions (zones). Figure \ref{fig:scaling}a demonstrates the key challenge of this study: that no zone/belt remained unchanged over the 34 years under study, making it difficult to choose the right scaling region. However, the South South South Temperate Zone (S3TZ) between $\sim 46-48^\circ$ S and the South Temperate Belt (STB) at $\sim 24-28^\circ$ S, underwent the smallest temporal variability, with a relative brightness variability smaller than 5$\%$ and 10$\%$, respectively. A careful analysis of various potential scaling regions (e.g. the S3TZ, the STB, mid-latitudes above 50$^\circ$ N and the North Equatorial Belt, among others) show that the belts, as well as the mid-latitudes above 50$^\circ$ N, are not good scaling regions as the obtained scaled normalized-brightness displays a strong dependence on these scaling regions, differing the scaled-normalized brightness completely from one case to the next. Therefore, here we analyze the S3TZ and the STB as potential scaling regions. \\

Figure \ref{fig:scaling}b and Figure \ref{fig:scaling}c show the normalized average brightness scaled to the STB and S3TZ, computed by using equation 1, with $d$ averaged between 24$^\circ$ and 28$^\circ$ S and 46$^\circ$ and 48$^\circ$ S, respectively. Figure \ref{fig:scaling}e and Figure \ref{fig:scaling}f show the mean-centered brightness as a function of time. Note that the normalization of the mean-centered brightness is perform independently for each of the scaling region. The STB scaling region seems to be a good choice for all the dates except 2018, where the scaling drops the average brightness of 2018 significantly (see Figure \ref{fig:scaling}b). This is due to an unusual 5-$\mu$m-bright STB observed in the 5-$\mu$m data from 2018, potentially related to the convective storms that erupted at the cyclonic region of the STBn known as the STB 'Ghost', leaving a very turbulent region \citep{Inurrigarro_2019}. On the other hand, the S3TZ seems to be a valid scaling region at every epoch except 1992, where the Protocam data shows a bright S3TZ, not observed in other dates. Figure \ref{fig:scaling}b and Figure \ref{fig:scaling}c show that both scaling regions give similar results over most of the studied epochs, showing Jupiter's temporal overall variability. However, it is worth noting that the intensity of the normalized average brightness varies with the chosen scaling region, due to the normalization performed, where we normalized the maximum average brightness to 1, for each of the scaling regions. Therefore, we advise the reader to be cautious when comparing the intensity of the zones/belts.\\

Finally, in order to analyze the robustness of the results shown in this study, we perform the same analysis using each of these three scaling regions independently (i.e. EZ for the dates without a EZ disturbance, STB and S3TZ) and we then compare the results. We also analyzed different potential scaling regions. All figures in section 4 and 5 are computed by scaling the average brightness to the S3TZ (i.e. 46-48$^\circ$ S, see the supplemental material for the results obtained scaling at the STB and at the EZ).\\

\subsection{Lomb-Scargle Periodogram} \index{Lomb-Scargle Periodogram} \indent

The Lomb-Scargle periodogram \citep{Scargle_1982} is a widely used tool in the study of planetary science that characterizes periodic signals in unevenly sampled data, by computing least-square fits with sinusoidal functions. Previous studies of Jupiter have mainly used this technique to study possible wavenumbers of diverse tropospheric and stratospheric waves (e.g. \cite{Harrington_1996b, Ortiz_1998, Arregi_2006, Barrado-Izagirre_2009, Garcia_Melendo_2011, Simon_Miller_2012}), although it has also been used to analyze possible periodicities in the changes of the tropospheric zonal wind structure (e.g. \cite{Simon_Miller_2010, Tollefson_2017}) among others. In this study, we use the Lomb-Scargle analysis to study possible periodicities on the brightness variability of Jupiter's atmosphere below the ammonia cloud level, at the 1-4 bar region. \\

In particular, we use the "Scargle" function written in the Interactive Data Language (IDL). This algorithm computes possible periodicities automatically and allows the setting of a desired false alarm probability to control the significance of the different periodicities returned by the algorithm. In this study, we compute the Lomb-Scargle periodogram analysis for each latitude by setting a false alarm probability of 0.001, or 99.99\% significance, and a minimum period of 1 year periodicity. The uncertainty of the obtained periods is assumed to be equal to the full width at half maximum (FWHM) of the power spectrum peak. Results are shown in section 5. However, we urge caution in the interpretation of the results as the Lomb-Scargle analysis expects sinusoidal changes of the data, something not observed in our data.\\

\subsection{Wavelet Transform Analysis} \index{Wavelet Transform Analysis} \indent

In order to validate the results obtained from the Lomb-Scargle analysis and to be able to discard possible spurious results, we also analyze possible periodicities of the 5-$\mu$m brightness variability by performing a Wavelet Transform analysis. This is a powerful tool for analyzing temporal variations within a time series, which enables determination of the dominant modes of variability and their temporal change \citep{Torrence_1998}. This analysis has been widely used in geophysics because, unlike a Fourier analysis or Lomb-Scargle periodogram that expand functions in terms of sines and cosines, the wavelet analysis expands functions in terms of a set of functions (called wavelets) which are built by translating and dilating a fixed mother wavelet chosen by the user \citep{Lee_1994}. \\

The continuous Wavelet Transform of a real signal s(t) with respect to a mother wavelet $\phi$(t) is given by the convolution of the real signal with a set of wavelets:

\begin{equation}
S(b,a)=\frac{1}{\sqrt{a}} \thinspace \int_{-\infty}^\infty \thinspace \phi'  \left(\frac{t-b}{a} \right) s\left(t\right) dt 
\end{equation} 
 
\noindent \citep{Lee_1994}, where $\phi'$ is the complex conjugate of $\phi$, $a$ is the scale factor (dilating value) that controls the width of the window and the oscillation period of the real signal, and $b$ is the time shift (translation) of the mother wavelet along the time axis. The correct choice of the mother wavelet is essential,  as this choice influences the time and frequency resolution of the results. In this study, we use the Morlet wavelet as the mother wavelet, which consists on a plane wave modulated by a Gaussian \citep{Chapa_1998, Huang_1998}, and it is given by

\begin{equation}
\phi \left(t \right)=\pi ^{-1/4} \thinspace e ^{i \omega_{0} t}  \thinspace e^{-t^{2}/2}
\end{equation} 

\noindent where $\omega_{0}$ is a wavenumber. This is the most commonly used mother wavelet as it is very well localized in frequency, where the scale factors nearly represent the Fourier period of the real signal (the period obtained by computing a Fourier Transform to the Morlet wavelet of a specific scale '$a$', \cite{Torrence_1998}). \\

In this study, we compute the Wavelet Transform analysis using the function 'wv\_cwt.pro' written in IDL. This code dilates and translates the mother wavelet given in equation 3 to compute a set of Morlet wavelets with different scale values and then performs the convolution of the normalized average brightness and each wavelet. As the wavelet analysis requires the data to be evenly spaced in time, we interpolate the normalized average brightness by a linear interpolation, and then apply a boxcar average of 10 to smooth the interpolated normalized average brightness. Additionally, in order to minimize the errors introduced at the boundaries of our study we set the keyword PAD, which pads the normalized average brightness with enough zeros to make the total length of the normalized average brightness array equal to the next-higher power of 2. Finally, we compute the wavelet power spectrum as $| S \left(b,a \right) | ^{2}$. As in the Lomb-Scargle analysis, the uncertainty of the obtained periods is assumed to be equal to the full with at half maximum (FWHM) of the wavelet power spectrum peak. The results of this analysis are shown in section 6.

\subsection{Principal Component Analysis} \index{Principal Component Analysis} \indent

With the aim of analyzing possible correlations/anticorrelations in the 5-$\mu$m changes at different zones and belts, we also compute a Principal Component Analysis (PCA). This technique decomposes a dataset into a set of basis vectors or principal axes that recognize the major sources of variance within the dataset \cite[e.g.,][]{Jolliffe_2002}. With this technique, each point in a dataset can then be represented in terms of its components (i.e. principal components) along each of the principal axes. The advantage of this technique is that, in practice, the first few axes represent the majority of the variability within the dataset, and less important axes can be ignored, reducing the “dimensionality” of the data. \\

In this study, we perform PCA following the next steps: First, we compute the mean of all the 5-$\mu$m data used in this study to obtain a vector that represents the centre of the distribution of points in n-dimensional space. This mean is then subtracted from the data, such that the dataset is now 'mean-centered'. This allows us to analyze the variability of the 5-$\mu$m brightness with respect to the averaged state of Jupiter. Once the mean-centered brightness is obtained, we create an n x m matrix (where n=141 is the number of latitudes, and m corresponds to the number of dates in our dataset), which is then multiplied by its transpose, to form the n x n covariance matrix of the data. From this, the n principal axes computed as the eigenvectors of the covariance matrix, and the corresponding eigenvalues represent the contribution of each axis to the variance within the dataset. The principal components of each profile are the inner products of the corresponding vector with each of the principal axes. \\

The real dataset can then be reconstructed exactly as a linear combination of all the principal axes, or approximated by a linear combination of just the most important principal axes. A choice must be made as to the number of principal axes to retain, which depends on the amount of variance that should be represented in the approximation. In this study, we analyze the first three principal axes, as they display most of the major changes observed at 5 $\mu$m and described in section 4 and 5. The results of the PCA is shown in section 6.

\section{Results: Temporal Mean Brightness} 

At visible wavelengths the boundaries of Jupiter's whitish zones and brownish belts are usually colocated with the locations of maximum anticyclonic and cyclonic wind shear observed at the top of the ammonia cloud level at $\sim$ 700 mbar, being regions of upwelling and downwelling, respectively, in Jupiter's atmosphere. However, this is not always true at mid-latitudes, where the coloration of the belts and zones does not always follow the cyclonic and anticyclonic regions \citep{Rogers_1995}. \\

\begin{figure}[H]
	\centering
		\includegraphics[width=0.6\textwidth]{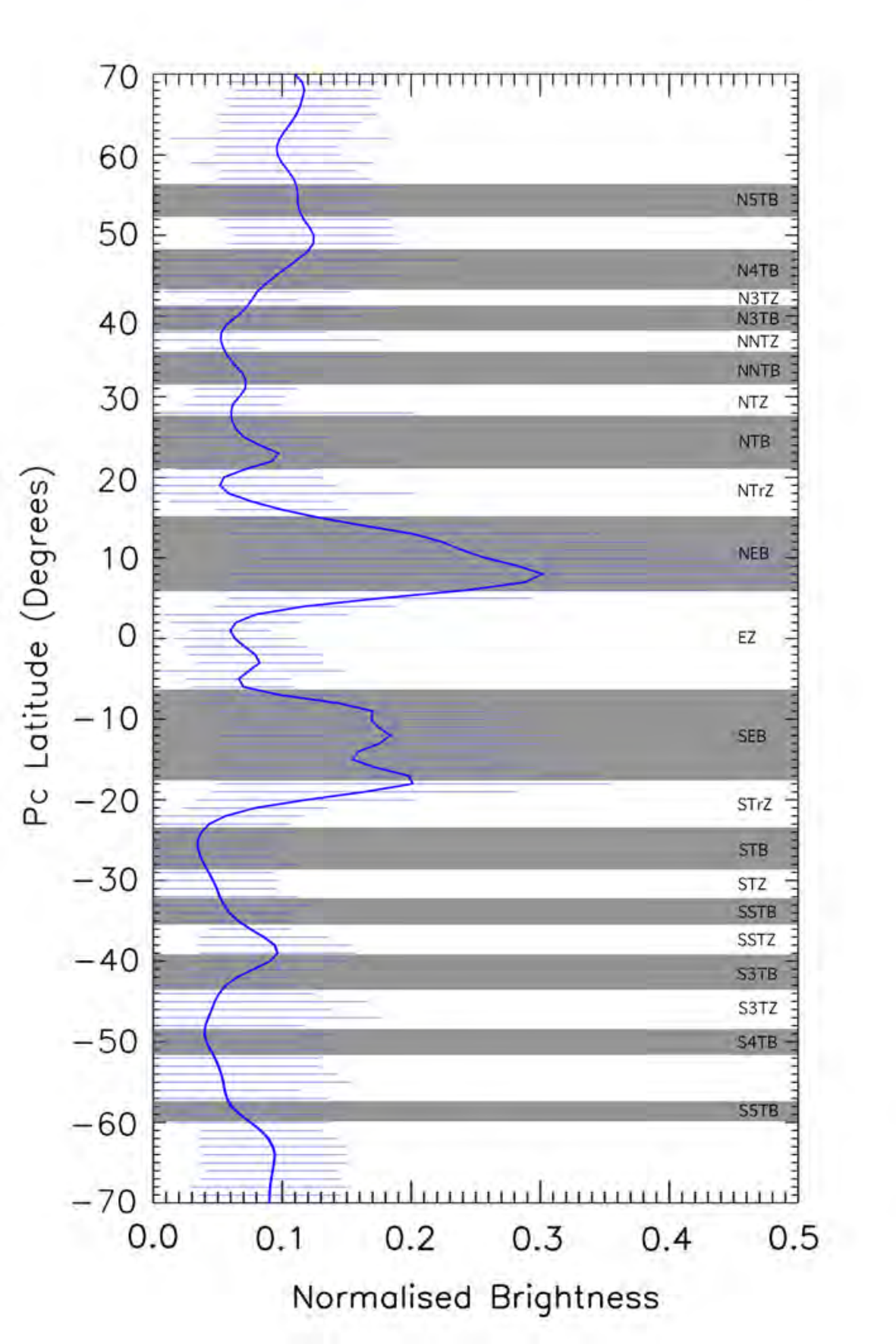}
	\begin{quote}\caption[Temporal Mean]{Temporal average (1984-2018) of the normalized and zonally averaged brightness scans at 5 $\mu$m as a function of latitude, scaled at the South South South Temperate Zone between $\sim 46^\circ$ S and $48^\circ$ S, showing the 5-$\mu$m brightness pattern of Jupiter's zones and belts (blue solid line). The light blue lines represent the standard deviation of the temporal average and the dark grey regions indicate the location of Jupiter's belts, according to \cite{Porco_2003}}
	\label{fig:temporal_mean}
	\index{Temporal Mean}
	\end{quote}
\end{figure}

In order to analyze the relationship between the observed banded structure at visible wavelengths and the 5-$\mu$m brightness, we compute the 34-year temporal mean of the normalized brightness, and compare the 5-$\mu$m-bright and -dark regions with the locations of the visible belts and zones, as defined by the zonal wind peaks at the cloud level by \cite{Porco_2003}. The results are shown in Figure \ref{fig:temporal_mean}, where the solid blue line represents the temporal average of the normalized 5-$\mu$m brightness over the 34 years (1984-2018) and the shadowed grey regions represent the belts observed at visible wavelengths. \\

Figure \ref{fig:temporal_mean} shows a clear asymmetry in the temporal mean 5-$\mu$m brightness between the northern and the southern hemisphere. In the northern hemisphere, the temporal mean of the 5-$\mu$m brightness displays a large meridional variability, mainly correlated with the observed belts and zones at equatorial and tropical latitudes. In the southern hemisphere, however, the picture is completely different. The correlation between the banded structure and the 5-$\mu$m brightness is only observed at the SEB and the South Tropical Zone (STrZ), while at the tropical and mid-latitudes the 5-$\mu$m brightness shows a bland meridional variability with a bright peak at $\sim 33^\circ$ S - $45^\circ$ S. This asymmetry between the northern and southern equatorial and tropical latitudes is also usually observed in the main cloud deck, where images captured at 8.6 $\mu$m with the TEXES instrument display a morphologically bland southern hemisphere compared to a much richer northern hemisphere, suggesting a hazier southern region \citep{Fletcher_2016}. This is in agreement with the more detailed meridional structure observed at 5 $\mu$m in the northern hemisphere than in the south. At mid- to high-latitudes an asymmetry between the northern and southern hemispheres is also observed, where the region poleward of $40^\circ$ N appears to be brighter than its southern counterpart.  \\
 
The simplistic picture of bright, cloud-free belts and dark, cloudy zones only seems to hold at the tropical and equatorial latitudes.  At mid- to high-latitudes the correspondence between the albedo, zonal jets, and 5-$\mu$m brightnesses breaks down, as stated in \cite{Rogers_1995}, and this point is often not considered properly in the available literature. We emphasize that the tropical and extra-tropical latitudes are observationally different from one another, and that inferences from the easy-to-observe tropics should not be blindly applied to the extra-tropics. We shall return to a discussion of the permanent hemispheric asymmetry in Section 6. In this study we define the locations and names of the belts and zones by using the zonal wind peaks from \cite{Porco_2003} as their boundaries (see shadowed grey regions in Figure \ref{fig:temporal_mean}).  

\section{Results: Brightness variability at Equatorial and Tropical Latitudes} 

The extended temporal coverage of Jupiter 5-$\mu$m ground-based data allows us to analyze the brightness temporal variability of Jupiter's zones and belts at the cloud-forming region (1-4 bar) over almost 3 Jovian years (1984-2018), essential to understand the nature and origin of the planetary scale  variability observed at Jupiter's banded structure. Below we analyze the temporal variability of a number of interesting regions at latitudes within $\pm$35$^\circ$ of the equator. A summary of all the changes described below is shown in Table \ref{tab:variations}. It is important to note that the differences between the peak brightness values of the different events might not be significant, as they change with the chosen scaling region (see section 2.3 and the appendix) and thus, one should be cautious when comparing the intensity of the events.\\

\subsection{North Temperate Region} \index{North Temperate Region} 

The North Temperate Region (NTR), located between 21$^\circ$ N and 31.5$^\circ$ N, is formed by the North Temperate Belt (NTB, i.e. 21-28$^\circ$ N) and by the North Temperate Zone (NTZ) at 28-31.5$^\circ$ N. Here we also include the North North Temperate Belt (NNBT, i.e. 31.5-35.5$^\circ$ N). These regions have been observed to undergo large temporal variations since the early 20$^{th}$ century, where the NTB fades, revives and drifts in latitude \citep{Peek_1958, Rogers_1995, Sanchez_Lavega_1991, Sanchez_Lavega_2008, Sanchez_Lavega_2017}. \\

\begin{figure}[H]
	\centering
		\includegraphics[width=0.6\textwidth]{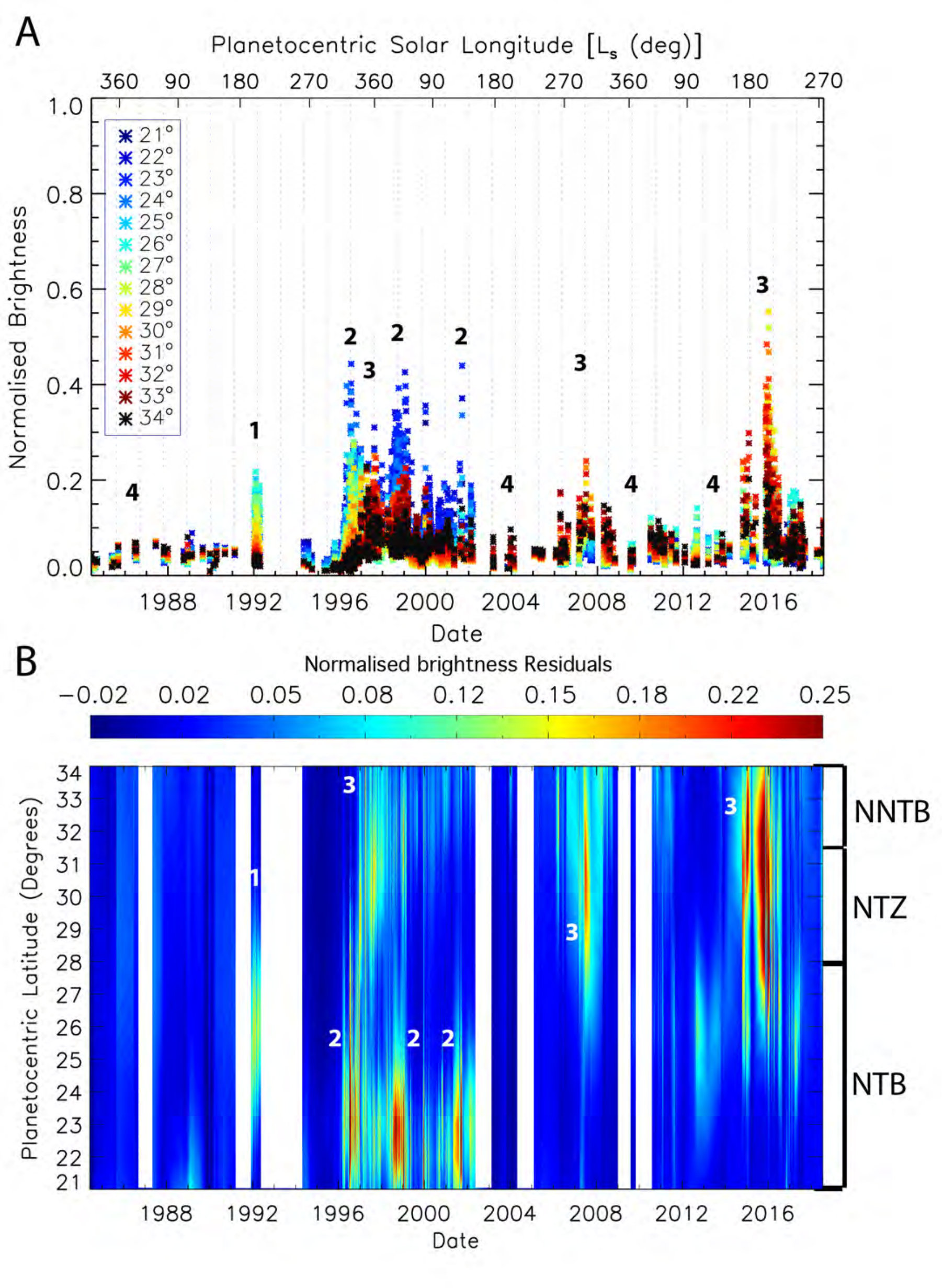}
	\begin{quote}
	\caption[Brightness_variability_NTB]{Normalized average brightness as a function of time (A) for the region between 21$^\circ$ N and 34$^\circ$ N, showing the temporal variability of the brightness of the North Temperate Belt at 5 $\mu$m, between 1984 and 2018. Normalized average brightness residuals (B) for the region between 21$^\circ$ N and 34$^\circ$ N, showing the increase of the normalized average brightness relative to the temporal mean of the smallest 10$\%$ brightness values of each latitude. The average brightness is scaled to the S3TZ (i.e. 46-48$^\circ$ S). Vertical dotted blue lines in B represent Jupiter's opposition dates. The numbers indicate the different changes described in the text. See the supplementary only material for results after scaling to the STB and EZ.}
	\label{fig:brightness_NTB}
	\index{Brightness variability}
	\end{quote}
\end{figure}

In Figure \ref{fig:brightness_NTB}a and Figure \ref{fig:brightness_NTB}b we represent the 5-$\mu$m brightness temporal variability of the NTR between April 1984 and July 2018 as a function of latitude and time, showing the large temporal variability of the 5-$\mu$m emission at the cloud-forming region during those epochs. In Figure \ref{fig:brightness_NTB}b, we show the relative variability of brightness at each latitude with respect to their darkest state, computed by comparing the normalized brightness at each latitude and date to the temporal mean of the smallest 10$\%$ of brightness values of each latitude. This technique allows us to show the brightness variability at each latitude independently in the same figure, with the changes of the darker latitudes not being suppressed by the changes of the brighter latitudes. In this figure the zero corresponds to the temporal mean of the smallest 10$\%$ of brightness values of each latitudes. From these figures we see that this region underwent at least four kinds of major changes (labeled by numbers in Figure \ref{fig:brightness_NTB}a and  Figure \ref{fig:brightness_NTB}b) in its brightness contrast in the last 34 years. The first major change (labeled as 1 in Figure \ref{fig:brightness_NTB}a and Figure \ref{fig:brightness_NTB}b) is observed between January and April 1992, where the NTR displays a wide increase of its brightness between 23$^\circ$ and 30$^\circ$ N, reaching its maximum brightness at 26$^\circ$ N. During this epoch, the NTB seems to be moved poleward, brightening part of the usually dark NTZ. The formation of this belt is quite unique, as it is the widest belt developed in the NTR between 1984 and 2018, as shown in Figure \ref{fig:brightness_NTB}a. The lack of infrared data between February 1991 and January 1992 does not allow us to analyze the formation of this belt, however, it might be related to the eruption of two outbreaks at the zonal jet at $\sim$21$^\circ$ N observed in early 1990, which started a revival of the previously faded NTB, creating low and high albedo regions at the southern edge of the NTB \citep[NTB(S),][] {Sanchez_Lavega_1991, Rogers_1992, Garcia_Melendo_2005}. \\

\begin{figure}[H]
	\centering
		\includegraphics[width=0.9\textwidth]{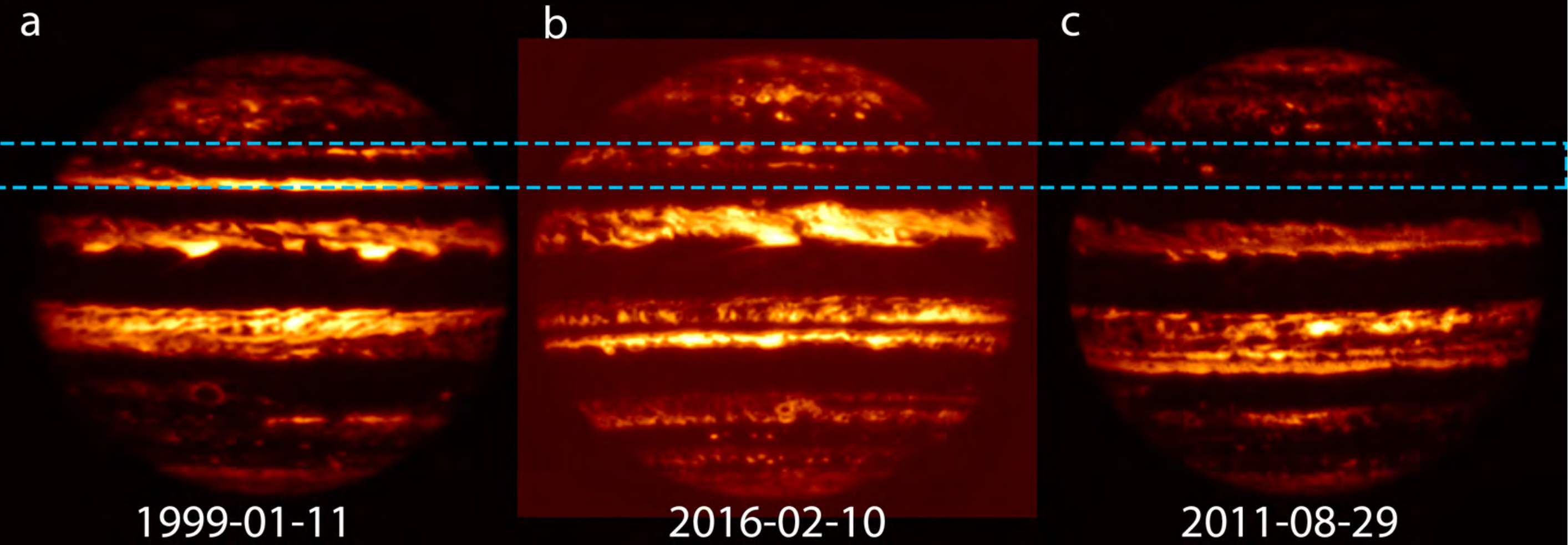}
	\begin{quote}
	\caption[Images_NTR]{5-$\mu$m images of Jupiter captured by SpeX (b, c) and NSFCam (a) instruments mounted on the Infrared Telescope Facility (IRTF) on Maunakea, showing the presence of narrow and bright band at the southern edge of the NTB (NTB(S)) in 1999 (a), the presence of a non-homogeneous bright region known as the North Temperate Zone Belt in 2016 (b) and a partially faded North Temperate Belt in August 2011 (c). The North Temperate Region is highlighted by the blue dashed box. The 5-$\mu$m radiances in these images are not calibrated nor scaled and therefore one must be cautious when comparing the contrast.}
	\label{fig:images_NTR}
	\index{Images NTR}
	\end{quote}
\end{figure}

Additional major changes are observed in 1996, 1998-1999 and 2001, when a notable increase of the 5-$\mu$m emission is observed at the NTB(S), shown in Figure \ref{fig:brightness_NTB}a and Figure \ref{fig:brightness_NTB}b by green/yellow colors and labeled as number 2. Unlike the formation of the NTB observed in 1992, these brightness increases were not preceded by outbreaks at the NTBs. During these epochs, the normalized brightness of the NTB(S) and of the NNTB at 31-35$^\circ$ N seem to increase and decrease contemporaneously, in time intervals of two years, as shown in Figure \ref{fig:brightness_NTB}b (blue colors representing the NTB(S) and red/black colors representing the NNTB), suggesting that the NTB and the NNTB are somehow connected to each other. 5-$\mu$m images of these epochs show that at the times of normalized average brightness maxima at this region, the NTB appears as a narrow bright region at 21-25$^\circ$ N, followed by a dark region between $\sim$ 25$^\circ$ N and $\sim$ 30$^\circ$ N (NTB(N) and NTZ) and accompanied by a non-homogenous bright NNTB between $\sim$ 30$^\circ$ N and $\sim$ 34$^\circ$ N. An example of this is shown in Figure \ref{fig:images_NTR}a. At the epochs of the brightness minima, a slightly bright narrow band is observed at 21-23$^\circ$ N. These 5-$\mu$m changes were not apparent at visible wavelengths, where the NTB appeared normal (i.e. broad and dark grey) during those years.\\

During 1997, 2007 and 2016 the NTR displays a completely different appearance at 5 $\mu$m, with a dark (cloud-covered) region between 21$^\circ$ N and 25$^\circ$ N (blue crosses in Figure \ref{fig:brightness_NTB}b) and a 5-$\mu$m-bright region spanning from the usually dark North Temperate Zone up to the NNTB  (i.e. 28-34$^\circ$ N, orange and red crosses in Figure \ref{fig:brightness_NTB}b), forming a new North Temperate Zone Belt (NTZB) as shown by the number 3. Figure  \ref{fig:images_NTR}b shows that during those epochs the NTZB presents a chain of 5-$\mu$m-bright ovals, not present at the times where the NTZ is observed dark at 5 $\mu$m. The formation of these NTZB seems to occur in time intervals of 8-10 years, between 1984 and 2018, in agreement with previous studies \citep{Rogers_1995, Fletcher_2017a}, with a missing NTZB in the BOLO-1 data from 1987. The exact periodicity of these changes is analyzed in section 6. \\

Finally, during 1984-1989, 2003-2005, 2009 and 2011-2014 the entire region was mainly dark at 5 $\mu$m, displaying a remarkable low brightness at all latitudes (number 4 in Figure \ref{fig:brightness_NTB}b). Jupiter images at 5 $\mu$m from those dates show that the NTB was partially faded (5-$\mu$m-dark), with some 5-$\mu$m-bright spots of various sizes found at different latitudes intermittently, especially during 2003-2005 and 2009-2012 (see Figure  \ref{fig:images_NTR}c). The lower resolution of the BOLO1 data between 1984 and 1989, does not allow us to observe any bright spots at the NTR. These fading events do not seem to be related to the outbreaks that occur at $\sim$21$^\circ$ N creating a planetary-scale disturbance known as the North Temperate Belt Disturbance \citep[NTBD][]{Peek_1958, Sanchez_Lavega_1991, Sanchez_Lavega_2008, Sanchez_Lavega_2017} (see Table \ref{tab:outbreaks} where a list of observed NTB outbreaks is given). Finally, the contemporaneous fading of the NTB and the NNTB suggests that these two belts are connected to each other. Both the fading of the NTB and NNTB and their lifetimes appear to be random with no clear periodicities. This is discussed further in section 6.

\begin{table} 

	\centering
	\begin{tabular}{p{1.5cm}  p{5cm}}
     
	\hline
        &\\
	\textbf{Year} & \textbf{Reference}\\
	&\\
	\hline

	&\\
	1985$^*$ &   \cite{Rogers_1995}\\
  &\\
	1990 &   \cite{Sanchez_Lavega_1991, Rogers_1992,Garcia_Melendo_2005}\\
  &\\
	2007  & \cite{Sanchez_Lavega_2008, Rogers_2008}\\
	    &\\
         2012   &  \cite{Rogers_2019}\\
          &\\
          2017 &  \cite{Sanchez_Lavega_2017}\\
           &\\
          \hline
           &\\
         	
	\end{tabular}
	\begin{quote}
	\caption[Outbreaks]{Summary of the observed NTB outbreaks between 1984 and 2018. $*$ In early 1984 the NTB was quite faint at visible wavelengths and did not revive until 1985, when the coloration changed strongly. However, no outbreaks where clearly observed in 1985.}
	\label{tab:outbreaks}
	\end{quote}
	
    \end{table}

\subsection{The North Tropical Region}\index{The North Tropical Region}

The North Tropical Region (NTrR), formed by the usually brownish (5-$\mu$m-bright) North Equatorial Belt (NEB) at 7-15$^\circ$ N and by the typically 5-$\mu$m-dark North Tropical Zone (NTrZ) at 15-21$^\circ$ N, displays a large variety of cloud morphology and wave phenomena that changes the appearance of this region, both at the cloud tops ($\sim$700 mbar) and at the cloud-forming region at 1-4 bar \citep{Rogers_1995, Fletcher_2017b}. Visible observations have shown that the boundary between the NEB and the NTrZ varies in latitude from year to year, due to cloud-clearing events at the NTrZ, called NEB expansions (NEEs) that take place every 3-5 years \citep{Rogers_1995, Simon_Miller_2001, Garcia_Melendo_2001, Rogers_2017, Rogers_2019, Fletcher_2017b}. \\

\begin{figure}[H]
	\centering
		\includegraphics[width=0.60\textwidth]{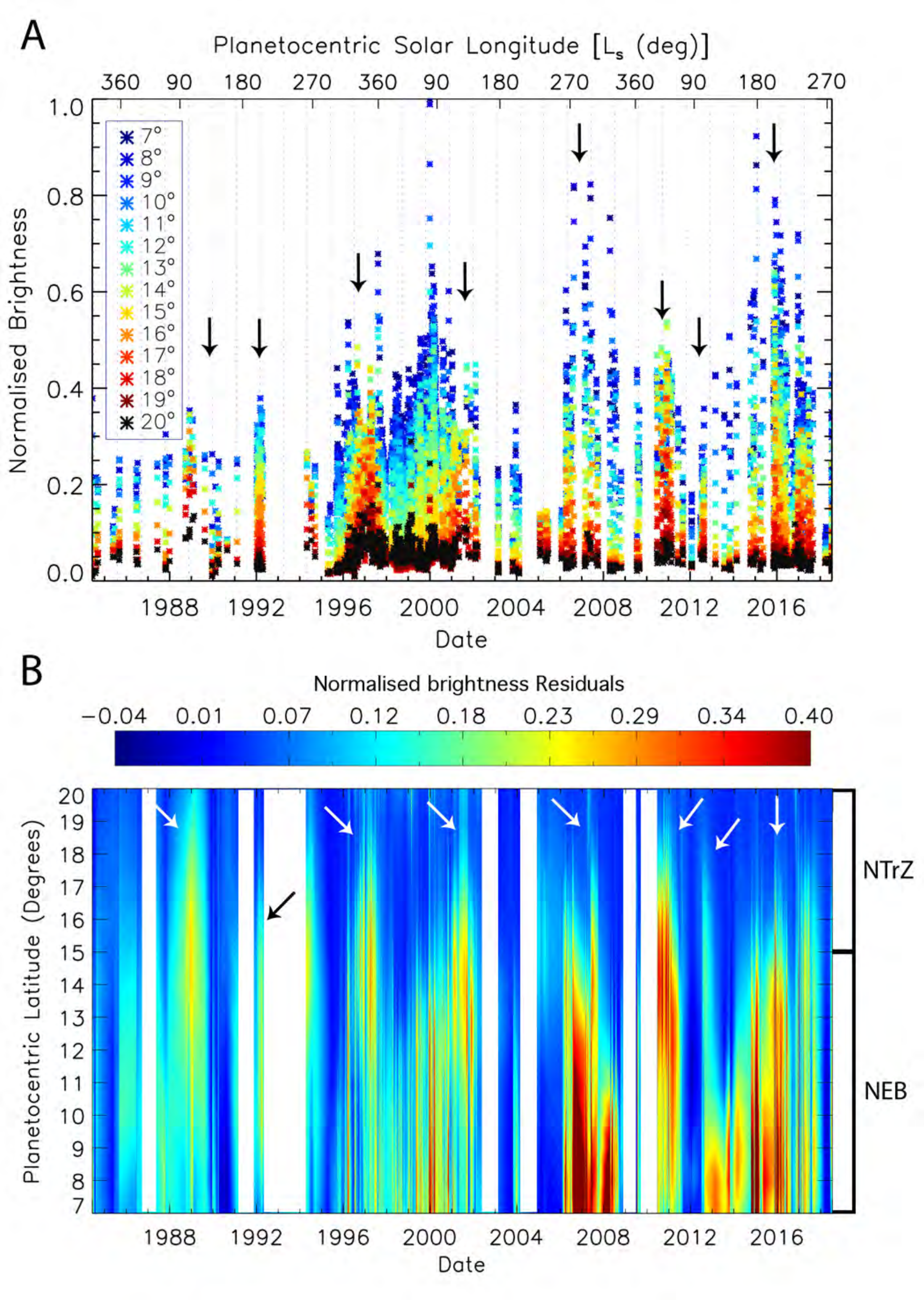}
	\begin{quote}
	\caption[Brightness_variability_NEB]{Same as Figure \ref{fig:brightness_NTB}, but for the North Tropical region between 7$^\circ$ and 20$^\circ$ N. Different color crosses in (A) represent different latitudes and the vertical dotted blue lines in A represent Jupiter's opposition dates. The white regions in (B) indicate dates where no data is available during 9 months or longer. The arrows point the NEB expansions (NEE). See the supplementary only material for results after scaling to the STB and EZ.}
	\label{fig:brightness_NEB}
	\index{Brightness variability}
	\end{quote}
\end{figure}

In Figure \ref{fig:brightness_NEB}, we represent the normalized average brightness of the NTrR, between 7$^\circ$ N and 20$^\circ$ N over the over the same 34 years as discussed for the NTR (section 4.1). As in Figure \ref{fig:brightness_NTB}a, different colors indicate different latitudes in Figure \ref{fig:brightness_NEB}a. In Figure \ref{fig:brightness_NEB}b we represent the normalized brightness residuals by showing the deviation of the normalized brightness from the temporal mean of the smallest 10$\%$ brightness values of each latitude. We observe that the southern edge of the NEB, between 7$^\circ$ N and $\sim 9^\circ$ N (represented by dark blue crosses in Figure \ref{fig:brightness_NEB}a), displays the largest brightness intensity of the entire region, potentially due to the absence of the 5-$\mu$m-dark ovals and rifts (turbulent convective regions) that are usually present at higher latitudes \citep{Rogers_1995} and the presence of the 5-$\mu$m hotspots located at $\sim$7$^\circ$ N \citep{Keay_1973, Terrile_1977, Beebe_1989, Rogers_1995, Ortiz_1998, Hueso_2017}. Poleward of these latitudes, the brightness contrast decreases with latitude, reaching its minimum value at the northern edge of the NTrZ. \\

\begin{figure}[H]
	\centering
		\includegraphics[width=0.65\textwidth]{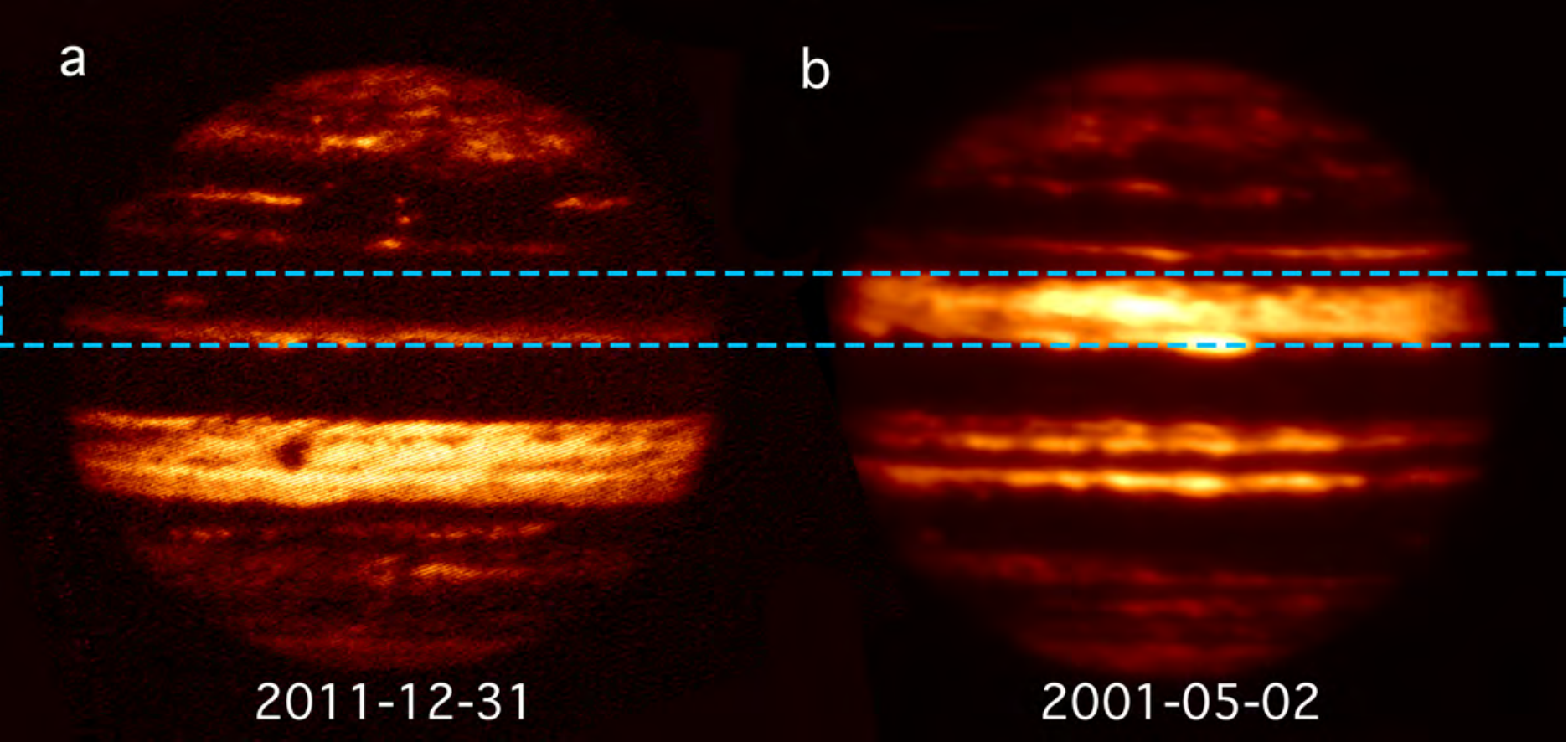}
	\begin{quote}
	\caption[Images_NTR]{5-$\mu$m images of Jupiter captured by SpeX (a) and NSFCam (b) instruments, showing a broadening of North Equatorial Belt in May 2001 (b), compared to a narrow and almost faded NEB in December 2011 (a). The North Tropical region (i.e. 7-21$^\circ$ N) is highlighted by the blue dashed box. The 5-$\mu$m radiances in these images are not calibrated nor scaled and therefore one must be cautious when comparing the contrast.}
	\label{fig:images_NEB}
	\index{Images NEB}
	\end{quote}
\end{figure}

The temporal variability of the 5-$\mu$m emission due to the latitudinal expansions/contractions of the NEB (i.e cloud clearing/covering at the southern edge of NTrZ) is clearly observed in our results, where the brightness between 15$^\circ$ N and 18$^\circ$ N is observed to increase and decrease significantly every 3-5 years (see arrows in Figure \ref{fig:brightness_NEB}a and Figure \ref{fig:brightness_NEB}b), in agreement with previous studies \citep{Rogers_1995, Rogers_2004, Rogers_2013, Rogers_2017, Fletcher_2017b}. The latitudinal coverage of the NEB expansion (i.e. brightness increases of the NEB(N) and NTrZ(S)) are observed to vary from event to event, with the events from 1989, 1992 and 2001 expanding from 7$^\circ$ N to $\sim$20$^\circ$ N (see Figure \ref{fig:images_NEB}b), while the 1996, 2006-2007, 2009-2010, 2012 and 2015-2017 events expanded from 7$^\circ$ N to $\sim$18$^\circ$ N. This is in agreement with the latitudinal coverage of the NEB expansion from 2015-2016 given by \cite{Fletcher_2017b, Fletcher_2017}. Interestingly, during the events from 1992, 2007-2008 and 2015-2017 not only was the NEB observed to be expanded, but it was also observed to be brighter than usual and also brighter than the rest of the NEB expansions. This suggests that the NEB expansions differ from one event to the next, with some producing 'stronger' cloud clearings of the NTrZ than others. \\

Finally, during 1984-1988, 1995, 1998, 2005, 2009 and 2012-2013 the boundary between the NEB and the NTrZ at 5 $\mu$m seems to be moved equatorward, varying between 10$^\circ$ N and 15$^\circ$ N from event to event. During these epochs, the NEB was not only narrower, but it was also darker at 5 $\mu$m than during the expansions. This is particularly remarkable in 2005 and 2011-2012, where Figure \ref{fig:brightness_NEB}b shows that during these events the NEB displayed the lowest brightness in the 34 years (i.e. the brightness did not deviate from the temporal average brightness of the smallest 10$\%$ brightness values). Additionally, in late 2011 and early 2012 the NEB seemed to be bright at 5 $\mu$m only between 7$^\circ$ N and 10$^\circ$ N (see Figure \ref{fig:images_NEB}a), in agreement with the analysis of the NEB fade in 2012 at visible wavelengths reported by \cite{Rogers_2018}, where the NEB seemed to be the narrowest and faintest it had ever been since 1920. 

\subsection{Equatorial Zone}\index{Equatorial Zone}

Jupiter's Equatorial Zone (EZ), covering latitudes within 7$^\circ$ of the equator, displays unique atmospheric phenomena at its northern and southern boundaries, where regions of depleted volatiles and aerosols known as 5-$\mu$m hotspots \citep{Keay_1973, Terrile_1977, Beebe_1989, Ortiz_1998, Wong_2004, de_Pater_2016, Fletcher_2016, Rogers_2017, Hueso_2017} and small periodic chevron-like features \citep{Simon_Miller_2012} are observed at its northern and southern edge, respectively. Additionally, a recent study by \cite{Antunano_2018} has shown that the usually 5-$\mu$m-dark EZ undergoes strong variations every 6-8 years, where its appearance changes completely from the darkest region on Jupiter to a region with a bright band south of the equator and large bright filaments connecting the North Equatorial Belt and this new bright band (see Figure \ref{fig:image_EZ}a). \\

\begin{figure}[H]
	\centering
		\includegraphics[width=0.6\textwidth]{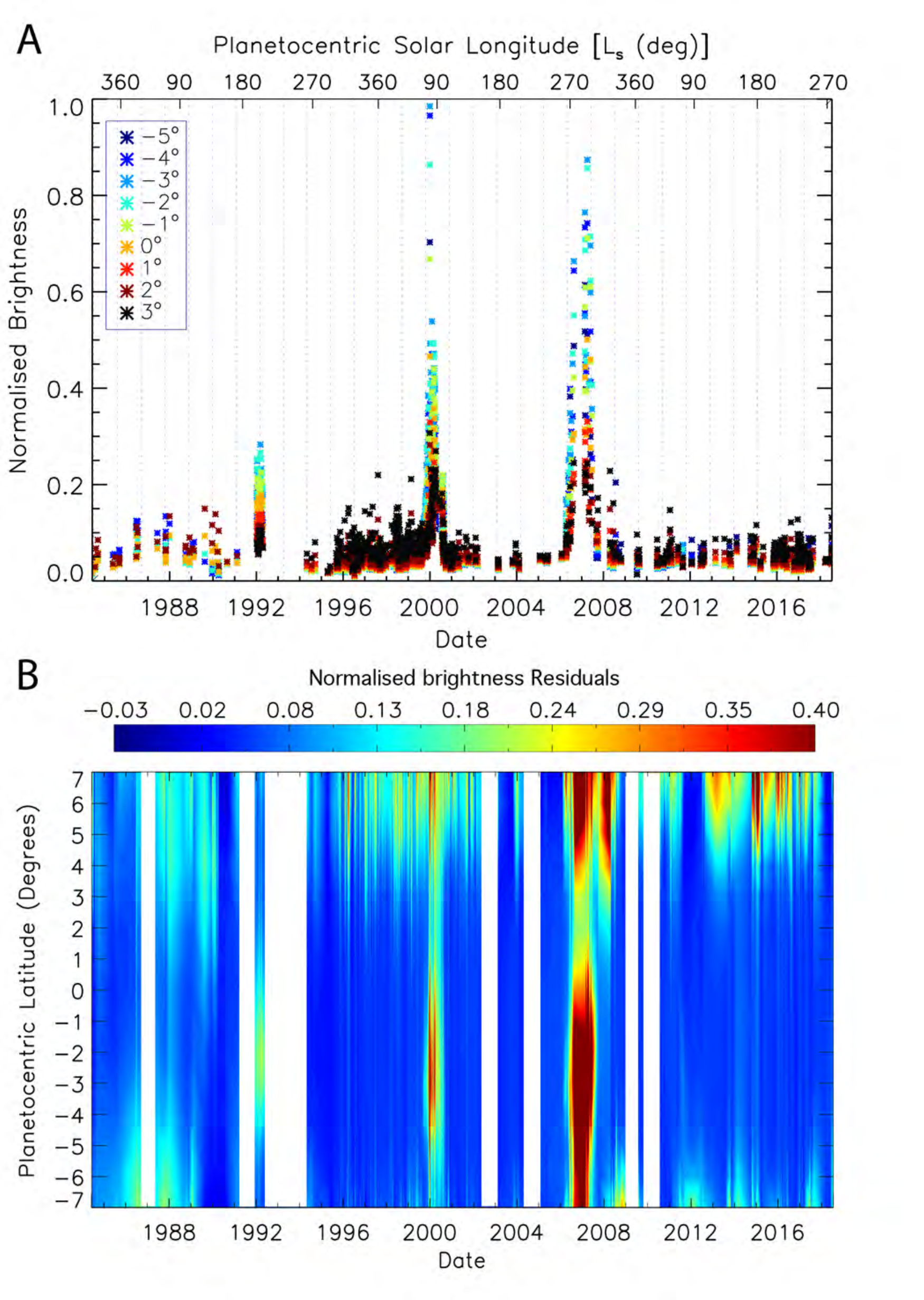}
	\begin{quote}
	\caption[Brightness_variability_EZ]{Same as Figure \ref{fig:brightness_NTB}, but for the North Equatorial region between 7$^\circ$ S and 7$^\circ$ N.}
	\label{fig:brightness_EZ}
	\index{Brightness variability}
	\end{quote}
\end{figure}

\begin{figure}[H]
	\centering
		\includegraphics[width=0.9\textwidth]{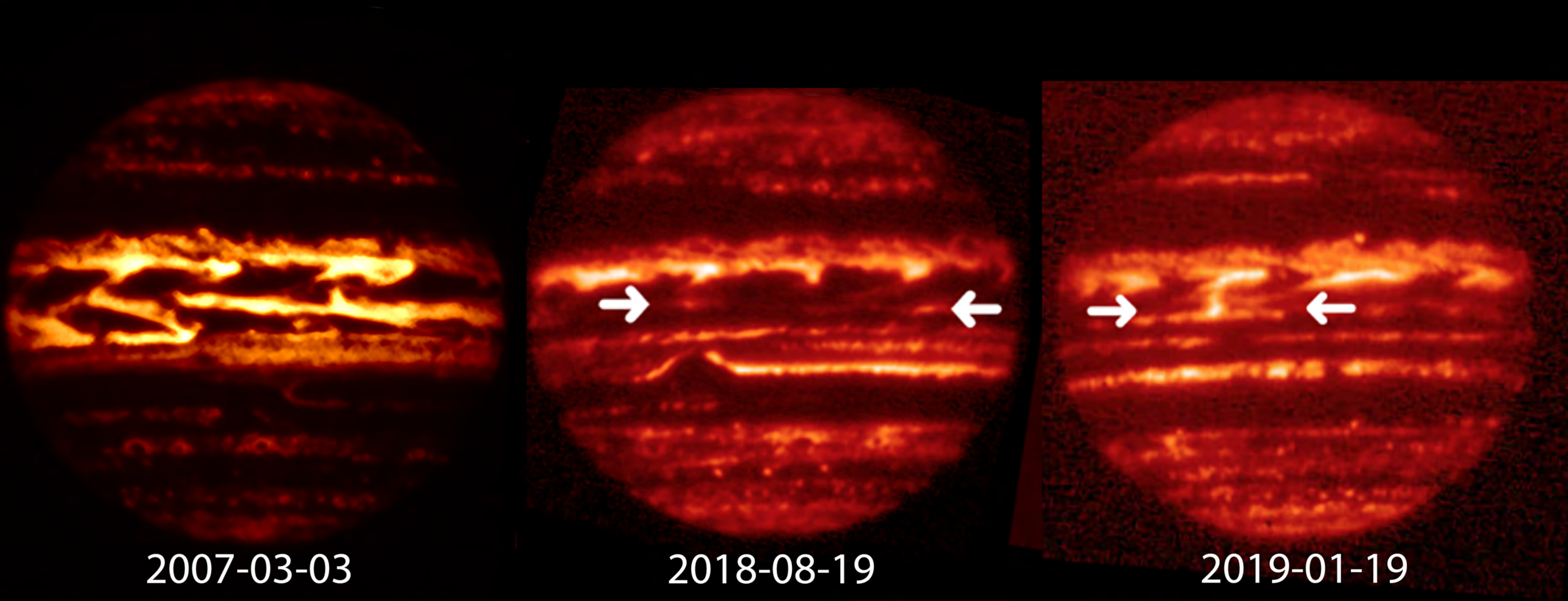}
	\begin{quote}
	\caption[Image_EZ]{5-$\mu$m images of Jupiter from (a) 3 March 2007 captured with NSFCam2, (b) 19 August 2018 and (c) 19 January 2019 captured with SpeX, showing the EZ disturbance from 2006-2007 and the latest appearance of the EZ, where a new EZ disturbance is developing. The 5-$\mu$m radiances in these images are not calibrated nor scaled and therefore one must be cautious when comparing the contrast. The arrows indicate the emerging disturbance.}
	\label{fig:image_EZ}
	\index{Image EZ}
	\end{quote}
\end{figure}

In Figure \ref{fig:brightness_EZ}a we represent the 5-$\mu$m brightness of the EZ, expanding Figure 2c in \cite{Antunano_2018} to show the new data between August 2017 and July 2018. In Figure \ref{fig:brightness_EZ}b the normalized brightness residuals are shown. Both figures show the rapid 5-$\mu$m brightness increases during the EZ disturbances of 1992, 1999-2000 and 2006-2007. Figure \ref{fig:brightness_EZ}b also shows strong brightness increases/decreases at 6-7$^\circ$ N, related to the presence/absence of the 5-$\mu$m hotspots. Most of these brightness increases seem to be related to periods when an EZ disturbance was occurring. However, a strong increase is also observed at 6-7$^\circ$ N in 1990, 2013 and 2015-2017 when no EZ disturbance was observed. Finally, Figure \ref{fig:image_EZ}b and Figure \ref{fig:image_EZ}c show that as predicted by \cite{Antunano_2018} a new EZ disturbance has been developing since August 2018, with the latest images (February 2019) suggesting that the EZ disturbance is yet not completely developed. 

\subsection{South Tropical Region}\index{South Tropical Region}

The South Tropical Region (STrR), formed by the SEB at 6$^\circ$-17$^\circ$ S and by the South Tropical Zone (STrZ) at 17$^\circ$-25$^\circ$ S \citep{Rogers_1995}, undergoes the most spectacular variations of all the regions, with the SEB changing in a very turbulent way from being one of the broadest dark brown belts of Jupiter to a whitish zone like appearance \citep[e.g.][]{Peek_1958, Sanchez_Lavega_1989, Rogers_1995, Sanchez_Lavega_1996, Fletcher_2011, Perez_Hoyos_2012, Fletcher_2017}. \\

\begin{figure}[H]
	\centering
		\includegraphics[width=0.6\textwidth]{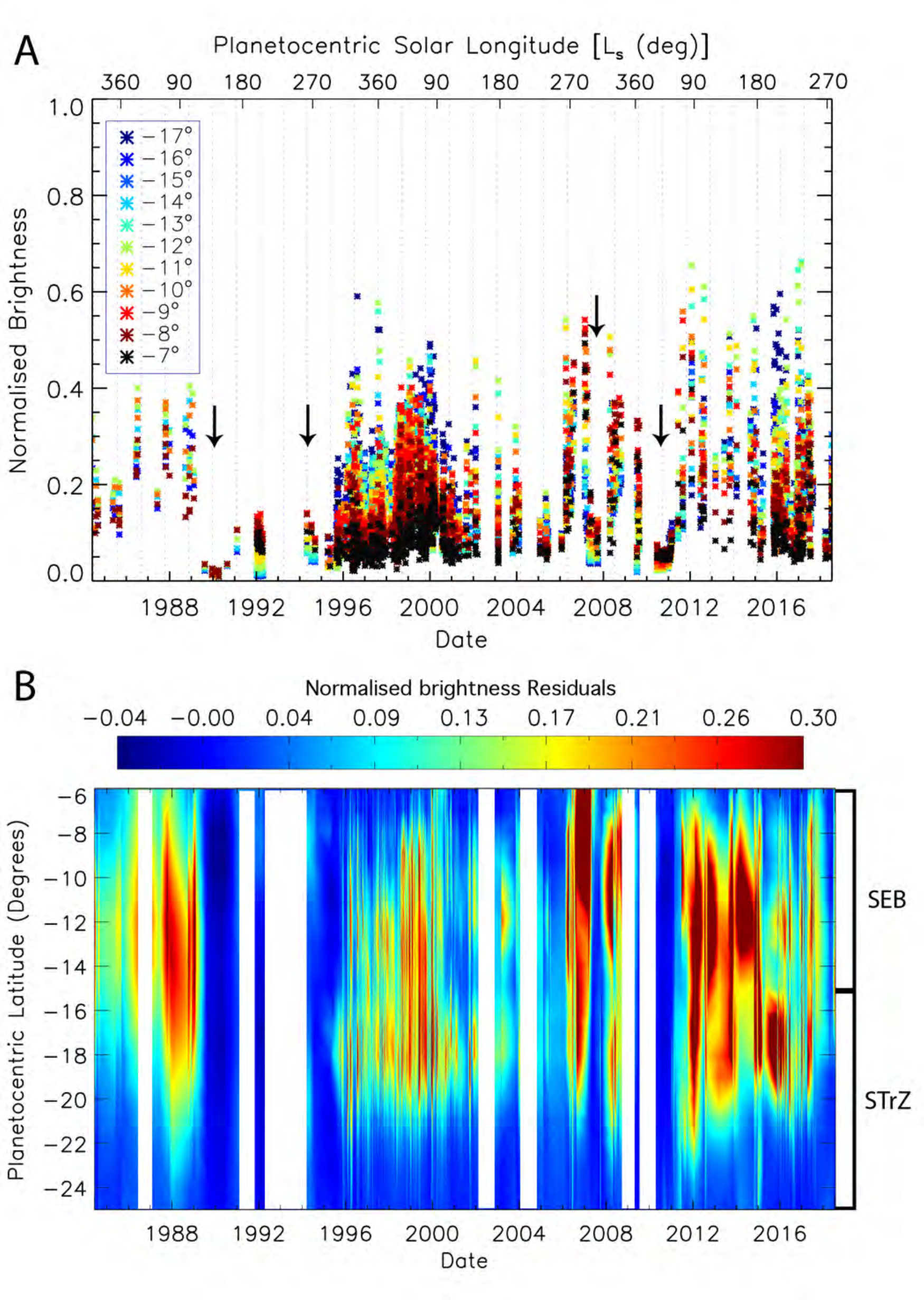}
	\begin{quote}
	\caption[Brightness_variability_SEB]{Same as Figure \ref{fig:brightness_NTB}, but for the South Tropical Region between 7$^\circ$ S and 25$^\circ$ S. The black arrows indicate the dates of the SEB fading. See the supplementary only material for results after scaling to the STB and EZ.}
	\label{fig:brightness_SEB}
	\index{Brightness variability}
	\end{quote}
\end{figure}

\begin{figure}[H]
	\centering
		\includegraphics[width=0.95\textwidth]{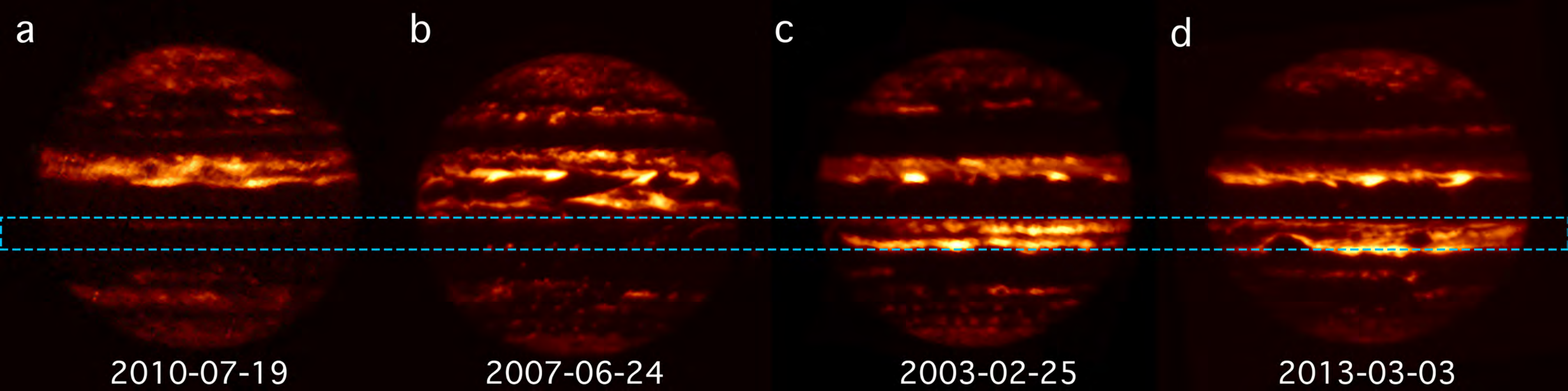}
	\begin{quote}
	\caption[Images_SEB]{5-$\mu$m images of Jupiter from (a) July 2010 captured by SpeX instrument, (b) June 2007 by the NSFCam instrument, (c) February 2003 and (d) March 2013, showing diverse appearances of the South Equatorial Belt. The South Tropical Region is highlighted by the blue dashed box. The 5-$\mu$m radiances in these images are not calibrated nor scaled and therefore one must be cautious when comparing the contrast.}
	\label{fig:images_SEB}
	\index{Images SEB}
	\end{quote}
\end{figure}

In the last 34 years, four fade-and-revival cycles have been observed at the SEB in (i) 1989-1990 \citep{Yanamandra_Fisher_1992, Kuehn_1993, Satoh_1994}, (ii) 1992-1993 \citep{Sanchez_Lavega_1996, Moreno_1997}, (iii) 2007 \citep{Reuter_2007, Baines_2007, Rogers_2007a, Rogers_2007b}, and (iv) 2009-2011 \citep{Fletcher_2011, Perez_Hoyos_2012, Fletcher_2017, Rogers_2017b, Rogers_2017c}. These fading episodes of the SEB are also observed in our analysis of the 5-$\mu$m brightness variability, indicated by strong decreases on the brightness during 1989-1990, second half of 2007 and 2010, due to the formation of ammonia ice clouds in this region. The gap in the 5-$\mu$m data between mid 1992 and early 1994, does not allow us to observed the 1992-1993 fading event. However, our data shows a low 5-$\mu$m brightness in 1994, right after the SEB revival, corresponding to a quiescence period of the SEB that seems to occur rapidly after the SEB revivals \citep{Rogers_2017c}. All these brightness decreases are shown in Figure \ref{fig:brightness_SEB} indicated by black arrows. An example of what Jupiter looked like during the 2010 SEB fading event is shown in Figure \ref{fig:images_SEB}a. In the case of the 2007 fading event, the 5-$\mu$m brightness at 7$^\circ$-9$^\circ$ S did not decrease as much as during the other fading events (where all the latitudes present a strong decrease in their brightness), indicating that in 2007 the SEB did not fade completely, displaying a bright SEB(N) (see Figure \ref{fig:images_SEB}b). This is in agreement with observations reported in \cite{Rogers_2007b}, where a violent revival was observed before the SEB was completely faded. It is important to note that the brightness decrease observed in 2005 in Figure \ref{fig:brightness_SEB} depends on the scaling region and it is not related to a faded event of the SEB (see the appendix for the brightness scaled at the STB).\\

The normalized brightness residuals shown in Figure \ref{fig:brightness_SEB}b, show that the brightness of the SEB varies from date to date when it is not faded. In particular, it is observed that at some particular epochs (1996-1998, 2000-2001, 2008, 2015-2016) the SEB displayed bright regions at $\sim$ 8$^\circ$-12$^\circ$ S and 16$^\circ$-18$^\circ$ S, separated by a low brightness and cool region named South Equatorial Belt Zone \citep[SEBZ,][]{Fletcher_2011}, while in 1998-2000,  2004 and 2014 the SEB was observed bright continuously between 8$^\circ$ S and $\sim$18$^\circ$ S. Figure \ref{fig:images_SEB}c and Figure \ref{fig:images_SEB}d show two examples of how the SEB looked like during these epochs. Periodicities of these changes are analyzed in section 6.\\

\subsection{South Temperate Region} \index{South Temperate Region} 

The South Temperate Region (STR) comprises the South Temperate Belt (STB) at 25-29$^\circ$ S and the South Temperate Zone (STZ) at 29-32$^\circ$ S and we also include the South South Temperate Belt (SSTB) at 32-35.5$^\circ$ S. As shown in Figure \ref{fig:temporal_mean} overall this region displays a completely different appearance at visible and 5-$\mu$m wavelengths between 1984 and 2018, when the belt and zone locations at visible wavelengths do not follow the 5-$\mu$m-bright and -dark regions. An example of this is the STB that, as described in section 2.3 and shown in Figure \ref{fig:scaling}, is one of the darkest regions of Jupiter at 5 $\mu$m (after the EZ), with a very small temporal variability.  \\

\begin{figure}[H]
	\centering
		\includegraphics[width=0.6\textwidth]{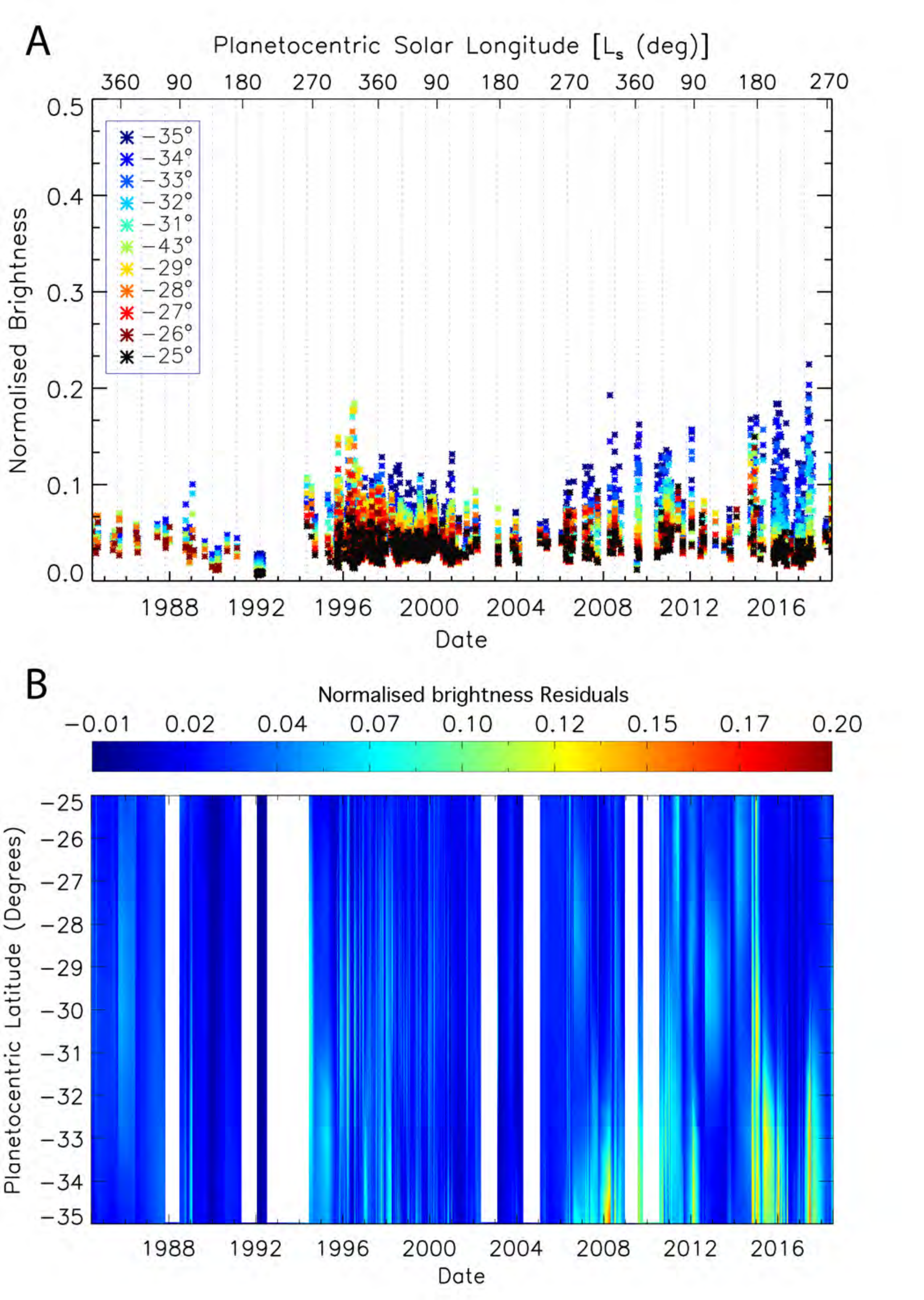}
	\begin{quote}
	\caption[Brightness_variability_STR]{Same as Figure \ref{fig:brightness_NTB}, but for the South Temperate Region between 25$^\circ$ S and 35$^\circ$ S. Note the difference in the scale in the normalized brightness.}
	\label{fig:brightness_STR}
	\index{Brightness variability STR}
	\end{quote}
\end{figure}

\begin{figure}[H]
	\centering
		\includegraphics[width=0.6\textwidth]{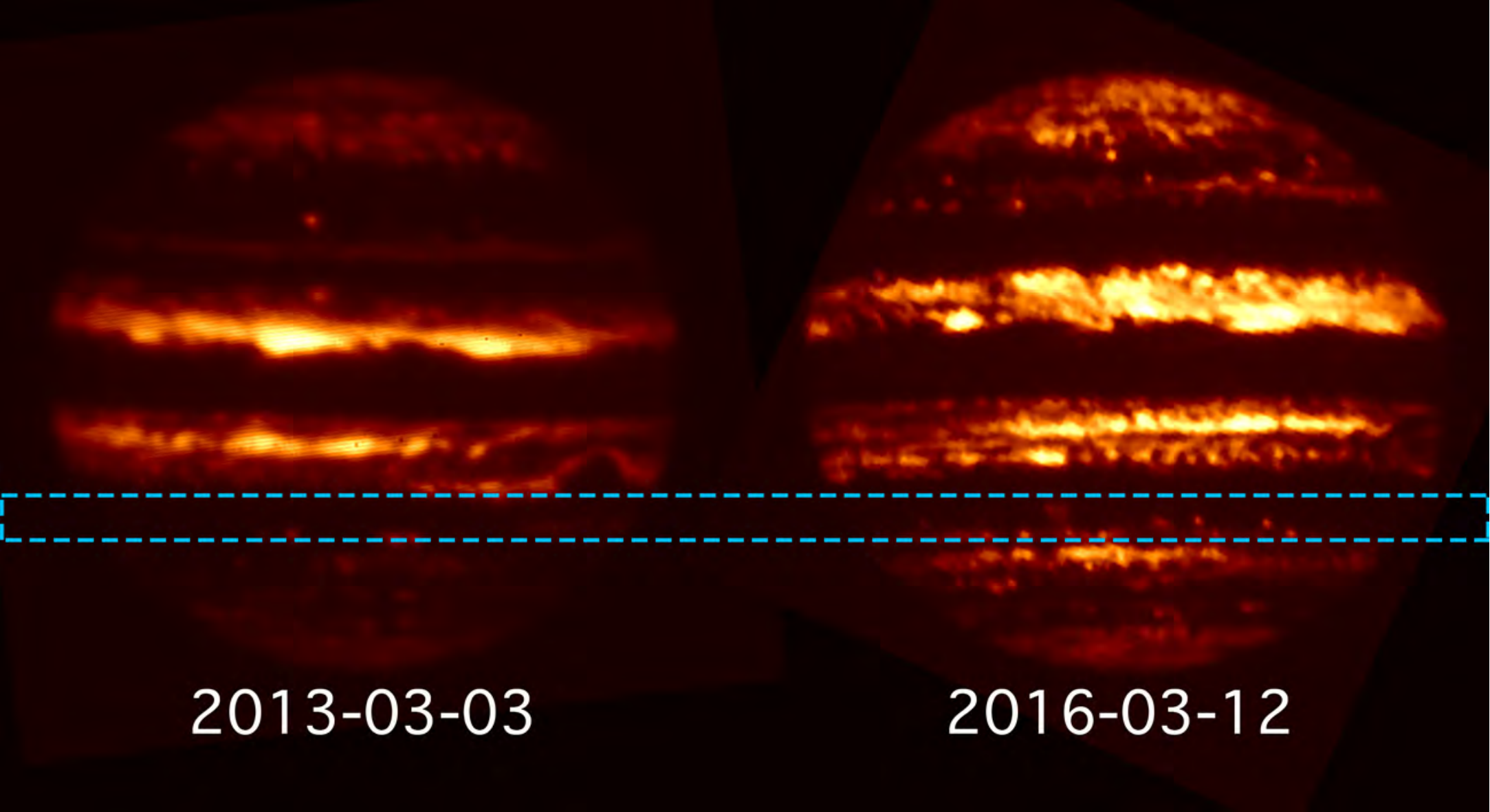}
	\begin{quote}
	\caption[Images_STR]{5-$\mu$m images of Jupiter from 03 March 2013 (left) and 12 March 2016 (right) captured by SpeX, showing diverse appearances of the South Temperate Region, highlighted by the blue dashed box. The 5-$\mu$m radiances in these images are not calibrated nor scaled and therefore one must be cautious when comparing the contrast.}
	\label{fig:images_STR}
	\index{Images STR}
	\end{quote}
\end{figure}

In Figure \ref{fig:brightness_STR} we represent the normalized brightness of the STR as a function of time and the normalized brightness residuals. Overall the entire region displays a low 5-$\mu$m brightness over the 34-years under study and compared to the other regions described in this section, the South Temperate Region is one of the least variable. The largest brightness variability is found at the SSTB (blue crosses in Figure \ref{fig:brightness_STR}a), whose brightness increases and decreases gradually. Figure \ref{fig:images_STR} shows two 5-$\mu$m images of Jupiter from March 2013 and 2016, showing that the SSTB sometimes displays a number of small 5-$\mu$m-bright ovals \citep[not related to the regular white anticyclones at ~40$^\circ$ S,][]{Rogers_1995, dePater_2010}. These ovals are not present at all the longitudes, making the belt artificially bright at the dates where the observations capture them. As the normalized brightness shown in Figure \ref{fig:brightness_STR} corresponds to the zonally averaged brightness of each date, the increases and decreases of the SSTB observed in our results need to be interpreted with caution, as dates where a complete global map of Jupiter is not available would result in a totally different zonal mean brightness than for dates where we have global maps available. 

\begin{landscape}

\begin{table} 

	\centering
	\begin{tabular}{p{1cm} p{2cm} p{11cm} p{6cm} }
         &&\\
	\hline
	\hline
        &&\\
	\textbf{Band} & \textbf{Latitude} &  \textbf{Event} &  \textbf{Dates}\\
	&&\\
	\hline
	\hline
	&&\\
	NTR &   21$^\circ$ - 35.5$^\circ$ N &     NTB expansion brightening the 23$^\circ$-30$^\circ$ N region  & Apr-Jan 1992 \\
	&&\\
	  &   & Brightness increase of NTB(S) between 21$^\circ$ and 25$^\circ$ N & 1996, 1998-1999, 2001 \\
	  &&\\
           &    &  Poleward translation of the NTB brightening the 28$^\circ$-35$^\circ$ N region & 1997, 2007, 2016 \\
           &&\\
           &  &   Brightening of the entire NTR & 1984-1989, 2003-2005, 2009-2011 \\
           &&\\
           \hline
           &&\\
         NTrR  &  7$^\circ$ - 21$^\circ$ N &   NEB expansion (NEE) brightening the 7$^\circ$-20$^\circ$ N region & 1996, 2007-2008, 2010, 2015-2017 \\
         &&\\
         &  &  NEB expansion (NEE) brightening the 7$^\circ$-18$^\circ$ N region & 2011-2012, 2015-2016 \\
          &&\\
          &  &  Partially fading of the NEB & 2011-2012 \\
          &&\\
          \hline
           &&\\
         EZ	& 7$^\circ$ S - 7$^\circ$ N & EZ disturbance & 1992, 1999-2000, 2006-2007 \\
        &&\\
          \hline
           &&\\
           STrR	& 7$^\circ$ - 25$^\circ$ N & Fading of the SEB & 1989-1990, 1994, 2007, 2010 \\
           &&\\
           	&  & Formation of a South Equatorial Belt Zone & 996-1998, 2000-2001, 2008, 2015-2016 \\

         &&\\
        \hline
        &&\\
        STR	& 25$^\circ$ - 35.5$^\circ$ N & SSTB brightness increases* at 32$^\circ$-35$^\circ$ S  & 1992, 2007-2010, 2015-2017 \\
           &&\\
           \hline
	\hline
	
	\end{tabular}
	\begin{quote}
	\caption[Dataset]{Summary of the 5-$\mu$m changes described in section 3 as a function of latitude. *These changes need to be interpreted with caution.}
	\label{tab:variations}
	\end{quote}
	
    \end{table}
    \end{landscape}

\section{Results: Brightness variability at Mid-Latitudes} 

\begin{figure}[H]
	\centering
		\includegraphics[width=0.98\textwidth]{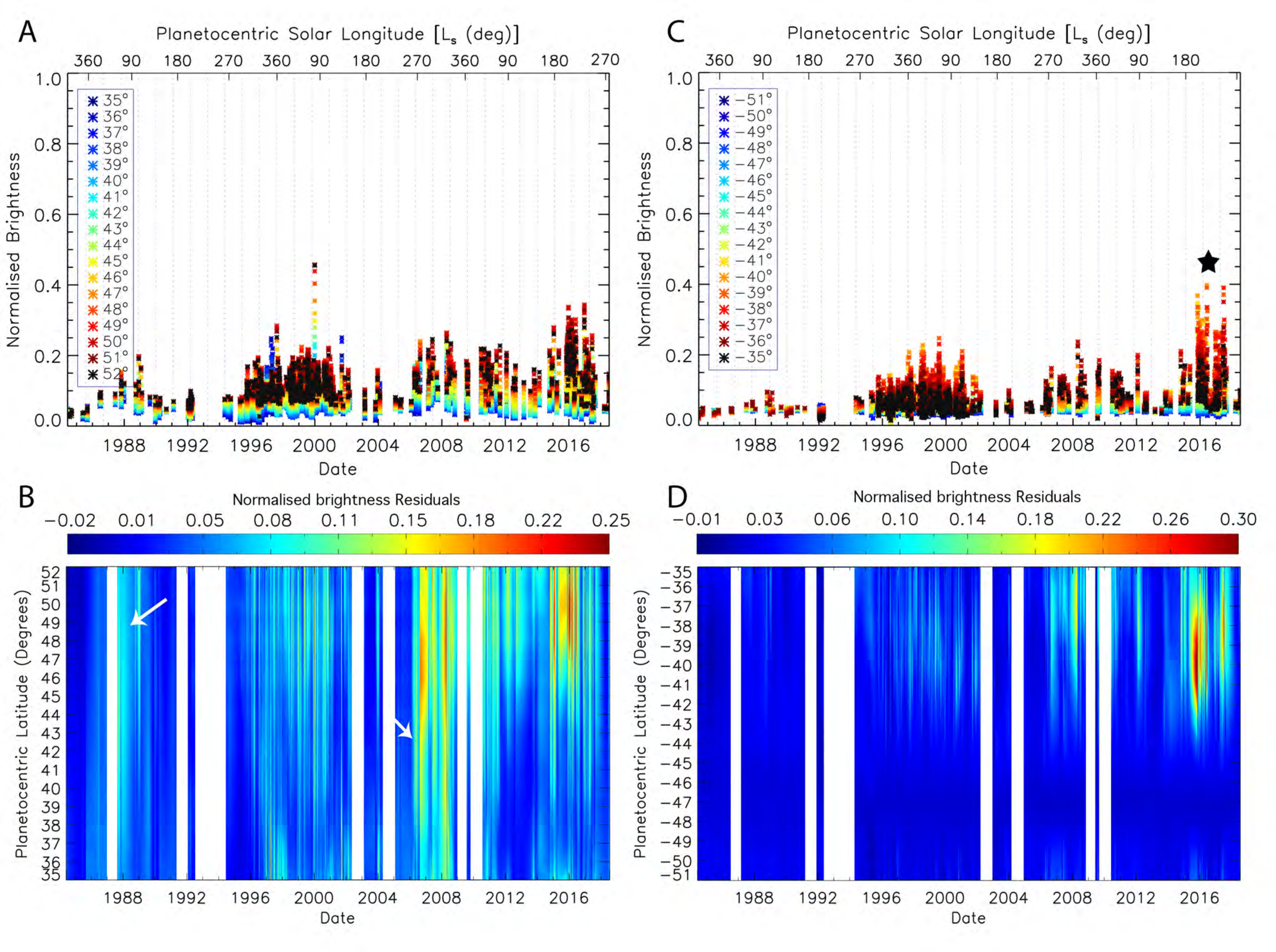}
	\begin{quote}
	\caption[Brightness_variability_NNTR]{Figure \ref{fig:brightness_NTB}, but for the North North Temperate Region between 35$^\circ$ N and 52$^\circ$ N (A, B) and the South South Temperate Region between 35$^\circ$ S and 52$^\circ$ S (C, D). The star in (C) indicate the brightness maxima of the changes observed at 36$^\circ$ - 39$^\circ$ S (location of small anticyclonic ovals). Arrows in (B) indicate the epochs when the entire mid-latitudes are observed bright at the northern hemisphere. See the supplementary only material for results after scaling to the STB and EZ.}
	\label{fig:brightness_NNTR}
	\index{Brightness variability NNTR}
	\end{quote}
\end{figure}

As shown in Figure \ref{fig:temporal_mean}, Jupiter's northern and southern mid-latitudes look completely different at 5 $\mu$m, with the northern region showing a larger meridional brightness variability than the south. In this section, we describe the brightness variability between 35$^\circ$ and $\sim$50$^\circ$ latitudes shown in Figure \ref{fig:brightness_NNTR}. From this figure it is clear that the overall brightness variability is smaller at mid-latitudes compared to the equatorial and tropical latitudes. This is especially true between 42$^\circ$ and 51$^\circ$ S, where the 5-$\mu$m brightness is very low and remains mainly constant at all the studied epochs, displaying the smallest temporal brightness variability in Jupiter (see Figure \ref{fig:brightness_NNTR}c and Figure \ref{fig:brightness_NNTR}d). Equatorward of these latitudes, the South South Temperate Zone (SSTZ) located between $\sim$36$^\circ$ and 39$^\circ$ S displays small brightness variations, reaching its brightness maximum in 2016-2017 (highlighted by a star in \ref{fig:brightness_NNTR}c). This is a particularly interesting region of Jupiter as it is the region where the highest number of small anticyclonic ovals are found \citep[e.g.][]{Rogers_1995}. These usually white ovals, which are observed to display a ring-like shape at 5 $\mu$m with a dark center surrounded by a narrow bright circle \citep{dePater_2010}, are the primary reason for the observed 5-$\mu$m variability at this region due to mergers and formations of the ovals. \\

Figure \ref{fig:brightness_NNTR}a shows that, unlike the southern hemisphere, the northern hemisphere displays a brightness that increases gradually with latitude between 35$^\circ$ and 52$^\circ$ N at all the studied epochs. Additionally, the normalized brightness residuals in Figure \ref{fig:brightness_NNTR}b show that the brightness varies overall more strongly between $\sim$46$^\circ$ and 52$^\circ$ N, contrary to what it is observed in the south. Interestingly, in 1992 and 2007-2008 the entire mid-latitudes between 35$^\circ$ and 52$^\circ$ N seem to be brighter than the rest of the dates (see arrows in Figure \ref{fig:brightness_NNTR}b). This is also observed at the southern hemisphere, although less clearly. The brightening of the mid-latitudes from 2007-2008 coincide with the period when there was also significant activity at other latitudes \citep{Rogers_1995, Rogers_2007a, Rogers_2007b}, where the SEB was faded/partially faded (\cite{Reuter_2007, Baines_2007, Rogers_2007a, Rogers_2007b}, see section 4.4), the EZ disturbance was occurring  (\cite{Antunano_2018} see section 4.3), the NEB was observed to be expanded (see section 4.2) and the North Temperate Region was observed to display a broad 5-$\mu$m-bright belt (see section 4.1). Protocam data from 1992 show higher brightness values at all latitudes, which might be not real. Finally, the brightness increases observed in 2015-2017 in Figure \ref{fig:brightness_NNTR}b  and \ref{fig:brightness_NNTR}c, seem to be contemporaneous with significant activity at other latitudes, as described in previous sections, where the SEB and the NNTB were observed to be brighter than usual (see section 4.4 and section 4.1), and the NEB was undergoing an expansion event (\cite{Fletcher_2017c}, see section 4.2). The relationship between all the significant changes occurring contemporaneously are analyzed and discussed in section 6.

\section{Discussion} 

The previous sections identified significant variability in Jupiter's 5-$\mu$m brightness at a range of latitudes, and these were discussed in qualitative terms. We now seek to quantify any periodicities observed in the data via the three techniques explained in Section 2 - a Lomb-Scargle periodogram analysis, a Wavelet Transform analysis, and a Principal Component Analysis.  By comparing and contrasting these three approaches, we test the robustness of any observed patterns in the data.  Furthermore, we remind the reader that this analysis was repeated for all three potential scaling regions - the EZ, STB, and S3TZ as shown below.

\subsection{Periodogram Analysis}\index{Periodogram Analysis}

\begin{figure}[H]
	\centering
		\includegraphics[width=0.98\textwidth]{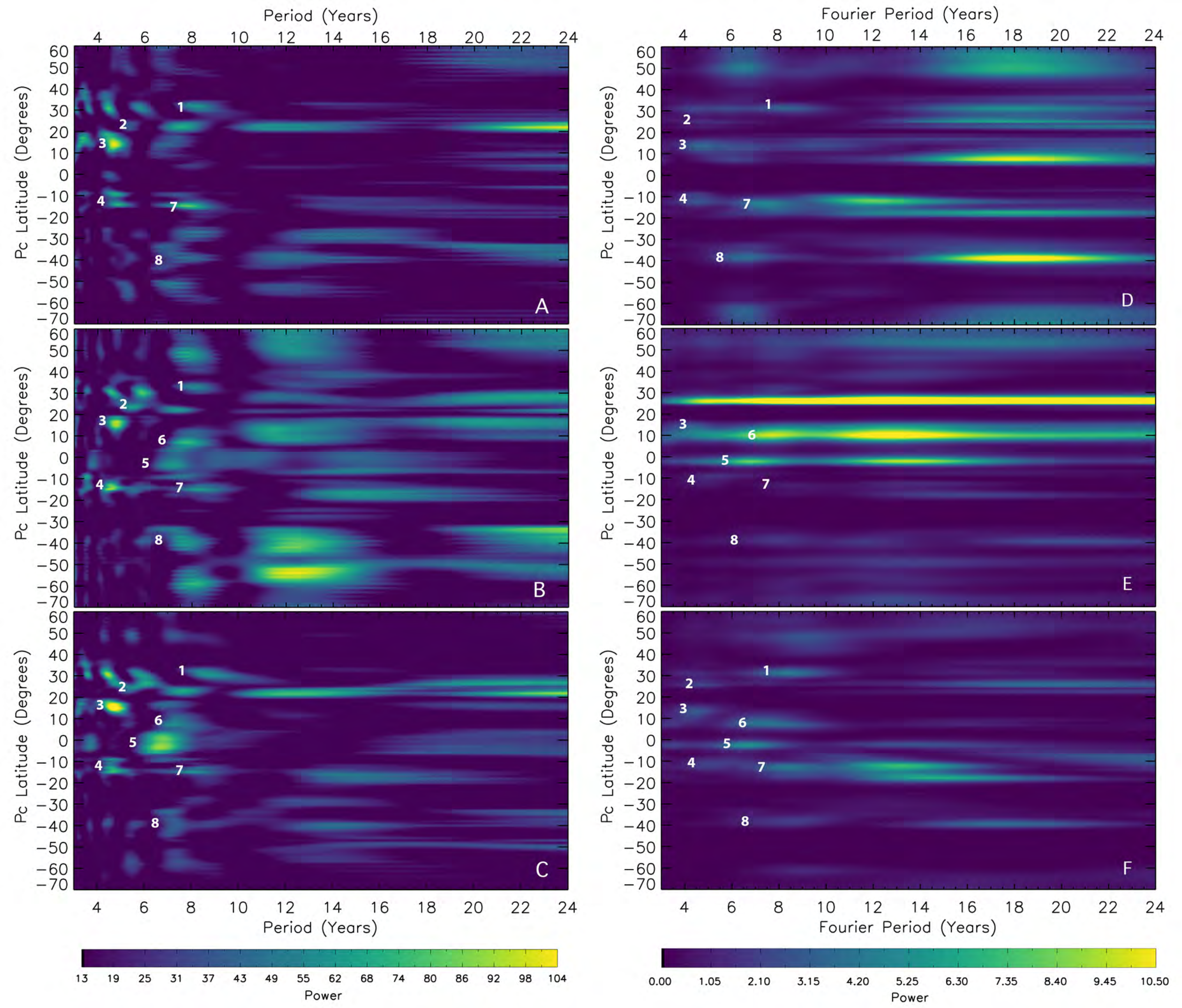}
	\begin{quote}
	\caption[Periodogram]{Lomb-Scargle periodogram and Wavelet Transform periodogram of the 5-$\mu$m emission variability between April 1984 and July 2018, for the data scaled to the equatorial zone between $\pm$5$^\circ$ (A, D respectively), to the South Tropical Zone at 24-28$^\circ$ S (B, E respectively) and to the South Temperate Zone at 46-48$^\circ$ S (C, F respectively). See section 2.3 for the different scaling regions. All periodicities with a false alarm probability smaller than 0.01$\%$ (i.e. power of 14-15) are neglected and a value of 13 is given to all of them to help to the visualization of the periodogram. The numbers in white represent the periodicities that do not depend on the scaling region, showing up in every panel with a similar periodicity. The periodicities that do not appear in all the panels with a similar periods are neglected from our analysis. }
	\label{fig:Periodogram}
	\index{Lomb-scargle}
	\end{quote}
\end{figure}

Figure \ref{fig:Periodogram} shows the Lomb-Scargle periodogram \citep[A-C panels,][]{Scargle_1982} and the Wavelet Transform Analysis periodogram (D-F panels) of the 5-$\mu$m brightness variability between April 1984 and July 2018, for the three scaling regions analyzed in this study. Below we analyze the most significant peaks with a false alarm periodicity of 0.01, or a confidence greater than 99.99$\%$, that remain invariant irrespective the scaling region and the analysis type (i.e. appear both in the Lomb-Scargle periodogram and the Wavelet Transform analysis for the three scaling regions with similar periods). These are indicated by numbers and represent the most trustworthy periodicities. Periodicities larger than 9 years are not discussed in this study as they strongly depend on the scaling region and technique. The Lomb-Scargle periodogram and the Wavelet Transform Analysis are described in sections 2.4 and 2.5, respectively. The uncertainties on the periodicities shown below are assumed to be equal to the FWHM of the power spectrum peak. \\

\begin{figure}[H]
	\centering
		\includegraphics[width=0.95\textwidth]{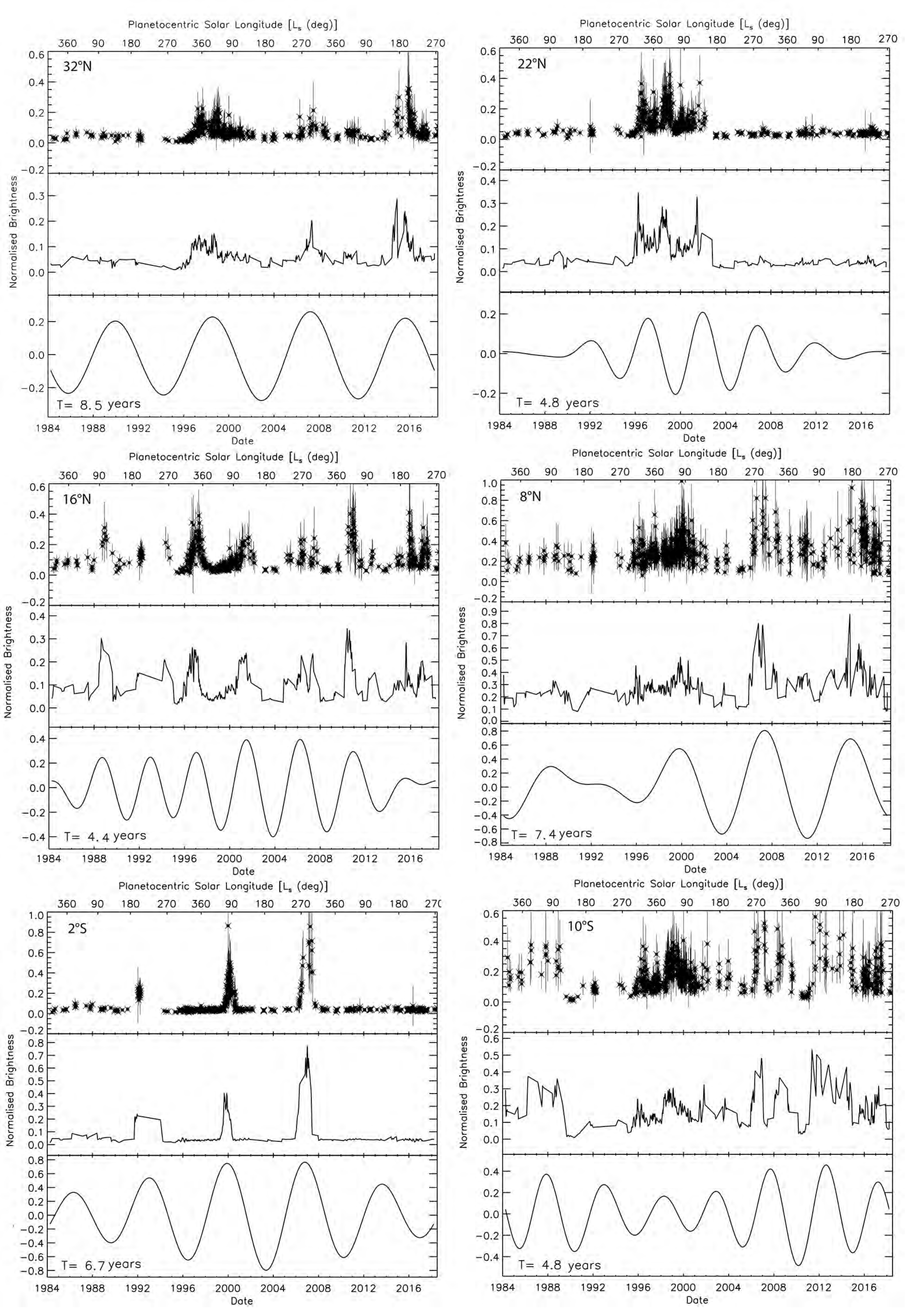}
	\begin{quote}
	\caption[Convolution]{Time series of the 5-$\mu$m brightness variability between 1984 and 2018 for 32$^\circ$ N (top left), 22$^\circ$ N (top right), 16$^\circ$ N (middle left), 8$^\circ$ N (middle right), 2$^\circ$ S (bottom left) and 10$^\circ$ S (bottom right), compared to the linear interpolation of the 5-$\mu$m brightness data (middle plots in all the panels) and the convolution between the interpolated 5-$\mu$m brightness and the selected wavelet (bottom plots in all the panels).}
	\label{fig:Convolution}
	\index{Convolution}
	\end{quote}
\end{figure}

Overall, all the indicated peaks appear at periods between 4 and 8 years. A peak at 8.5 $\pm 0.5$ years is observed at 30-33$^\circ$ N, in the southern edge of the NNTB and the northern edge of the NTZ. As described in section 4.1, these variations correspond to the northward motion of the NTB, brightening the usually dark NTZ to form a continuous belt that spans the latitude range between the NTB and the NNTB. \cite{Rogers_1995} described these kind of events observed during the 20$^{th}$ century and reported a $\sim$ 10-year periodicity, similar to the one observed in this study. This periodicity does not seem to be related to NTB outbreaks at 21$^\circ$ N, where a 4.8$\pm$0.6-years period is found in this study (number 2). The convolution between the brightness data at 32$^\circ$ N and the selected wavelet with an 8.5-year periodicity, represented in Figure \ref{fig:Convolution}, shows high correlation between 1995 and 2018, confirming the 8.5-year periodicity obtained in this study. However, this is not true between 1984 and 1995, where our periodogram analysis suggests a brightness increase at the NTZ in 1990, which is not observed in the BOLO-1 data. Nevertheless, this might be due to the lower spatial resolution of the BOLO-1 data (raster maps with a 1 arcsecond step and a circular field of view of 2 arcseconds, rather than 2D images) compared to the newer data. Our analysis suggests a new 5-$\mu$m brightening of the NTZ in 2024-2025. At 21-22$^\circ$ N there is no correlation between the convolution and the real data (see Figure \ref{fig:Convolution}) implying that the derived 4.8$\pm$0.6-year periodicity is not real and should not be taken into account. This result is particularly intriguing as NTBs outbreaks tend to occur every $\sim$5 years and implies that NTBs outbreaks do not directly change the 1-4 bar level. This is also observed in Figure \ref{fig:brightness_NTB}, where no particular change is observed at 5 $\mu$m at the epochs of NTBs outbreaks. A continuation of the 5-$\mu$m brightness time series will allow us to determine the cyclicity of this perturbation more accurately. \\

Peaks of 4.4$\pm$0.4 and 4.5$\pm$0.8 years are observed at the NEB (number 3 in Figure \ref{fig:Periodogram}) and at the SEB (number 4), respectively. The former is in agreement with \cite{Fletcher_2017b}, who suggested that the NEB expansions occur every 3 to 5 years. \cite{Tollefson_2017} and \cite{Simon_Miller_2007} also observed a 4-5-year periodicity in the zonal wind profile at 18$^\circ$ N and $\sim$10$^\circ$ S, respectively, in agreement with the 4.4$\pm$0.4 and 4.5$\pm$0.8 years found at the NEB and SEB in this study, as changes in the 5-$\mu$m brightness (e.g. cloud clearing or formation of higher clouds, thus brightening or darkening at 5 $\mu$m) affect the altitude of the wind measurements. The convolutions between the brightness data at 16$^\circ$ N (NEB) and 10$^\circ$ S (SEB) and the selected wavelet with their respective periodicities are represented in Figure \ref{fig:Convolution}. The comparison between the brightness data and its convolution with the selected wavelet gives us the trustworthiness of the derived period (i.e. a similar trend of these two confirms the veracity of the derived period). A high correlation at 16$^\circ$ N is observed at all the studied epochs confirming the 4.4$\pm$0.4-year periodicity obtained. At 10$^\circ$ S this correlation is not as clearly seen, however, the convolution follows the overall trend of the brightness variability at this latitude. \\

The observed periodicities of the NEB and SEB coincide with the $\sim$4.5 years variability found in the equatorial stratospheric temperatures, known as the Quasi Quadrennial Oscillation \citep[QQO,][]{Leovy_1991, Cosentino_2017}, where a vertical alternating pattern in the zonally-averaged temperature and winds is observed between 3 and 20 mbar at the equatorial and off-equatorial latitudes. This phenomenon resembles the quasi-biennial oscillation (QBO) in Earth's equatorial stratospheric winds and temperatures \citep[e.g.][]{Baldwin_2001} and Saturn's quasi-periodic equatorial oscillation \citep[QPO,][]{Fouchet_2008, Orton_2008}, which has been observed to span almost down to Saturn's tropopause \citep{Fletcher_2017c}. The relationship between the QQO and the 5-$\mu$m brightness variability observed at the NEB and SEB is not yet understood, as no direct relation is observed so far. \cite{Simon_Miller_2006} reported noticeable variations on the thermal wind profile in the upper troposphere at $\pm$15-20$^\circ$ latitude and the equator, retrieved from temperature measurements using Voyager IRIS and Cassini CIRS data. These authors suggested that these changes may reflect a QQO response, suggesting that like the QPO on Saturn, the QQO could also extend down as far as the upper troposphere. \\

Establishing whether the variations observed at the NEB, SEB, EZ and at 7-9$^\circ$ N are related to the QQO is a challenge. \cite{Leovy_1991}, \cite{Orton_1991} and \cite{Cosentino_2017} showed a clear anticorrelation between the equatorial and off-equatorial latitude temperatures at the upper stratosphere, where the main equatorial features of the QQO have off-equatorial secondary circulations that descend at about the same rate. If the periodicities observed in this study at the NEB and SEB were related to the QQO, one would also expect to see such an anticorrelation at 5 $\mu$m too. However, this relationship is not observed in our results, where both periodograms show a periodicity of 6.6$\pm$0.5 years south of the equator (number 5 in Figure \ref{fig:Periodogram} corresponding to EZ disturbance events, in agreement with \cite{Antunano_2018}), and a peak at 7.4$\pm$1 years at the southern edge of the NEB, at 7-9$^\circ$ N, in agreement with the $\sim$7-year periodicity observed on the changes of the equatorial wind profile at cloud level \citep{Simon_Miller_2007, Simon_Miller_2010, Tollefson_2017}. Nevertheless, \cite{Cosentino_2017} and \cite{Simon_Miller_2006} suggested the existence of two different periodicities of the QQO, with the period at $\sim$4 mbar being shorter than the period at $\sim$14 mbar, although more observational data would be needed to verify this. If the QQO indeed extends to the troposphere as it appears to do on Saturn, longer periods than 4.5 years might be then expected in the troposphere. Future observations and numerical simulations are essential to understand the relationship between the observed tropospheric variations and the QQO, and in the case of a connection between these two, to identify whether the QQO is responsible for the tropospheric variations (i.e. the warm and cool regions descending from the stratosphere into the upper troposphere could be influencing the condensation of volatiles in this region), or on the contrary, the tropospheric variations are the cause of the QQO (i.e. deeper processes such as changes in the gravity-wave forcing affect the stratosphere). \\

Finally, the $\sim$7-year peak observed at 36-41$^\circ$ S might not be real. An example of the 5-$\mu$m brightness variability is shown in Figure \ref{fig:brightness_NNTR}. The data show a high variability of the brightness in very short periods of time (days to weeks) with relatively large error bars. This is a clear sign of large longitudinal variations of the 5-$\mu$m brightness at this latitude, related to the small white ovals usually observed at 38-40$^\circ$ S. Therefore, one must be cautious as the brightness increases that the convolution shows (see Figure \ref{fig:Figure_S8} in the appendix) are related to dates with the largest amount of data available, making the temporal variation noisier. A summary of the measured periodicities described here are shown in Table \ref{tab:periodicities}.\\

The overall 4-8-year periods found in this study could hint at some moist convective process, where convective accumulated potential energy (CAPE) could be periodically released. \cite{Sugiyama_2014} performed numerical simulations of moist convection in Jupiter, showing periods of intermittency between convective eruptions of the order of tens of days. This is in agreement with some observed short-term scales events in Jupiter (e.g. the eruption of plumes in the SEB during its revival process), but cannot reproduce the long-term intervals found in this study. \cite{Li_2015} analyzed moist convection in Saturn as an explanation of the observed frequency in Saturn's giant storms. In their study they show that the cooling phase of the atmosphere due to thermal radiation at the top affects the stable-stratified layers causing convective inhibition. The 4-8-year periodicities found here are comparable to the radiative time constant ($\tau_{r}$) from \cite{Conrath_1990}, who suggested a $\tau_{r}$ between 4 and 6 years at 1 bar, and to the $\tau_{r} \sim$ 4 years at 5 bar given by \cite{Li_2018}. The similarities between the observed periodicities and the radiative time constant are therefore compelling, but not conclusive, that changes described here could also be related to moist convection. Future numerical simulations will be needed to probe this hypothesis. \\ 

\begin{table} [H]
	\centering
	\begin{spacing}{0.9}
	\begin{tabular}{p{1cm} p{2cm} p{2cm} p{2.5cm}}
         &&\\
	\hline
        &&\\
	\textbf{Label} & \textbf{Region} &  \textbf{Latitude} &  \textbf{Period (years)}\\
	&&\\
	\hline
	&&\\
	1 &   NTZ &     30$^\circ$-33$^\circ$ N & 8.5$\pm$0.5 \\
	&&\\
	2  &   NTB &   21$^\circ$-23$^\circ$ N & 4.8$\pm$0.6 \\
	&&\\
         3  &  NEB  &  14$^\circ$-18$^\circ$ N & 4.4$\pm$0.4 \\
         &&\\
         4  & SEB  &   12$^\circ$-14$^\circ$ S & 4.5$\pm$0.8 \\
         &&\\
         5  &  EZ &   0$^\circ$-5$^\circ$ S & 6.6$\pm$0.5 \\
         &&\\
         6 & NEB &  7$^\circ$-9$^\circ$ N & 7.4$\pm$1\\
         &&\\
         7  & SEB &  12$^\circ$-15$^\circ$ S & 8.2$\pm$0.5*$^1$\\
         &&\\
         8 & SSTB & 36$^\circ$-41$^\circ$ S & 7$\pm$0.4*$^2$ \\
        
         &&\\
	\hline
	
	\end{tabular}
	\begin{quote}
	\caption[Dataset]{Summary of the measured periodicities. The labels are chosen to be the same as those in Figure \ref{fig:Periodogram}. \
	*$^1$ Although both the Lomb-Scargle periodogram and the Wavelet Transform analysis return this period, it does not match the variability observed in the real data. \
	*$^2$ This results needs to be interpreted with caution as this region displays a large inhomogeneity in the longitude and therefore, it could just be showing a spurious variability. }
	\label{tab:periodicities}
	\end{quote}
	
\end{spacing}
    \end{table}

\subsection{Principal Component Analysis}\index{Principal Component Analysis}

\begin{figure}[H]
	\centering
		\includegraphics[width=0.95\textwidth]{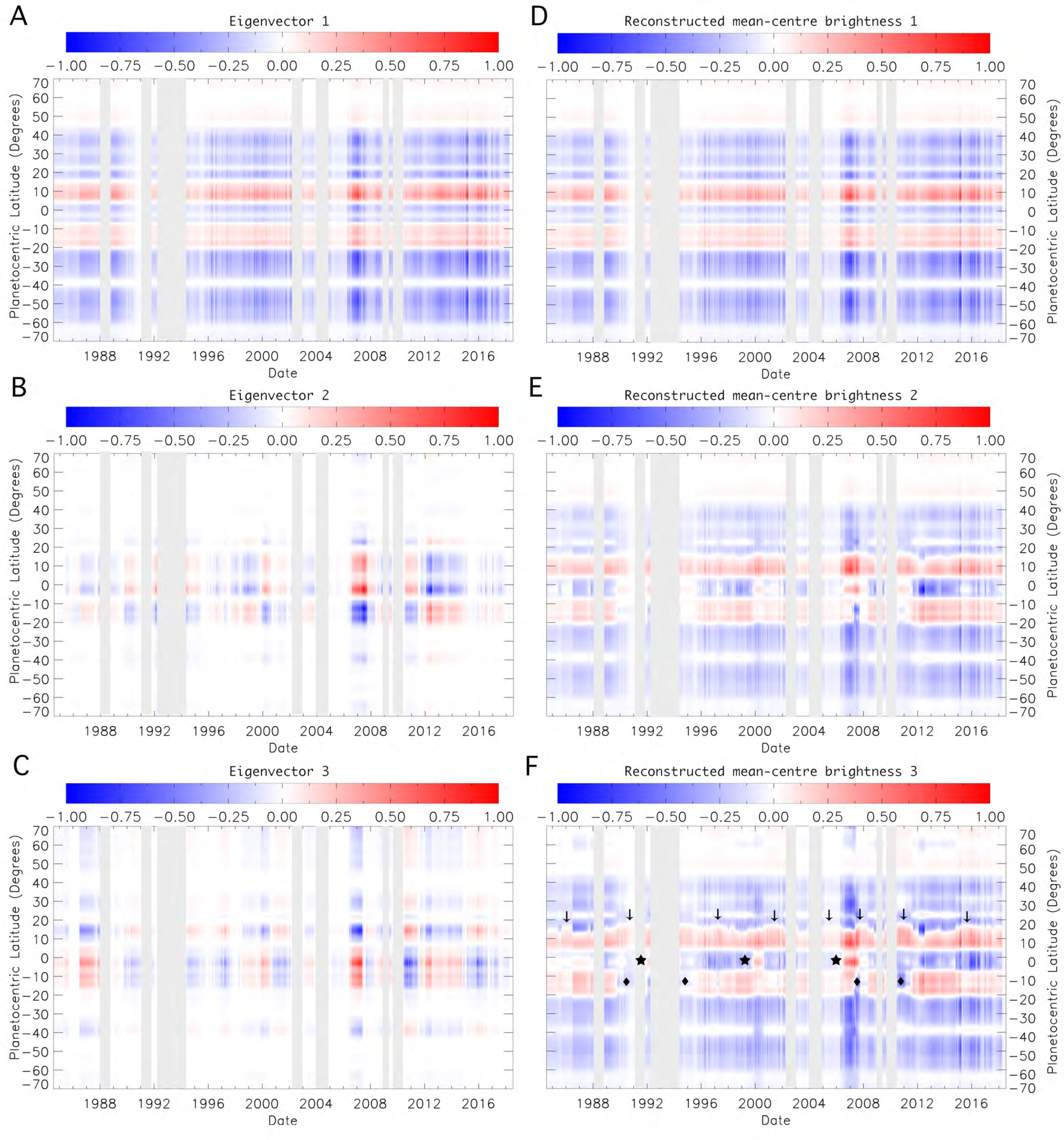}
	\begin{quote}
	\caption[PCA]{Principal component analysis results of the 5-$\mu$m brightness variability between 1984 and 2018, showing the first three eigenvectors (A to C) and the reconstructed mean-centered brightness corresponding to each eigenvector (D to F). The reconstructed mean-centre brightness of each eigenvector is computed by adding the changes shown in each eigenvector to the reconstructed mean-brightness of the previous eigenvector, the first one being equal to the first eigenvector.}
	\label{fig:PCA}
	\index{PCA}
	\end{quote}
\end{figure}

The analyses of the Lomb-Scargle and Wavelet transform periodograms show a very similar periodicity in the 5-$\mu$m brightness variability of the NEB and SEB (see section 5.1). However, these analysis do not give us any information on whether these changes occur simultaneously or with a temporal shift and whether a correlation/anticorrelation exists in their brightness variations, information that is needed to understand the relation and the origin of these changes. Here, we assess the latter question by analyzing the results of a Principal Component Analysis performed (described in section in 2.6) shown in Figure \ref{fig:PCA}. \\

Figure \ref{fig:PCA} shows the first three eigenvectors of the Principal Component analysis (a-c) and the reconstructed mean-centered brightness for each eigenvector (d-f). The red colors correspond to brightnesses larger than the meridional-mean brightness at each date, while blue colors represent the contrary (see section in 2.6 for a detailed description of the methodology of the PCA). In the PCA one would need as many vectors as the number of data points to be able to reconstruct the signal entirely. However, in this study we only show and analyze the first three eigenvectors, which display most of the major changes described in section 4, like the EZ disturbances, NEB expansion, SEB fades (highlighted by stars, arrows and diamonds, respectively, in Figure \ref{fig:PCA}f), enough to study the correlation/anticorrelation between the NEB and SEB. Higher eigenvectors give us information of the finer-scale variations, like the NTZ and NTB variations described in section 4.\\ 

\begin{figure}[H]
	\centering
		\includegraphics[width=0.75\textwidth]{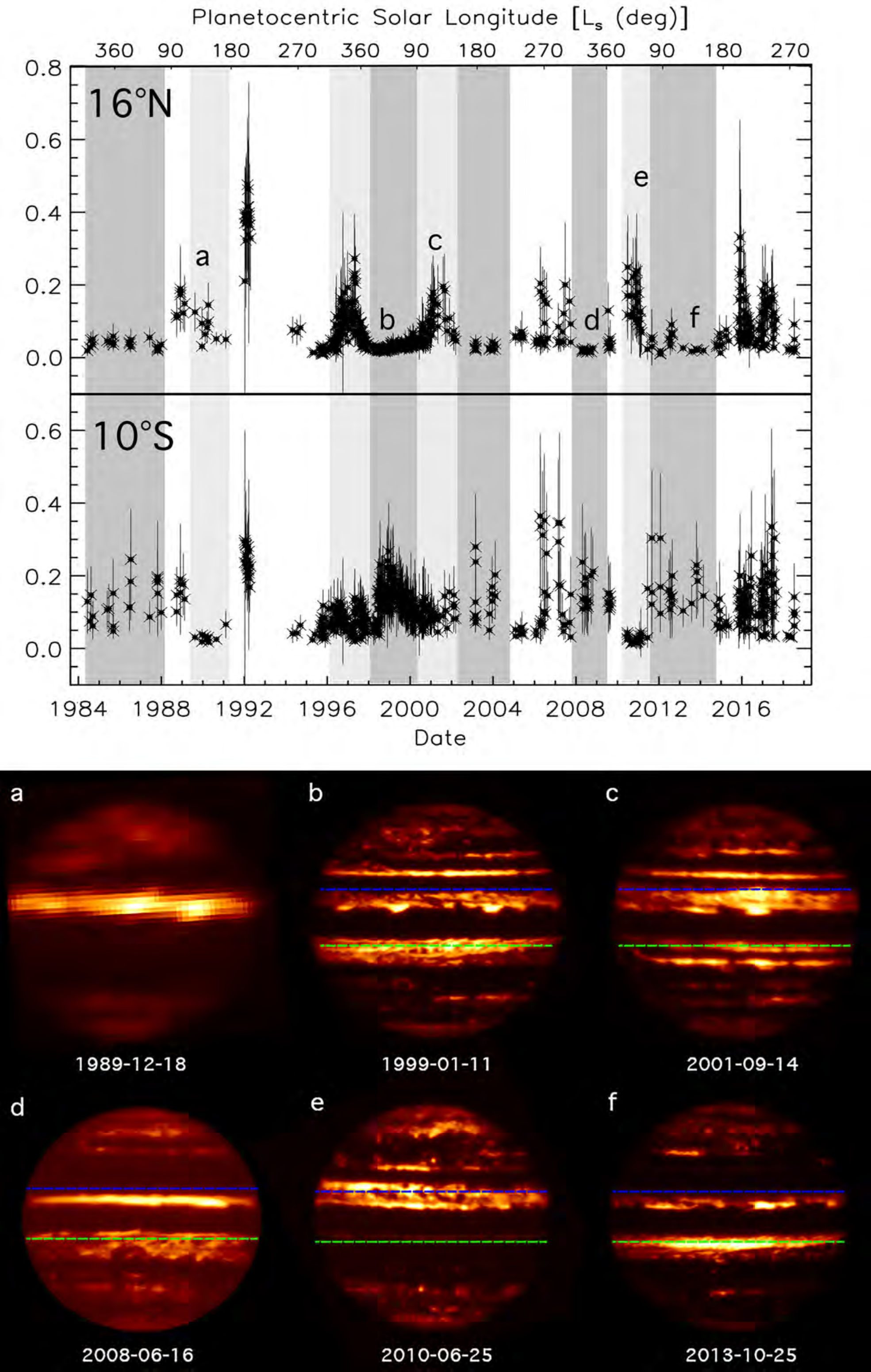}
	\begin{quote}
	\caption[Anticorrelation]{ 5-$\mu$m brightness variability at 16$^\circ$ N and 10$^\circ$ S between 1984 and 2018 (top) and six 5-$\mu$m Jupiter images showing its appearance at six chosen dates where the NEB and SEB brightness anticorrelation is observed (bottom). The shadowed dark grey regions in the top figure indicate the dates where a strong brightness decrease at the NEB and a brightness decrease at the SEB are observed simultaneously. The shadowed light grey regions represent the dates where the opposite happens. The regions with white backgrounds indicate the dates at which no anticorrelation is found. The dashed blue and green lines in the bottom figure indicate the 16$^\circ$ N and 10$^\circ$ S latitude, respectively.}
	\label{fig:Anticorrelation}
	\index{Anticorrelation}
	\end{quote}
\end{figure}

Figure \ref{fig:PCA}b and Figure \ref{fig:PCA}c show a clear anticorrelation in the brightness of the NEB and the SEB at all the studied epochs, suggesting that brightness increases and decreases of these belts are related. This is also observed in Figure \ref{fig:Anticorrelation} where we present the 5-$\mu$m brightness at 16$^\circ$ N (NEB) and 10$^\circ$ S (SEB) between 1984 and 2018. As indicated by the shadowed dark grey regions, at least five of the six brightness decreases observed at 16$^\circ$ N (that is when the NEB is not expanded) are related with strong brightness increases at 10$^\circ$ S. Contrary, four of the seven brightness increases at 16$^\circ$ N are observed contemporaneously with brightness decreases at 10$^\circ$ S (indicated by the shadowed light grey bands). In Figure \ref{fig:Anticorrelation} we also show what Jupiter looked like at 5 $\mu$m at six particular dates, where the NEB and SEB anticorrelation is clearly visible. These results, combined with the periodogram analysis, suggest a cyclic anticorrelation of the 5-$\mu$m brightness at the tropical belts of Jupiter. This anticorrelation differs from 'Global Upheaval' events, which correspond to episodes where the SEB fade/revival cycle, the strong EZ coloration and the NTBs outbreaks occur contemporaneously, spreading the coloration from one hemisphere to the other \citep{Rogers_1995, Rogers_2007a, Rogers_2007b}. These occurred in 1990 and 2007 and, unlike the anticorrelation we observed, they are not cyclic.  So far, the physical processes underlying the tropical belts anticorrelation and global upheavals are not understood and no circulation model can explain this phenomenon. \cite{Fletcher_2017c} observed that a storm that erupted at Saturn's northern mid latitudes strongly affected the stratospheric temperatures at the equatorial latitudes, significantly perturbing the wind structure in the middle-atmosphere. These authors suggested that this could be originated by westward momentum being injected into the equatorial winds from waves created by the storm. Something similar could also happen in Jupiter, where outbreaks at tropical and mid-latitudes could influence the equatorial latitudes via meridional wave propagation. The development of new General Circulation Models (GCM) and numerical simulations will be essential to understand this phenomenon.\\

Finally, Figure \ref{fig:PCA}a also shows a clear asymmetry between the northern and southern hemispheres at latitudes greater than $\sim$45$^\circ$, where the northern region appears to be brighter at 5 $\mu$m than the southern region during the entirety of the 34 years analyzed in this study. This is also observed at all the reconstructed mean-centre brightness maps shown in Figure \ref{fig:PCA} (the first reconstructed mean-centered brightness represented in Figure \ref{fig:PCA}d it is the same as Figure \ref{fig:PCA}a) and in Figure \ref{fig:temporal_mean}. Previous studies by \cite{Irwin_2004}, \cite{Fletcher_2009b} and \cite{Giles_2015} show a north/south asymmetry in phosphine distribution at mid-latitudes, with higher amounts of phosphine detected at the northern mid-latitudes above 40$^\circ$ than in the south. Similarly, \cite{Zhang_2013} and \cite{Achterberg_2006} reported north/south asymmetries in the aerosol and ammonia cloud distribution, respectively, with higher amounts at the northern mid-latitudes. This north/south asymmetry observed in phosphine, aerosols and ammonia distributions cannot explain the observed differences at 5 $\mu$m in the mid-latitudes, where a brighter southern mid-latitudes at 5 $\mu$m would be expected due to the phosphine, aerosols and ammonia blocking more of the 5-micron radiance from escaping in the northern mid-latitudes. Future spectroscopy analysis using JIRAM data will be essential to understand the nature of the 5-$\mu$m north/south asymmetry.

\section{Conclusions} 

In this study we use ground-based 5-$\mu$m infrared data spanning almost 3 Jovian years, captured between April 1984 and July 2018 with diverse instruments, to characterize the long-term variability of Jupiter's banded structure between 70$^\circ$ S and 70$^\circ$ N at the cloud-forming region between 1 and 4 bar pressure levels. We also analyze possible periodicities of these variations, which could help us to understand their origin, the physical processes behind these changes and predict future events. The conclusions of this study are summarized below:

\begin{itemize}

\item A clear asymmetry is observed in the 5-$\mu$m temporal mean brightness between Jupiter's northern and southern hemispheres. The northern hemisphere displays a large meridional brightness variability, mainly correlated to the observed belts and zones at visible wavelengths at the equatorial and tropical latitudes. The southern hemisphere however, shows a bland meridional brightness variability with a low brightness between 20$^\circ$ S and 50$^\circ$S that does not follow the banded structure found at visible wavelengths. This is in agreement with a hazier southern hemisphere compared to the northern hemisphere, blocking more of the 5-micron radiance from escaping. 

\item An asymmetry is also observed at latitudes poleward of $\pm$40$^\circ$, where the northern latitudes are observed to be brighter than in the south during the 34 years under study. The nature of this asymmetry is still not understood as previous studies analyzing the phosphine, aerosol and ammonia distributions at these latitudes suggest the contrary. 

\item The equatorial and tropical latitudes up to $\pm$35$^\circ$ display the largest temporal brightness variability, where the brightness of the zones and belts does not only vary during disturbed dates but also from one quiescent epoch to the next. The 5-$\mu$m brightness at mid-latitudes between 35$^\circ$ and 52$^\circ$ north and south is less variable than the equatorial and tropical latitudes, with the South South South Temperate Zone at 43$^\circ$ - 49$^\circ$ S being the least variable region in Jupiter's atmosphere with only a 5$\%$ brightness variability. 

\item The Lomb-Scargle periodogram and the Wavelet Transform analysis show that some of the changes observed in the banded structure are periodic/cyclic. An 8.5 $\pm$ 0.5 years timescale is found at variations at 30-33$^\circ$ N, where the North Temperate Zone brightens to form a broad bright belt that connects the North Temperate Belt and the North North Temperate Belt. We predict a new NTZ brightening in 2024-2025. These brightenings of the NTZ are not related to NTBs outbreaks, which occur cyclically every $\sim$5 years. Our long-term analysis shows that the NTBs outbreaks do not directly affect the cloud-forming region at 1-4 bar level.

\item Periods of 4.4 $\pm$ 0.4 and 4.5 $\pm$ 0.8 years are observed in the brightness changes of the North Equatorial Belt and the South Equatorial Belt. This is in agreement with previous studies of the NEB expansions, which suggested 3-to-5-year periodicity, and with the periods observed on the zonal wind profiles at the NEB and SEB. Taking these results into account, we predict a new NEB expansion to occur in 2021-2022. Predictions of a new SEB fade are not possible from this study due to the large variability of the SEB fade-revival phenomena. The periods of the NEB and SEB coincide with the 4.5-year observed variability in the equatorial and off-equatorial stratospheric temperatures, known as the Quasi Quadrennial Oscillations. The relationship between the changes observed at 5 $\mu$m and the QQO is not yet understood, although previous studies by \cite{Tollefson_2017} and \cite{Simon_Miller_2007} showing noticeable changes in the thermal wind profile in the upper troposphere suggests that the QQO could affect at least the upper troposphere. 

\item A 6.6 $\pm$ 0.5-year periodicity is found at the equatorial latitudes, where three cloud-clearing events, plus a coloration event in 2012 that did not achieve cloud clearance detectable at 5 $\mu$m, have been observed in the last 34 years. This is in agreement with previous studies. New 5-$\mu$m ground-based images from February 2019 have shown that a new equatorial zone disturbance is underway at the time of writing, but it is not entirely developed yet.   

\item The Principal Component analysis performed in this study shows a clear anticorrelation between brightness changes in the North Equatorial Belt and in the South Equatorial Belt, suggesting that variations in these belts are somehow connected to each other. This anticorrelation is cyclic with a periodicity of 4.5 years and differs from global upheavals. There is no circulation model that could explain this relationship, however observations of Saturn have indicated that meridional transport of momentum by waves can serve to connect regions of the atmosphere that are geographically separated (a process called teleconnection). We suggest something similar could happen in Jupiter. \\

\end{itemize}

So far, the origin and the vertical extent of the EZ disturbances, their relationship with the QQO or moist convection events and the relationship between the brightness anticorrelation observed at the NEB and SEB and the possible 'global upheavals' are not understood. In this study, we have attempted to place observations of periodic variations in Jupiter's belts and zones into a robust, multi-year framework. It has confirmed the cyclic behavior of some processes (equatorial disturbances, NEB expansions, NTB outbreaks) and revealed some that are more subtle. It has also shown that Jupiter remains hard to predict, as some phenomena “skip a beat”, such as equatorial disturbances or expansions that fail to reach completion. So each event is not identical, and some are more severe than others. This study also reveals a potential connection to the cyclic behavior of stratospheric oscillations, although whether the QQO structure is influencing the troposphere, or tropospheric variations are influencing the stratosphere, is not clear from our data alone. Most intriguingly, this study supports the idea of a deep interconnection between phenomena at different latitudes, with events in one location seemingly influencing events in another. This could simply be coincidence, and ongoing monitoring is required to confirm or refute these ideas. Furthermore, we hope that this suggestion can be tested via next-generation numerical simulations, which are ultimately required to understand these phenomena.\\

Finally, we note that this study has used only a single wavelength to study the changes in aerosol opacity in Jupiter’s main cloud-forming regions. The next key question is how the aerosols are related to atmospheric temperatures and chemical constituents. To address this issue, temporal studies must be extended to other wavelengths, both longward of 5 $\mu$m to determine temperatures and ammonia distributions, and shortward of 5 $\mu$m to study cloud and haze reflectivity. Finally, the changes observed in the cloud-forming region must be related to those at deeper, higher pressures, being observed at the time of writing by NASA’s Juno spacecraft. Connecting these myriad datasets will be the topic of future studies.
 
\section*{References}

\bibliography{Bibliography_Jupiter}

\begin{thebibliography}{}

\bibitem[Achterberg et~al., 2006]{Achterberg_2006}
Achterberg, R.~K., Conrath, B.~J., and Gierasch, P.~J. (2006).
\newblock Cassini {CIRS} retrievals of ammonia in {J}upiter's upper
  troposphere.
\newblock {\em Icarus}, 182(1):169 -- 180.

\bibitem[Antu{\~n}ano et~al., 2018]{Antunano_2018}
Antu{\~n}ano, A., Fletcher, L.~N., Orton, G.~S., Melin, H., Rogers, J.~H.,
  Harrington, J., Donnelly, P.~T., Rowe-Gurney, N., and Blake, J. S.~D. (2018).
\newblock Infrared characterization of jupiter's equatorial disturbance cycle.
\newblock {\em Geophysical Research Letters}.

\bibitem[Arregi et~al., 2006]{Arregi_2006}
Arregi, J., Rojas, J.~F., S{\'a}nchez-Lavega, A., and Morgado, A. (2006).
\newblock Phase dispersion relation of the 5-micron hot spot wave from a
  long-term study of {J}upiter in the visible.
\newblock {\em Journal of Geophysical Research}, 111(E9).

\bibitem[Baines et~al., 2007]{Baines_2007}
Baines, K.~H., Simon-Miller, A.~A., Orton, G.~S., Weaver, H.~A., Lunsford, A.,
  Momary, T.~W., Spencer, J., Cheng, A.~F., Reuter, D.~C., Jennings, D.~E., and
  et~al. (2007).
\newblock Polar lightning and decadal-scale cloud variability on {J}upiter.
\newblock {\em Science}, 318(5848):226--229.

\bibitem[Baldwin et~al., 2001]{Baldwin_2001}
Baldwin, M., Gray, L., Dunkerton, T., Hamilton, K., Haynes, P., Randel, W.,
  Holton, J., Alexander, M., Hirota, I., Horinouchi, T., et~al. (2001).
\newblock The quasi-biennial oscillation.
\newblock {\em Reviews of Geophysics}, 39(2):179--229.

\bibitem[Barrado-Izagirre et~al., 2009]{Barrado-Izagirre_2009}
Barrado-Izagirre, N., P{\'e}rez-Hoyos, S., and S{\'a}nchez-Lavega, A. (2009).
\newblock Brightness power spectral distribution and waves in jupiter's upper
  cloud and hazes.
\newblock {\em Icarus}, 202(1):181 -- 196.

\bibitem[Beebe et~al., 1989]{Beebe_1989}
Beebe, R., Orton, G., and West, R. (1989).
\newblock Time-variable nature of the jovian cloud properties and thermal
  structure: An observational perspective.
\newblock {\em NASA Special Publication}, 494.

\bibitem[Chapa et~al., 1998]{Chapa_1998}
Chapa, S.~R., Rao, V.~B., and Prasad, G. S. S.~D. (1998).
\newblock Application of wavelet transform to meteosat-derived cold cloud index
  data over south america.
\newblock {\em Monthly weather review}, 126(9):2466--2481.

\bibitem[Conrath et~al., 1990]{Conrath_1990}
Conrath, B.~J., Gierasch, P.~J., and Leroy, S.~S. (1990).
\newblock Temperature and circulation in the stratosphere of the outer planets.
\newblock {\em Icarus}, 83(2):255--281.

\bibitem[Cosentino et~al., 2017]{Cosentino_2017}
Cosentino, R.~G., Butler, B., Sault, B., Morales-Juber{\'\i}as, R., Simon, A.,
  and de~Pater, I. (2017).
\newblock Atmospheric waves and dynamics beneath {J}upiter's clouds from radio
  wavelength observations.
\newblock {\em Icarus}, 292:168--181.

\bibitem[de~Pater et~al., 2016]{de_Pater_2016}
de~Pater, I., Sault, R.~J., Butler, B., DeBoer, D., and Wong, M.~H. (2016).
\newblock Peering through {J}upiter's clouds with radio spectral imaging.
\newblock {\em Science}, 352(6290):1198--1201.

\bibitem[de~Pater et~al., 2010]{dePater_2010}
de~Pater, I., Wong, M.~H., Marcus, P., Luszcz-Cook, S., {\'A}d{\'a}mkovics, M.,
  Conrad, A., Asay-Davis, X., and Go, C. (2010).
\newblock Persistent rings in and around jupiter's anticyclones -- observations
  and theory.
\newblock {\em Icarus}, 210(2):742 -- 762.

\bibitem[Deutsch et~al., 2003]{Deutsch_2003}
Deutsch, L.~K., Hora, J.~L., Adams, J.~D., and Kassis, M. (2003).
\newblock {MIRSI}: a {M}id-{I}nfra{R}ed {S}pectrometer and {I}mager.
\newblock {\em Instrument Design and Performance for Optical/Infrared
  Ground-based Telescopes}.

\bibitem[Fletcher et~al., 2009a]{Fletcher_2009b}
Fletcher, L., Orton, G., Teanby, N., and Irwin, P. (2009a).
\newblock Phosphine on jupiter and saturn from cassini/cirs.
\newblock {\em Icarus}, 202(2):543 -- 564.

\bibitem[Fletcher et~al., 2009b]{Fletcher_2009}
Fletcher, L., Orton, G., Yanamandra-Fisher, P., Fisher, B., Parrish, P., and
  Irwin, P. (2009b).
\newblock Retrievals of atmospheric variables on the gas giants from
  ground-based mid-infrared imaging.
\newblock {\em Icarus}, 200(1):154 -- 175.

\bibitem[Fletcher, 2017]{Fletcher_2017a}
Fletcher, L.~N. (2017).
\newblock Cycles of activity in the {J}ovian atmosphere.
\newblock {\em Geophysical Research Letters}, 44(10):4725--4729.

\bibitem[Fletcher et~al., 2016]{Fletcher_2016}
Fletcher, L.~N., Greathouse, T., Orton, G., Sinclair, J., Giles, R., Irwin, P.,
  and Encrenaz, T. (2016).
\newblock Mid-infrared mapping of {J}upiter's temperatures, aerosol opacity and
  chemical distributions with {IRTF/TEXES}.
\newblock {\em Icarus}, 278:128--161.

\bibitem[Fletcher et~al., 2017a]{Fletcher_2017c}
Fletcher, L.~N., Guerlet, S., Orton, G.~S., Cosentino, R.~G., Fouchet, T.,
  Irwin, P. G.~J., Li, L., Flasar, F.~M., Gorius, N., and
  Morales-Juber{\'\i}as, R. (2017a).
\newblock Disruption of saturn's quasi-periodic equatorial oscillation by the
  great northern storm.
\newblock {\em Nature Astronomy}, 1(11):765--770.

\bibitem[Fletcher et~al., 2018]{Fletcher_2018}
Fletcher, L.~N., Melin, H., Adriani, A., Simon, A., Sanchez-Lavega, A.,
  Donnelly, P., Antu{\~n}ano, A., Orton, G., Hueso, R., Kraaikamp, E., et~al.
  (2018).
\newblock Jupiter's mesoscale waves observed at 5 $\mu$m by ground-based
  observations and juno jiram.
\newblock {\em The Astronomical Journal}, 156(2):67.

\bibitem[Fletcher et~al., 2010]{Fletcher_2010}
Fletcher, L.~N., Orton, G., Mousis, O., Yanamandra-Fisher, P., Parrish, P.,
  Irwin, P., Fisher, B., Vanzi, L., Fujiyoshi, T., Fuse, T., Simon-Miller, A.,
  Edkins, E., Hayward, T., and Buizer, J.~D. (2010).
\newblock Thermal structure and composition of {J}upiter's {G}reat {R}ed {S}pot
  from high-resolution thermal imaging.
\newblock {\em Icarus}, 208(1):306 -- 328.

\bibitem[Fletcher et~al., 2017b]{Fletcher_2017}
Fletcher, L.~N., Orton, G., Rogers, J., Giles, R., Payne, A., Irwin, P., and
  Vedovato, M. (2017b).
\newblock Moist convection and the 2010--2011 revival of {J}upiter's {S}outh
  {E}quatorial {B}elt.
\newblock {\em Icarus}, 286:94--117.

\bibitem[Fletcher et~al., 2011]{Fletcher_2011}
Fletcher, L.~N., Orton, G., Rogers, J., Simon-Miller, A., de~Pater, I., Wong,
  M., Mousis, O., Irwin, P., Jacquesson, M., and Yanamandra-Fisher, P. (2011).
\newblock Jovian temperature and cloud variability during the 2009--2010 fade
  of the {S}outh {E}quatorial {B}elt.
\newblock {\em Icarus}, 213(2):564 -- 580.

\bibitem[Fletcher et~al., 2017c]{Fletcher_2017b}
Fletcher, L.~N., Orton, G.~S., Sinclair, J.~A., Donnelly, P., Melin, H.,
  Rogers, J.~H., Greathouse, T.~K., Kasaba, Y., Fujiyoshi, T., Sato, T.~M., and
  et~al. (2017c).
\newblock Jupiter's {N}orth {E}quatorial {B}elt expansion and thermal wave
  activity ahead of {J}uno's arrival.
\newblock {\em Geophysical Research Letters}, 44(14):7140--7148.

\bibitem[Fouchet et~al., 2008]{Fouchet_2008}
Fouchet, T., Guerlet, S., Strobel, D., Simon-Miller, A., B{\'e}zard, B., and
  Flasar, F. (2008).
\newblock An equatorial oscillation in saturn's middle atmosphere.
\newblock {\em Nature}, 453(7192):200.

\bibitem[Garc{\'\i}a-Melendo et~al., 2011]{Garcia_Melendo_2011}
Garc{\'\i}a-Melendo, E., Arregi, J., Rojas, J., Hueso, R., Barrado-Izagirre,
  N., G{\'o}mez-Forrellad, J., P{\'e}rez-Hoyos, S., Sanz-Requena, J., and
  S{\'a}nchez-Lavega, A. (2011).
\newblock Dynamics of {J}upiter's equatorial region at cloud top level from
  {C}assini and {HST} images.
\newblock {\em Icarus}, 211(2):1242 -- 1257.

\bibitem[Garc{\'\i}a-Melendo and S{\'a}nchez-Lavega, 2001]{Garcia_Melendo_2001}
Garc{\'\i}a-Melendo, E. and S{\'a}nchez-Lavega, A. (2001).
\newblock A study of the stability of jovian zonal winds from hst images:
  1995--2000.
\newblock {\em Icarus}, 152(2):316 -- 330.

\bibitem[Garc{\'\i}a-Melendo et~al., 2005]{Garcia_Melendo_2005}
Garc{\'\i}a-Melendo, E., S{\'a}nchez-Lavega, A., and Dowling, T.~E. (2005).
\newblock Jupiter's 24$\,^{\circ}$ n highest speed jet: Vertical structure
  deduced from nonlinear simulations of a large-amplitude natural disturbance.
\newblock {\em Icarus}, 176(2):272 -- 282.

\bibitem[Giles et~al., 2015]{Giles_2015}
Giles, R., Fletcher, L., and Irwin, P. (2015).
\newblock Cloud structure and composition of {J}upiter's troposphere from 5-μm
  {C}assini {VIMS} spectroscopy.
\newblock {\em Icarus}, 257:457 -- 470.

\bibitem[Harrington et~al., 1996a]{Harrington_1996}
Harrington, J., Dowling, T.~E., and Baron, R.~L. (1996a).
\newblock Jupiter's tropospheric thermal emission {I}. {O}bservations and
  techniques.
\newblock {\em Icarus}, 124(1):22 -- 31.

\bibitem[Harrington et~al., 1996b]{Harrington_1996b}
Harrington, J., Dowling, T.~E., and Baron, R.~L. (1996b).
\newblock Jupiter's tropospheric thermal emission {II}. {P}ower spectrum
  analysis and wave search.
\newblock {\em Icarus}, 124(1):32 -- 44.

\bibitem[Huang et~al., 1998]{Huang_1998}
Huang, N.~E., Shen, Z., Long, S.~R., Wu, M.~C., Shih, H.~H., Zheng, Q., Yen,
  N.-C., Tung, C.~C., and Liu, H.~H. (1998).
\newblock The empirical mode decomposition and the hilbert spectrum for
  nonlinear and non-stationary time series analysis.
\newblock In {\em Proceedings of the Royal Society of London A: mathematical,
  physical and engineering sciences}, volume 454, pages 903--995. The Royal
  Society.

\bibitem[Hueso et~al., 2017]{Hueso_2017}
Hueso, R., S{\'a}nchez-Lavega, A., I{\~n}urrigarro, P., Rojas, J.~F.,
  P{\'e}rez-Hoyos, S., Mendikoa, I., G{\'o}mez-Forrellad, J.~M., Go, C., Peach,
  D., Colas, F., and et~al. (2017).
\newblock Jupiter cloud morphology and zonal winds from ground-based
  observations before and during {J}uno's first perijove.
\newblock {\em Geophysical Research Letters}, 44(10):4669--4678.

\bibitem[Inurrigarro et~al., 2019]{Inurrigarro_2019}
Inurrigarro, P., Hueso, R., Legarreta, J., Sanchez-Lavega, A., Eichstadt, G.,
  Rogers, J.~H., Orton, G.~S., Gomez-Forrellad, J.~M., Perez-Hoyos, S., and
  Rojas, J.~F. (2019).
\newblock Observations and numerical modelling of a convective disturbance in a
  large-scale cyclone in jupiter's south temperate belt.
\newblock {\em Submitted to Icarus}.

\bibitem[Irwin et~al., 2004]{Irwin_2004}
Irwin, P., Parrish, P., Fouchet, T., Calcutt, S., Taylor, F., Simon-Miller, A.,
  and Nixon, C. (2004).
\newblock Retrievals of jovian tropospheric phosphine from cassini/cirs.
\newblock {\em Icarus}, 172(1):37--49.

\bibitem[Jolliffe, 2002]{Jolliffe_2002}
Jolliffe, I. (2002).
\newblock {\em Principal component analysis}.
\newblock Springer.

\bibitem[Keay et~al., 1973]{Keay_1973}
Keay, C., Low, F., Rieke, G., and Minton, R. (1973).
\newblock High-resolution maps of jupiter at 5 microns.
\newblock {\em The Astrophysical Journal}, 183:1063--1074.

\bibitem[Kuehn and Beebe, 1993]{Kuehn_1993}
Kuehn, D. and Beebe, R. (1993).
\newblock A study of the time variability of jupiter's atmospheric structure.
\newblock {\em Icarus}, 101(2):282 -- 292.

\bibitem[Lacy et~al., 2002]{Lacy_2002}
Lacy, J.~H., Richter, M.~J., Greathouse, T.~K., Jaffe, D.~T., and Zhu, Q.
  (2002).
\newblock {TEXES}: A sensitive high resolution grating spectrograph for the
  mid‐infrared.
\newblock {\em Publications of the Astronomical Society of the Pacific},
  114(792):153--168.

\bibitem[Lagage et~al., 2004]{Lagage_2004}
Lagage, P., Pel, J., Authier, M., Belorgey, J., Claret, A., Doucet, C.,
  Dubreuil, D., Durand, G., Elswijk, E., Girardot, P., et~al. (2004).
\newblock Successful commissioning of.
\newblock {\em The Messenger}, 117:12.

\bibitem[Lee and Yamamoto, 1994]{Lee_1994}
Lee, D.~T. and Yamamoto, A. (1994).
\newblock Wavelet analysis: theory and applications.
\newblock {\em Hewlett Packard journal}, 45:44--44.

\bibitem[Leovy et~al., 1991]{Leovy_1991}
Leovy, C.~B., Friedson, A.~J., and Orton, G.~S. (1991).
\newblock The quasiquadrennial oscillation of jupiter's equatorial
  stratosphere.
\newblock {\em Nature}, 354(6352):380.

\bibitem[Li and Ingersoll, 2015]{Li_2015}
Li, C. and Ingersoll, A.~P. (2015).
\newblock Moist convection in hydrogen atmospheres and the frequency of
  saturn's giant storms.
\newblock {\em Nature Geoscience}, 8(5):398--403.

\bibitem[Li et~al., 2018]{Li_2018}
Li, C., Ingersoll, A.~P., and Oyafuso, F. (2018).
\newblock Moist adiabats with multiple condensing species: A new theory with
  application to {G}iant-{P}lanet atmospheres.
\newblock {\em Journal of the Atmospheric Sciences}, 75(4):1063--1072.

\bibitem[Moreno et~al., 1997]{Moreno_1997}
Moreno, F., Molina, A., and Ortiz, J. (1997).
\newblock The 1993 south equatorial belt revival and other features in the
  jovian atmosphere: an observational perspective.
\newblock {\em Astronomy and Astrophysics}, 327:1253--1261.

\bibitem[Ortiz et~al., 1998]{Ortiz_1998}
Ortiz, J.~L., Orton, G.~S., Friedson, A.~J., Stewart, S.~T., Fisher, B.~M., and
  Spencer, J.~R. (1998).
\newblock Evolution and persistence of 5$\mu$m hot spots at the galileo probe
  entry latitude.
\newblock {\em Journal of Geophysical Research: Planets},
  103(E10):23051--23069.

\bibitem[Orton et~al., 1991]{Orton_1991}
Orton, G.~S., Friedson, A.~J., Baines, K.~H., Martin, T.~Z., West, R.~A.,
  Caldwell, J., Hammel, H.~B., Bergstralh, J.~T., MALCOM, M.~E., Golisch,
  W.~F., et~al. (1991).
\newblock Thermal maps of jupiter: Spatial organization and time dependence of
  stratospheric temperatures, 1980 to 1990.
\newblock {\em Science}, 252(5005):537--542.

\bibitem[Orton et~al., 1994]{Orton_1994}
Orton, G.~S., Friedson, A.~J., Yanamandra-Fisher, P.~A., Caldwell, J., Hammel,
  H.~B., Baines, K.~H., Bergstralh, J.~T., Martin, T.~Z., West, R.~A., Veeder,
  G.~J., and et~al. (1994).
\newblock Spatial organization and time dependence of {J}upiter's tropospheric
  temperatures, 1980-1993.
\newblock {\em Science}, 265(5172):625--631.

\bibitem[Orton et~al., 2008]{Orton_2008}
Orton, G.~S., Yanamandra-Fisher, P.~A., Fisher, B.~M., Friedson, A.~J.,
  Parrish, P.~D., Nelson, J.~F., Bauermeister, A.~S., Fletcher, L., Gezari,
  D.~Y., Varosi, F., et~al. (2008).
\newblock Semi-annual oscillations in saturn's low-latitude stratospheric
  temperatures.
\newblock {\em Nature}, 453(7192):196.

\bibitem[Peek, 1958]{Peek_1958}
Peek, B.~M. (1958).
\newblock {\em The Planet Jupiter}.
\newblock Faber and Faber, London.

\bibitem[P{\'e}rez-Hoyos et~al., 2012]{Perez_Hoyos_2012}
P{\'e}rez-Hoyos, S., Sanz-Requena, J., Barrado-Izagirre, N., Rojas, J., and
  S{\'a}nchez-Lavega, A. (2012).
\newblock The 2009--2010 fade of jupiter's south equatorial belt: Vertical
  cloud structure models and zonal winds from visible imaging.
\newblock {\em Icarus}, 217(1):256 -- 271.

\bibitem[Porco, 2003]{Porco_2003}
Porco, C.~C. (2003).
\newblock Cassini imaging of jupiter's atmosphere, satellites, and rings.
\newblock {\em Science}, 299(5612):1541--1547.

\bibitem[Rayner et~al., 2003]{Rayner_2003}
Rayner, J.~T., Toomey, D.~W., Onaka, P.~M., Denault, A.~J., Stahlberger, W.~E.,
  Vacca, W.~D., Cushing, M.~C., and Wang, S. (2003).
\newblock Spe{X}: {A} medium‐resolution 0.8--5.5 micron spectrograph and
  imager for the {NASA} {I}nfrared {T}elescope {F}acility.
\newblock {\em Publications of the Astronomical Society of the Pacific},
  115(805):362--382.

\bibitem[Reuter et~al., 2007]{Reuter_2007}
Reuter, D., Simon-Miller, A., Lunsford, A., Baines, K., Cheng, A., Jennings,
  D., Olkin, C., Spencer, J., Stern, S., Weaver, H., et~al. (2007).
\newblock Jupiter cloud composition, stratification, convection, and wave
  motion: a view from new horizons.
\newblock {\em Science}, 318(5848):223--225.

\bibitem[Rogers, 1992]{Rogers_1992}
Rogers, J. (1992).
\newblock Jupiter in 1989-90.
\newblock {\em Journal of the British Astronomical Association}, 102:135--150.

\bibitem[Rogers, 2007a]{Rogers_2007b}
Rogers, J. (2007a).
\newblock The climax of jupiter's global upheaval.
\newblock {\em Journal of the British Astronomical Association}, 117:226--230.

\bibitem[Rogers, 2007b]{Rogers_2007a}
Rogers, J. (2007b).
\newblock Jupiter embarks on a'global upheaval'.
\newblock {\em Journal of the British Astronomical Association}, 117:113--115.

\bibitem[Rogers, 2017a]{Rogers_2017c}
Rogers, J. (2017a).
\newblock Jupiter's south equatorial belt cycle in 2009-2011: I. the seb fade.
\newblock {\em Journal of the British Astronomical Association}, 127:146--158.

\bibitem[Rogers et~al., 2013]{Rogers_2013}
Rogers, J., Adamoli, G., Hahn, G., Jacquesson, M., Vedovato, M., and Mettig,
  H.-J. (2013).
\newblock Jupiter's north equatorial belt: An historic change in cyclic
  behaviour with acceleration of the north equatorial jet.
\newblock In {\em European Planetary Science Congress}, volume~8.

\bibitem[Rogers and Mettig, 2008]{Rogers_2008}
Rogers, J. and Mettig, H.-J. (2008).
\newblock Jupiter in 2007: Final numerical report.
\newblock {\em URL: http://www. britastro. org/jupiter/2007report20. htm}.

\bibitem[Rogers, 1995]{Rogers_1995}
Rogers, J.~H. (1995).
\newblock {\em The {G}iant {P}lanet {J}upiter}.
\newblock Cambridge University Press (CUP).

\bibitem[Rogers, 2017b]{Rogers_2017}
Rogers, J.~H. (2017b).
\newblock Jupiter's north equatorial belt and jet, i, cyclic expansions and
  planetary waves.
\newblock {\em arXiv preprint arXiv:1707.03343}.

\bibitem[Rogers, 2017c]{Rogers_2017b}
Rogers, J.~H. (2017c).
\newblock Jupiter's south equatorial belt cycle in 2009-2011: Ii, the seb
  revival.
\newblock {\em arXiv preprint arXiv:1707.03356}.

\bibitem[Rogers, 2018]{Rogers_2018}
Rogers, J.~H. (2018).
\newblock Jupiter's north equatorial belt and jet, ii, acceleration of the jet
  and the neb fade in 2011-12.
\newblock {\em arXiv preprint arXiv:1809.09719}.

\bibitem[Rogers and Adamoli, 2019]{Rogers_2019}
Rogers, J.~H. and Adamoli, G. (2019).
\newblock Jupiter's north equatorial belt and jet: Iii, the'great northern
  upheaval'in 2012.
\newblock {\em arXiv preprint arXiv:1809.09736}.

\bibitem[Rogers et~al., 2003]{Rogers_2003}
Rogers, J.~H., Mettig, H.-J., Peach, D., and Foulkes, M. (2003).
\newblock Jupiter in 1999/2000, part {I}: Visible wavelengths.
\newblock {\em J. Br. Astron. Assoc.}, 113(1):10--31.

\bibitem[Rogers et~al., 2004]{Rogers_2004}
Rogers, O.~H., Akutsu, T., and Orton, G.~S. (2004).
\newblock Jupiter in 2000/2001. part ii: Infrared and ultraviolet wavelengths.
\newblock {\em Journal of the British Astronomical Association}, 114:313.

\bibitem[S{\'a}nchez-Lavega, 1989]{Sanchez_Lavega_1989}
S{\'a}nchez-Lavega, A. (1989).
\newblock Great-scale changes in the belts and zones of jupiter: the outbursts
  of activity and disturbances in the seb and ntrz-ntbs regions.
\newblock {\em NASA Special Publication}, 494.

\bibitem[S{\'a}nchez-Lavega and G{\'o}mez, 1996]{Sanchez_Lavega_1996}
S{\'a}nchez-Lavega, A. and G{\'o}mez, J.~M. (1996).
\newblock The south equatorial belt of jupiter, i: Its life cycle.
\newblock {\em Icarus}, 121(1):1--17.

\bibitem[S{\'a}nchez-Lavega et~al., 1991]{Sanchez_Lavega_1991}
S{\'a}nchez-Lavega, A., Miyazaki, I., Parker, D., Laques, P., and Lecacheux, J.
  (1991).
\newblock A disturbance in {J}upiter's high-speed {N}orth temperate jet during
  1990.
\newblock {\em Icarus}, 94(1):92 -- 97.

\bibitem[S{\'a}nchez-Lavega et~al., 2008]{Sanchez_Lavega_2008}
S{\'a}nchez-Lavega, A., Orton, G.~S., Hueso, R., Garc{\'\i}a-Melendo, E.,
  P{\'e}rez-Hoyos, S., Simon-Miller, A., Rojas, J.~F., G{\'o}mez, J.~M.,
  Yanamandra-Fisher, P., Fletcher, L., and et~al. (2008).
\newblock Depth of a strong jovian jet from a planetary-scale disturbance
  driven by storms.
\newblock {\em Nature}, 451(7177):437--440.

\bibitem[S{\'a}nchez-Lavega et~al., 2017]{Sanchez_Lavega_2017}
S{\'a}nchez-Lavega, A., Rogers, J.~H., Orton, G.~S., Garc{\'\i}a-Melendo, E.,
  Legarreta, J., Colas, F., Dauvergne, J.~L., Hueso, R., Rojas, J.~F.,
  P{\'e}rez-Hoyos, S., and et~al. (2017).
\newblock A planetary-scale disturbance in the most intense {J}ovian
  atmospheric jet from {J}uno{C}am and ground-based observations.
\newblock {\em Geophysical Research Letters}, 44(10):4679--4686.

\bibitem[Satoh and Kawabata, 1994]{Satoh_1994}
Satoh, T. and Kawabata, K. (1994).
\newblock A change of upper cloud structure in jupiter's south equatorial belt
  during the 1989--1990 event.
\newblock {\em Journal of Geophysical Research: Planets}, 99(E4):8425--8440.

\bibitem[Scargle, 1982]{Scargle_1982}
Scargle, J.~D. (1982).
\newblock Studies in astronomical time series analysis {II}. statistical
  aspects of spectral analysis of unevenly sampled data.
\newblock {\em Astrophysical Journal}, 263:835--853.

\bibitem[Shure et~al., 1994]{Shure_1994}
Shure, M.~A., Toomey, D.~W., Rayner, J.~T., Onaka, P.~M., and Denault, A.~J.
  (1994).
\newblock {NSFCAM}: a new infrared array camera for the {NASA} {I}nfrared
  {T}elescope {F}acility.
\newblock {\em Instrumentation in Astronomy VIII}.

\bibitem[Simon-Miller et~al., 2001]{Simon_Miller_2001}
Simon-Miller, A.~A., Banfield, D., and Gierasch, P.~J. (2001).
\newblock Color and the vertical structure in jupiter's belts, zones, and
  weather systems.
\newblock {\em Icarus}, 154(2):459 -- 474.

\bibitem[Simon-Miller et~al., 2006]{Simon_Miller_2006}
Simon-Miller, A.~A., Conrath, B.~J., Gierasch, P.~J., Orton, G.~S., Achterberg,
  R.~K., Flasar, F.~M., and Fisher, B.~M. (2006).
\newblock Jupiter's atmospheric temperatures: From voyager iris to cassini
  cirs.
\newblock {\em Icarus}, 180(1):98 -- 112.

\bibitem[Simon-Miller and Gierasch, 2010]{Simon_Miller_2010}
Simon-Miller, A.~A. and Gierasch, P.~J. (2010).
\newblock On the long-term variability of {J}upiter's winds and brightness as
  observed from {H}ubble.
\newblock {\em Icarus}, 210(1):258 -- 269.

\bibitem[Simon-Miller et~al., 2007]{Simon_Miller_2007}
Simon-Miller, A.~A., Poston, B.~W., Orton, G.~S., and Fisher, B. (2007).
\newblock Wind variations in jupiter's equatorial atmosphere: A qqo
  counterpart?
\newblock {\em Icarus}, 186(1):192 -- 203.

\bibitem[Simon-Miller et~al., 2012]{Simon_Miller_2012}
Simon-Miller, A.~A., Rogers, J.~H., Gierasch, P.~J., Choi, D., Allison, M.~D.,
  Adamoli, G., and Mettig, H.-J. (2012).
\newblock Longitudinal variation and waves in {J}upiter's south equatorial wind
  jet.
\newblock {\em Icarus}, 218(2):817--830.

\bibitem[Sugiyama et~al., 2014]{Sugiyama_2014}
Sugiyama, K., Nakajima, K., Odaka, M., Kuramoto, K., and Hayashi, Y.-Y. (2014).
\newblock Numerical simulations of {J}upiter's moist convection layer:
  Structure and dynamics in statistically steady states.
\newblock {\em Icarus}, 229:71 -- 91.

\bibitem[Terrile and Westphal, 1977]{Terrile_1977}
Terrile, R.~J. and Westphal, J.~A. (1977).
\newblock Infrared imaging of jupiter in the 8--14-micrometer spectral region.
\newblock {\em Icarus}, 30(4):730--735.

\bibitem[Tollefson et~al., 2017]{Tollefson_2017}
Tollefson, J., Wong, M.~H., Pater, I.~d., Simon, A.~A., Orton, G.~S., Rogers,
  J.~H., Atreya, S.~K., Cosentino, R.~G., Januszewski, W.,
  Morales-Juber{\'\i}as, R., and et~al. (2017).
\newblock Changes in jupiter's zonal wind profile preceding and during the
  {J}uno mission.
\newblock {\em Icarus}, 296:163--178.

\bibitem[Toomey et~al., 1990]{Toomey_1990}
Toomey, D.~W., Shure, M.~A., Irwin, E.~M., and Ressler, M.~E. (1990).
\newblock Proto{CAM}: an innovative {IR} camera for astronomy.
\newblock {\em Instrumentation in Astronomy VII}.

\bibitem[Torrence and Compo, 1998]{Torrence_1998}
Torrence, C. and Compo, G.~P. (1998).
\newblock A practical guide to wavelet analysis.
\newblock {\em Bulletin of the American Meteorological society}, 79(1):61--78.

\bibitem[Wong et~al., 2004]{Wong_2004}
Wong, M.~H., Mahaffy, P.~R., Atreya, S.~K., Niemann, H.~B., and Owen, T.~C.
  (2004).
\newblock Updated {G}alileo probe mass spectrometer measurements of carbon,
  oxygen, nitrogen, and sulfur on {J}upiter.
\newblock {\em Icarus}, 171(1):153 -- 170.

\bibitem[Yanamandra-Fisher et~al., 1992]{Yanamandra_Fisher_1992}
Yanamandra-Fisher, P., Orton, G., and Friedson, J. (1992).
\newblock Time dependence of jupiter's tropospheric temperatures and cloud
  properties: The 1989 seb disturbance.
\newblock In {\em Bulletin of the American Astronomical Society}, volume~24,
  page 1039.

\bibitem[Zhang et~al., 2013]{Zhang_2013}
Zhang, X., West, R., Banfield, D., and Yung, Y. (2013).
\newblock Stratospheric aerosols on jupiter from cassini observations.
\newblock {\em Icarus}, 226(1):159 -- 171.

\end{thebibliography}

\section*{Acknowledgements}

Data are available upon reasonable request to Antu\~{n}ano and Fletcher, and are in the process of being archived with NASA's Planetary Data System. AA and LNF are supported by a European Research Council Consolidator Grant under the European Union's Horizon 2020 research and innovation program, grant agreement number 723890, at the University of Leicester. LNF is also supported by a Royal Society Research Fellowship. HR is supported by UK Science and Technology Facilities Council (STFC) Grant ST/N000749/1. PTD is supported by UK Science and Technology Facilities Council (STFC). A portion of this work was performed by GSO at the Jet Propulsion Laboratory, California Institute of Technology, under a contract with NASA. We are grateful to thank all those involved in the acquisition of these 5-$\mu$m data over many years, including but not limited to Kevin Baines, Jim Friedson, Tom Momary, Jose Luis Ortiz, John Spencer and Padma Yanamandra-Fisher. Part of this investigation is based on data acquired at the Infrared Telescope Facility, which is operated by the University of Hawaii under Cooperative Agreement no. NNX-08AE38A with the National Aeronautics and Space Administration, Science Mission Directorate, Planetary Astronomy Program. We recognize the significant cultural role of Maunakea within the indigenous Hawaiian community, and we appreciate the opportunity to conduct our Jupiter observations from this revered site.

\section*{Appendix}

 \begin{figure}[H]
	\centering
		\includegraphics[width=0.85\textwidth]{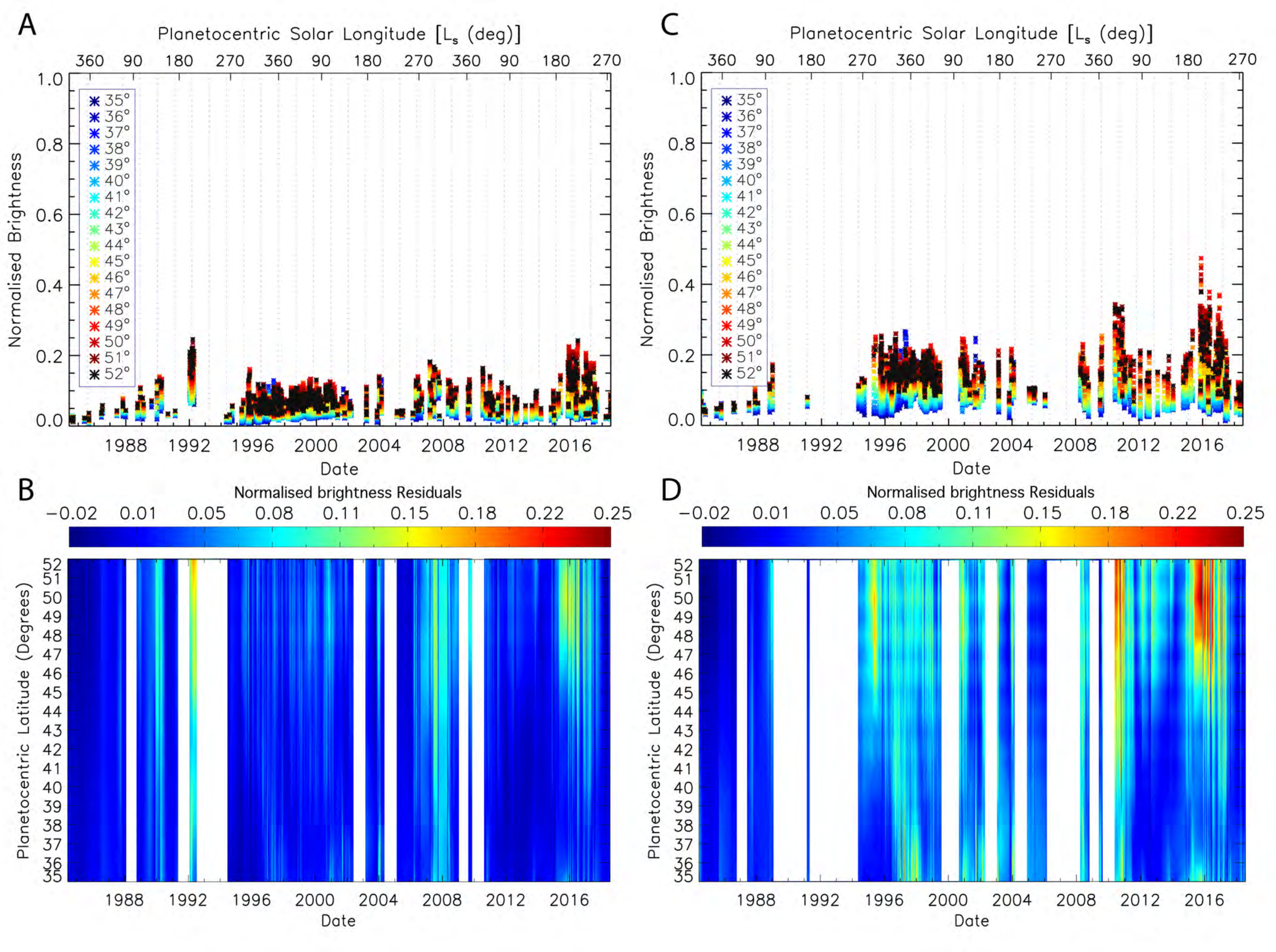}
	\begin{quote}
	\caption[S1]{Normalised average brightness as a function of time (A, C) and normalised brightness residuals computed relative to the temporal mean of the smallest 10$\%$ brightness values of each latitude (B, D) for the North North Temperate Region between 35$^\circ$ and 52$^\circ$ N latitude between 1984 and 2018, scaled at the South Temperate Zone (i.e. 24$^\circ$-28$^\circ$ S, A and B) and at the Equatorial Zone between $\pm$5$^\circ$ (C and D). Different colour crosses in A and C represent different latitudes and the vertical dotted blue lines in A and C represent Jupiter's opposition dates. The dotted blue vertical lines in A and C represent Jupiter opposition dates. The white regions in B and D indicate dates where no data is available during 9 months or longer.}
	\label{fig:Figure_S1}
	\index{Figure_s1}
	\end{quote}
\end{figure}

 \begin{figure}[H]
	\centering
		\includegraphics[width=0.85\textwidth]{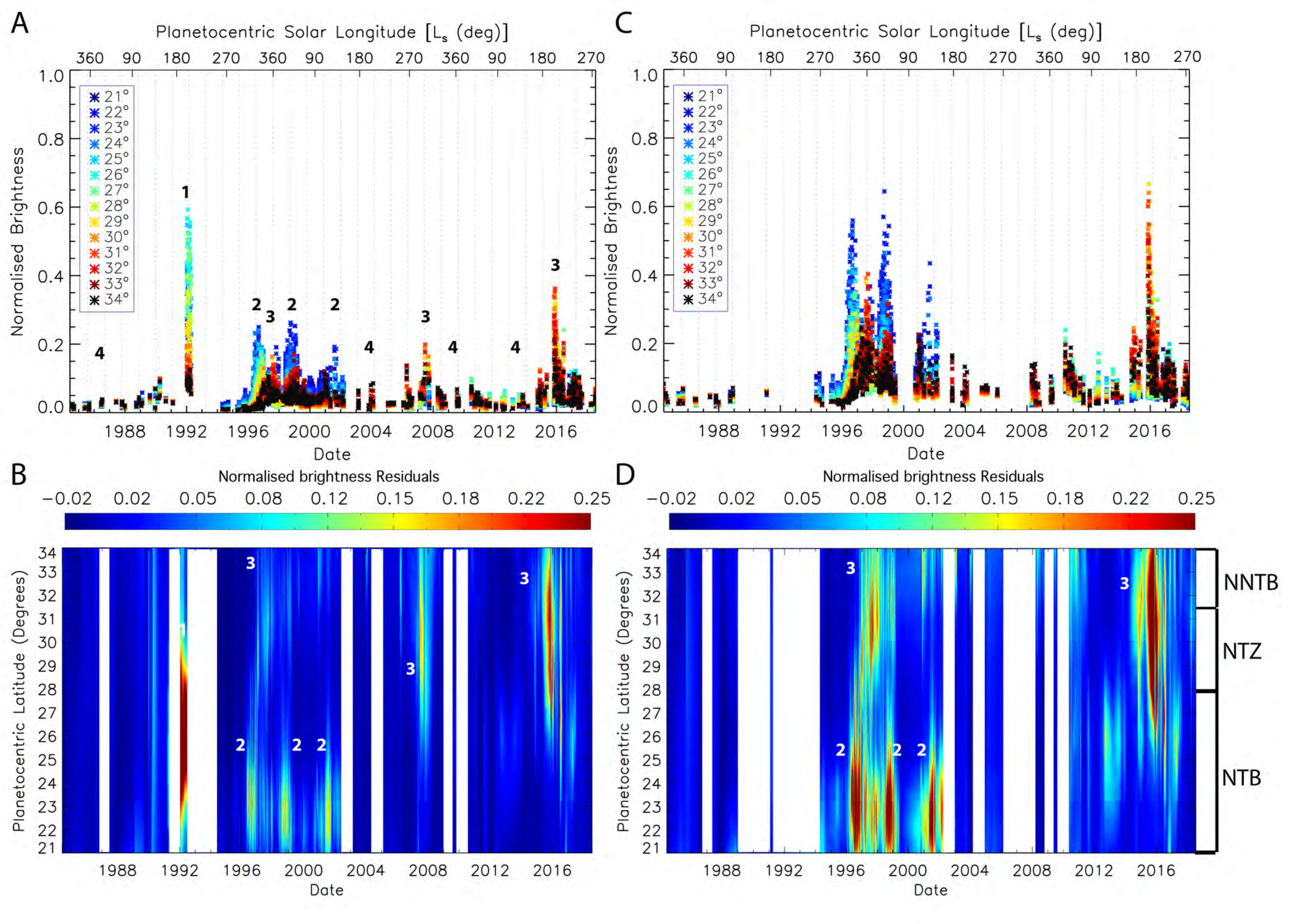}
	\begin{quote}
	\caption[S2]{Same as Figure \ref{fig:Figure_S1} but for the North Temperate Region between 21$^\circ$ and 34$^\circ$ N latitude. Numbers represent the changes described and labelled in the text (section 4.1). }
	\label{fig:Figure_S2}
	\index{Figure_S2}
	\end{quote}
\end{figure}

 \begin{figure}[H]
	\centering
		\includegraphics[width=0.85\textwidth]{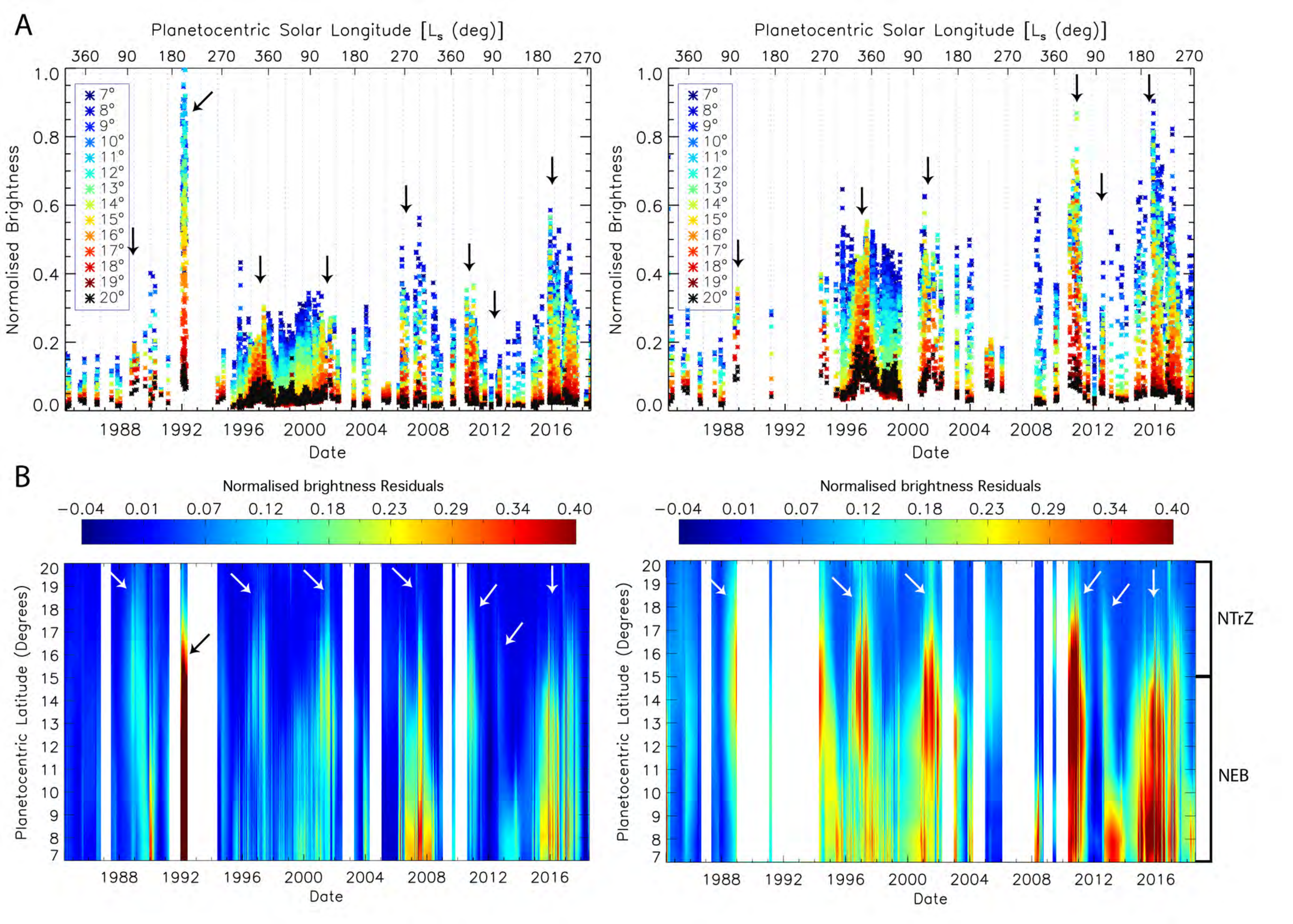}
	\begin{quote}
	\caption[S3]{Same as Figure \ref{fig:Figure_S1} but for the North Tropical Region between 7$^\circ$ and 20$^\circ$ N latitude. Arrows point the dates of NEB expansions (see section 4.2).}
	\label{fig:Figure_S3}
	\index{Figure_S3}
	\end{quote}
\end{figure}

 \begin{figure}[H]
	\centering
		\includegraphics[width=0.85\textwidth]{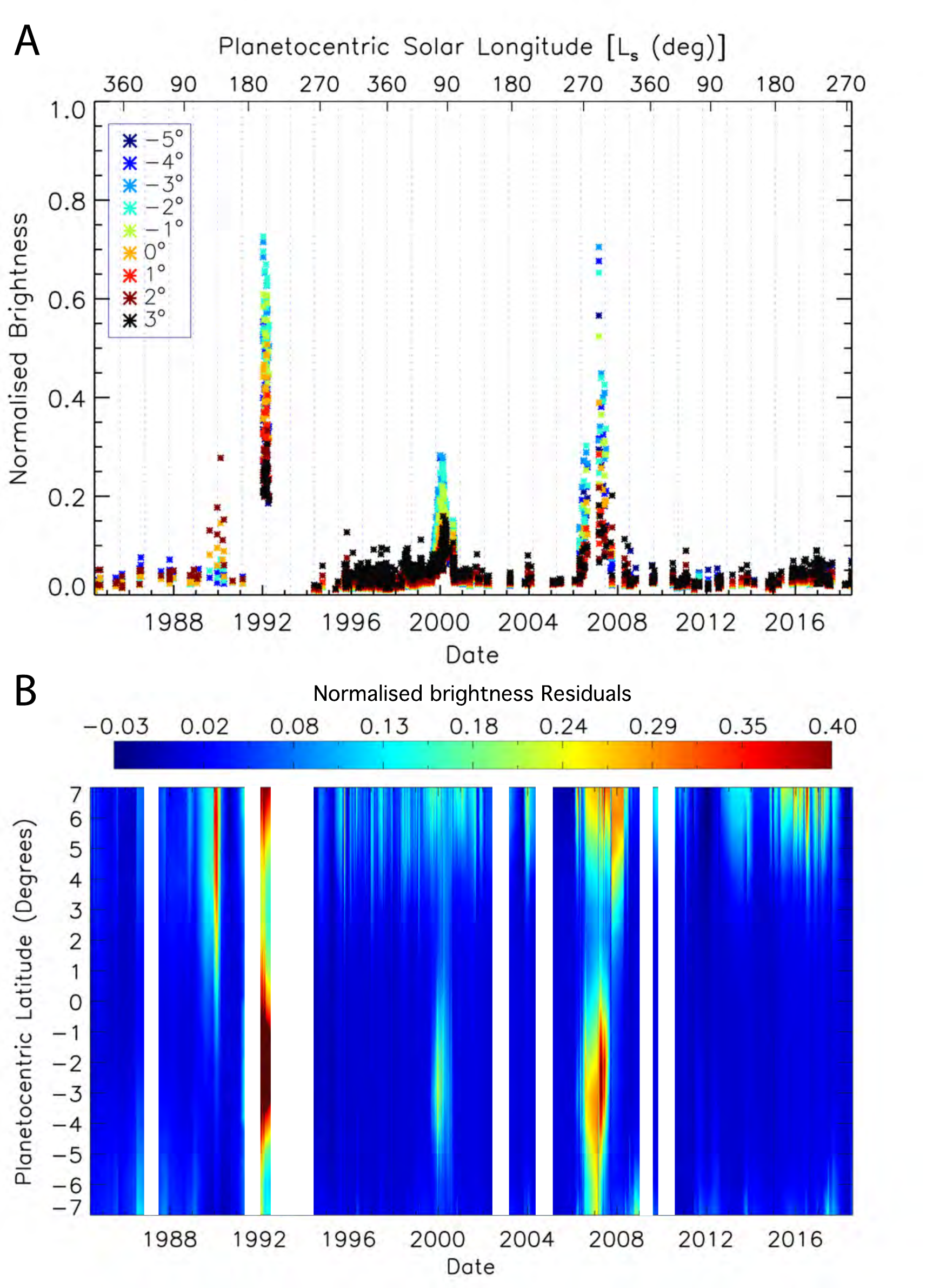}
	\begin{quote}
	\caption[S4]{Normalised average brightness as a function of time (A) and normalised brightness residuals computed respect to the temporal mean average brightness of each latitude (B) for the equatorial Zone between $\pm$7$^\circ$ N latitude between 1984 and 2018, scaled at the South Temperate Zone (i.e. 24$^\circ$-28$^\circ$ S). The dotted blue vertical lines in A represent Jupiter opposition dates. The white regions in B indicate dates where no data is available during 9 months or longer.}
	\label{fig:Figure_S4}
	\index{Figure_S4}
	\end{quote}
\end{figure}

\begin{figure}[H]
	\centering
		\includegraphics[width=0.85\textwidth]{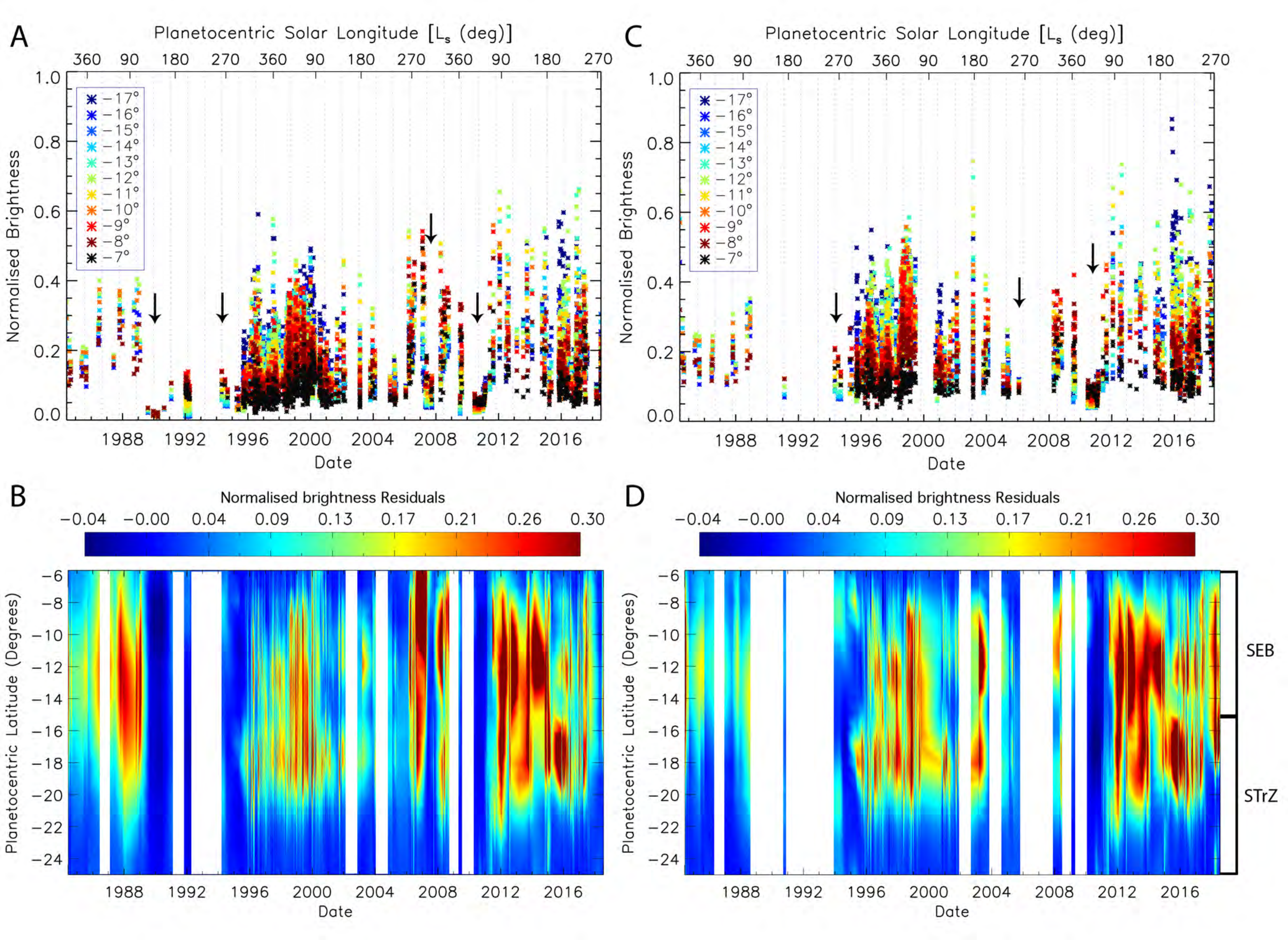}
	\begin{quote}
	\caption[S5]{Same as Figure \ref{fig:Figure_S1} but for the South Tropical Region between 6$^\circ$ S and 25$^\circ$ S latitude. Arrows point the dates of SEB fading (see section 4.4). }
	\label{fig:Figure_S5}
	\index{Figure_S5}
	\end{quote}
\end{figure}

\begin{figure}[H]
	\centering
		\includegraphics[width=0.85\textwidth]{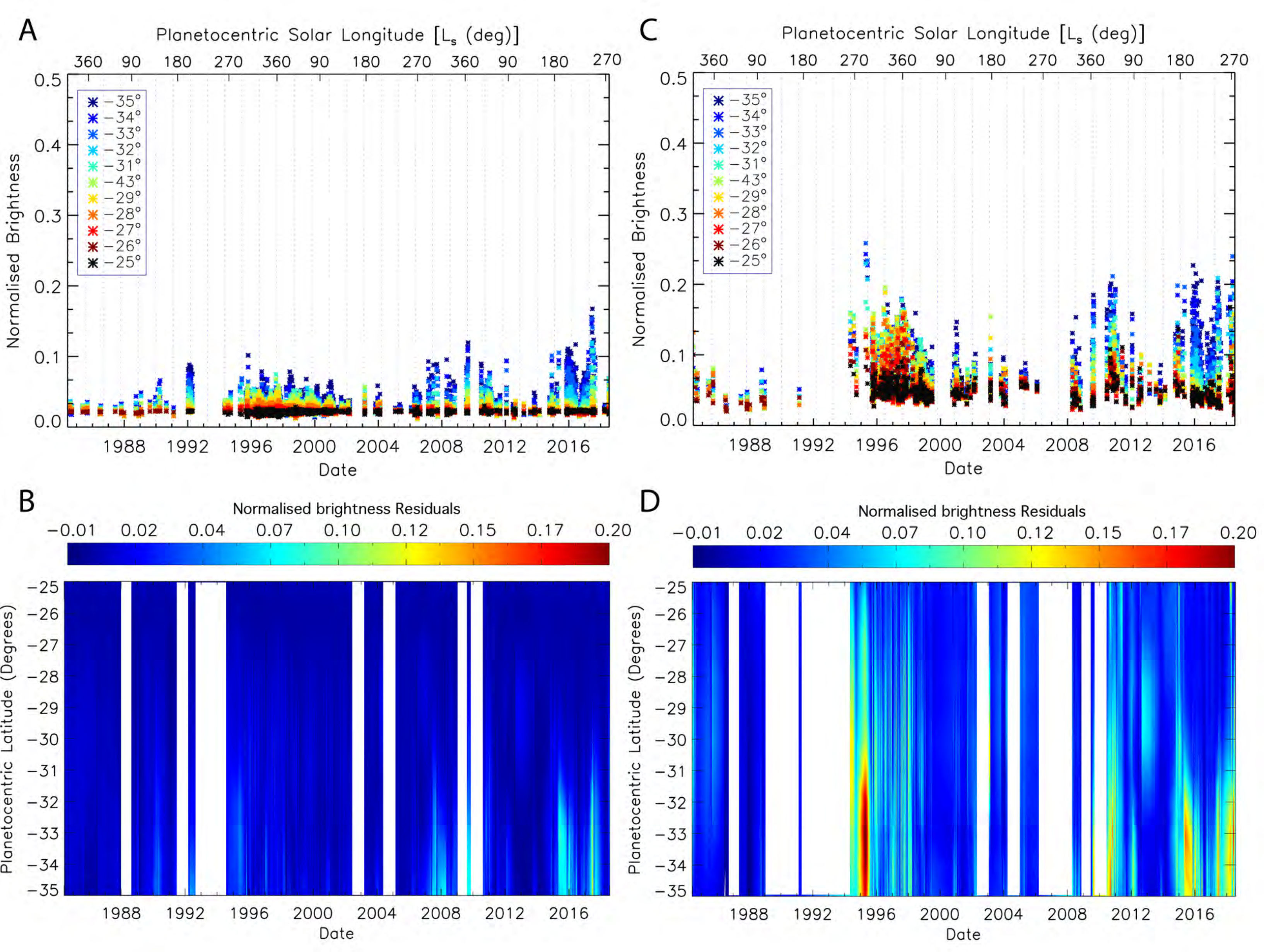}
	\begin{quote}
	\caption[S6]{Same as Figure \ref{fig:Figure_S1} but for the South Temperate Region between 25$^\circ$ S and 34$^\circ$ S latitude.}
	\label{fig:Figure_S6}
	\index{Figure_S6}
	\end{quote}
\end{figure}

\begin{figure}[H]
	\centering
		\includegraphics[width=0.85\textwidth]{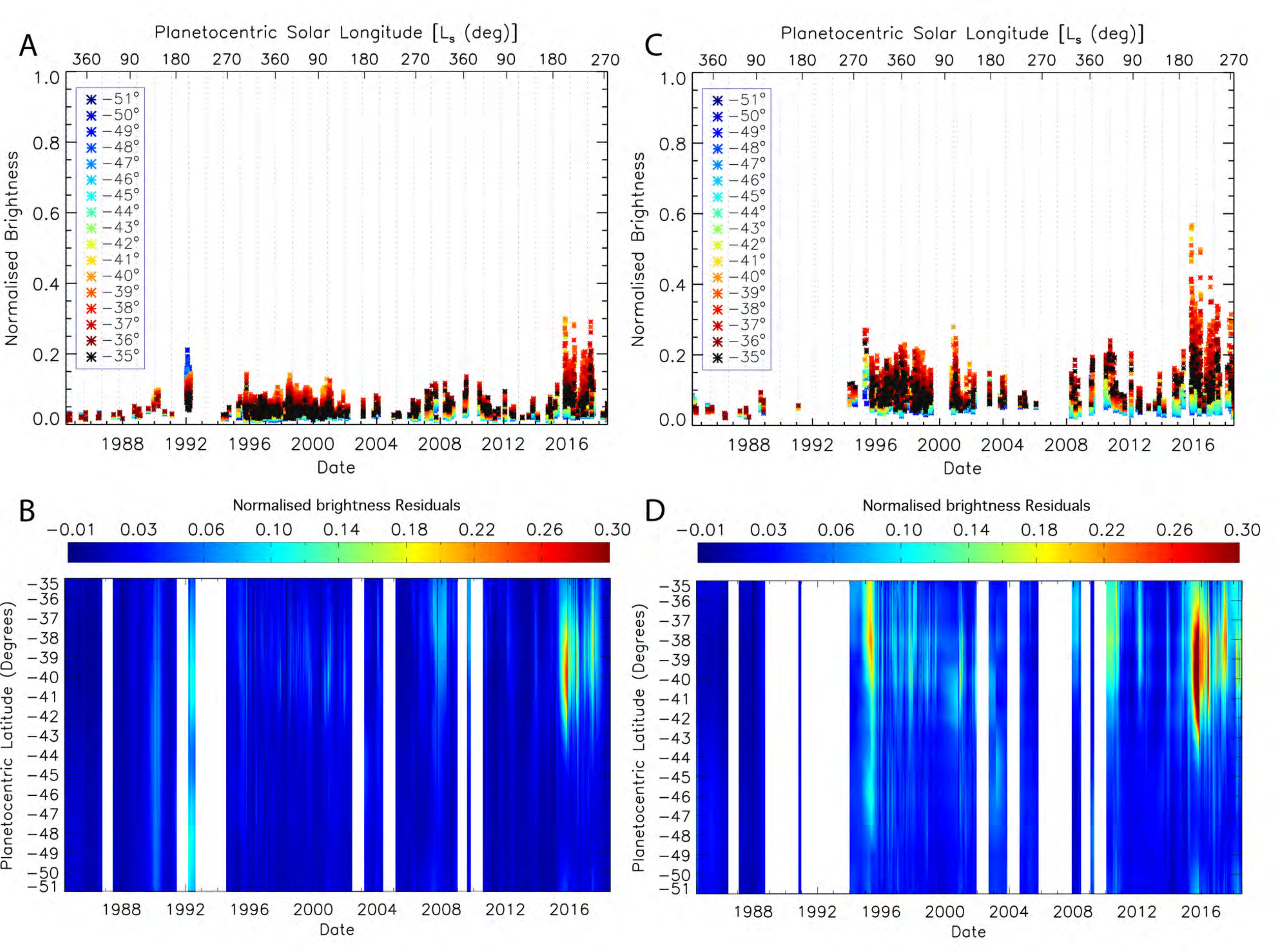}
	\begin{quote}
	\caption[S7]{Same as Figure \ref{fig:Figure_S1} but for the South South Temperate Region between 35$^\circ$ S and 51$^\circ$ S latitude.}
	\label{fig:Figure_S7}
	\index{Figure_S7}
	\end{quote}
\end{figure}

\begin{figure}[H]
	\centering
		\includegraphics[width=0.75\textwidth]{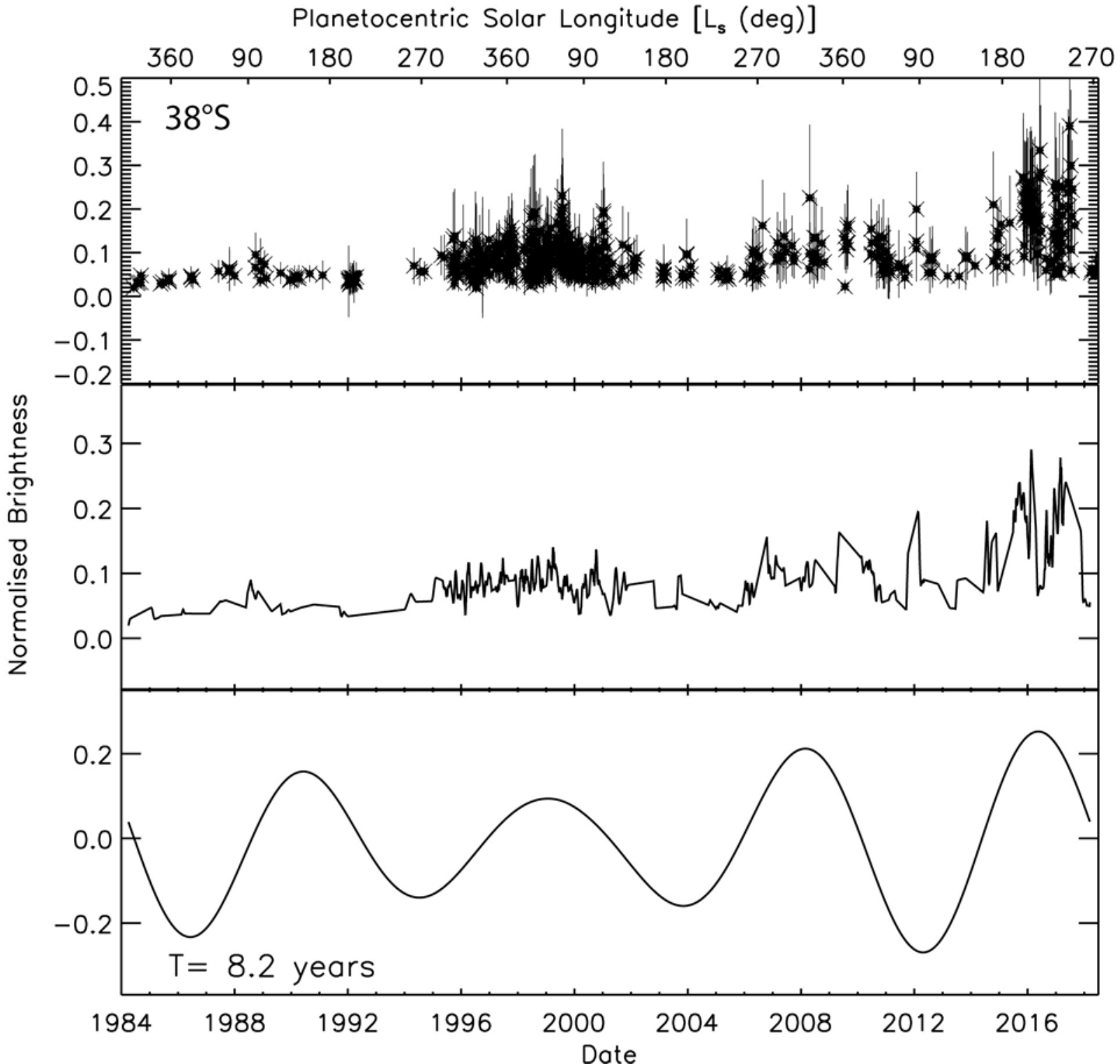}
	\begin{quote}
	\caption[S8]{5 $\mu$m brightness variability between 1984 and 2018 for 38$^\circ$ S (top), compared to the linear interpolation of the 5-$\mu$m brightness data (middle) and the convolution between the interpolated 5 $\mu$m and the selected wavelet (bottom), showing the correlation between the obtained periodicities and the real data.}
	\label{fig:Figure_S8}
	\index{Figure_S8}
	\end{quote}
\end{figure}

\end{document}